\documentclass[review]{elsarticle}

\usepackage{lineno,hyperref}
\usepackage{color}
\usepackage{amsfonts}
\usepackage{subfigure}

\makeatletter
\def\@xfootnote[#1]{%
  \protected@xdef\@thefnmark{#1}%
  \@footnotemark\@footnotetext}
\makeatother

\newcommand{\mm}[1]{\rm mm}

\newcommand{\mbf}[1]{\mathbf{#1}}
\newcommand{\mcal}[1]{\mathcal{#1}}

\newcommand{\beq}{\begin{equation}}
\newcommand{\eeq}{\end{equation}}
\newcommand{\bea}{\begin{eqnarray}}
\newcommand{\eea}{\end{eqnarray}}

\newcommand{\dt}{\Delta t}
\newcommand{\dx}{\Delta x}

\newcommand{\bit}{\begin{itemize}}
\newcommand{\eit}{\end{itemize}}
\newcommand{\ben}{\begin{enumerate}}
\newcommand{\een}{\end{enumerate}}

\newcommand{\bA}{\mathbf{A}}

\newcommand{\bF}{\mathbf{F}}
\newcommand{\bG}{\mathbf{G}}
\newcommand{\bH}{\mathbf{H}}

\newcommand{\bL}{\mathbf{L}}

\newcommand{\bR}{\mathbf{R}}

\newcommand{\bU}{\mathbf{U}}
\newcommand{\bV}{\mathbf{V}}
\newcommand{\bW}{\mathbf{W}}

\newcommand{\Z}{\mathbb{Z}}

\newcommand{\bl}{\mathbf{\ell}}

\newcommand{\mcalF}{\mathcal{F}}

\newcommand{\avg}{\overline}

\modulolinenumbers[5]

\journal{Journal of Computational Physics}

\begin{document}
\begin{frontmatter}

\title{The Piecewise Cubic Method (PCM) for Computational Fluid Dynamics}

\author[a]{Dongwook Lee\corref{mycorrespondingauthor}}
\cortext[mycorrespondingauthor]{Corresponding author}\ead{dlee79@ucsc.edu}
\author[a,c]{Hugues Faller}
\author[b]{Adam Reyes}

\address[a]{Applied Mathematics and Statistics, University of California, Santa Cruz, CA, U.S.A}
\address[b]{Department of Physics, University of California, Santa Cruz, CA, U.S.A}
\address[c]{D\'{e}partement de Physique, \'{E}cole Normale Sup\'{e}rieure, Paris, France}

\begin{abstract}
We present a new high-order finite volume reconstruction method for hyperbolic conservation laws.
The method is based on a piecewise cubic polynomial which provides its solutions 
a fifth-order accuracy in space. The spatially reconstructed solutions are evolved in time 
with a fourth-order accuracy by tracing the characteristics of the
cubic polynomials. As a result, our temporal update scheme provides a significantly simpler and
computationally more efficient approach in achieving
fourth order accuracy in time,
relative to the comparable fourth-order Runge-Kutta method.
We demonstrate that the solutions of PCM converges in fifth-order in solving 1D smooth flows
described by hyperbolic conservation laws.
We test the new scheme in a range of numerical experiments, including both
gas dynamics and magnetohydrodynamics applications in
multiple spatial dimensions.
\end{abstract}
\end{frontmatter}



\section{Introduction}
In this paper we are interested in solving multidimensional conservation laws of
the Euler equations and the ideal MHD equations, written as
\beq
\label{Eq:cons_law}
\frac{\partial \bU}{\partial t} + \nabla \cdot \mcalF(\bU) = 0,
\eeq
where $\bU$ is the vector of the conservative variables, and 
\beq
\mcalF(\bU)= [F(\bU), G(\bU), H(\bU)]^T = [\bF, \bG, \bH]^T
\eeq 
is the flux vector.

We present a new high-order piecewise cubic method (PCM) algorithm that is extended from 
the classical PPM and WENO schemes \cite{colella1984piecewise,jiang1996efficient}. 
These two algorithms, by far, have been extremely successful in various scientific fields 
where there are challenging computational needs for {\it{both}} high-order accuracy in smooth flows and 
well-resolved solutions in shock/discontinuous flows.
With the advent of high-performance computing (HPC) in recent years, such needs have been
more and more desired, and have become a necessary requirement in conducting
large scale, cutting edge simulations of gas dynamics and magnetohydrodynamics (MHD)
\cite{Dongarra2010, Dongarra2012, Subcommittee2014, Keyes2013}.

As observed in the success stories of the PPM and WENO methods,
discrete algorithms of data interpolation and reconstruction play 
a key role in numerical methods for PDE approximations 
\cite{LeVeque2002, leveque2007finite, Toro2009} 
within the broad framework of finite difference and finite volume discretization methods.
In view of this, computational improvements of such
interpolation and reconstruction schemes, particularly focused on the high-order property with
great shock-capturing capability,
take their positions at the center of HPC in modern computational fluid dynamics.

The properties of enhanced solution accuracy with lower numerical errors on a given grid resolution 
and faster convergence-to-solution rates are the key advantages in high-order schemes.
The advantage of using high-order methods in HPC is therefore clear:
one can obtain reproducible, admissible, and highly accurate numerical solutions
in a faster computational time at the expense of increased rate of floating point operations,
while at the same time, with the use of smaller size of grid resolutions.
This is by no means exceedingly efficient in high-performance computing (HPC),
in view of the fact that the increase of grid resolutions 
has a direct impact to an increase of memory footprints
which are bounded in all modern computing architecture.

In this regards, our goal in this paper is to lay down a mathematical foundation in 
designing a new high-order method using piecewise cubic polynomials.
We mainly focus on describing 
the detailed PCM algorithm in 1D finite volume framework for the scope of the current paper. 
For multidimensional problems,
we adopt the classical ``dimension-by-dimension'' approach for simplicity.
Although this approach has an advantage in its simplicity,
it unfortunately fails to retain the high-order accurate property of the 1D baseline algorithm
in multidimensional problems. 
Instead, it provides only a second-order accuracy in multidimensional nonlinear advection in finite volume method
due to the lack of accuracy in approximating a face-averaged flux function as a result of mis-using
an averaged quantity in place of a pointwise quantity, or vice versa
\cite{shu2009high,buchmuller2014improved,zhang2011order,mccorquodale2011high}. 
Although the baseline 1D PCM scheme can be extended to multiple spatial dimensions preserving its high-order accuracy
by following more sophisticated treatments 
\cite{shu2009high,buchmuller2014improved,zhang2011order,mccorquodale2011high},
more careful work is needed to carry out the detailed design, and this 
will be considered in our future research.


For a finite volume scheme in 1D we take the spatial average of
Eq.~(\ref{Eq:cons_law}) over the cell $I_i=[x_{i-1/2},x_{i+1/2}]$,
yielding a semi-discrete form,
\begin{equation}
  \label{Eq:1D_avg}
  \frac{\partial \avg{\bU}_i}{\partial t} = -\frac{1}{\Delta x}(\bF_{i+1/2}-\bF_{i-1/2})
\end{equation}
to get an equation for the evolution of the volume averaged variables,
$\avg{\bU}_i=\frac{1}{\dx}\int_{I_i}\bU(x,t)dx$. 
Typically to achieve high-order accuracy in time the
temporal update is done using a TVD Runge-Kutta scheme in
method-of-lines form \cite{shu1988tvd, mccorquodale2011high}. 
In this approach the high-order accuracy comes from taking
the {\it{multiple Euler stages}} of the RK time discretizations, which require
repeated reconstructions in a single time step, increasing the
computational cost.

Instead, as will be fully described in Section \ref{Sec:pcm}, 
one of the novel ideas in PCM is to employ
the simple  {\it{single stage}} predictor-corrector type temporal update formulation in which 
we take the time-average of Eq.~(\ref{Eq:1D_avg})
\begin{equation}
  \label{Eq:time_avg}
  \avg{\bU}^{n+1}_i = \avg{\bU}^n_i -\frac{\Delta t}{\Delta x}(\bF ^{n+1/2}_{i+1/2}-\bF^{n+1/2}_{i-1/2}).
\end{equation}
Here $\avg{\bU}^n_i = \avg{\bU}_i(t^n)$ is the volume averaged quantity at $t^n$,
and 
$\bF^{n+1/2}_{i \pm 1/2} = \frac{1}{\dt}\int_{t^n}^{t^{n+1}}\bF_{i\pm 1/2}(t)dt$ 
is the time average of the
interface flux from $t^n$ to $t^{n+1}$. 
In this way high-order in
space and time is accomplished with a single reconstruction in contrast to the
multiple Euler stages of the RK time discretizations, providing significant
benefits in computational efficiency per solution accuracy.
%

The organization of the paper is as follows: Section \ref{Sec:pcm}
describes the fifth-order accurate spatial reconstruction algorithm of PCM in 1D.
We highlight several desirable properties of the PCM scheme in terms of computational efficiency and solution accuracy.
Section \ref{Sec:chartracing} introduces the fourth-order accurate temporal updating scheme of PCM using a predictor-corrector type 
characteristic tracing, which is much simpler than the typical high-order Runge-Kutta ODE updates.
In Section \ref{Sec:pcmMultiD} we discuss how to extend the 1D scheme in Section  \ref{Sec:pcm}
to multiple spatial dimensions following the {\it{dimension-by-dimension approach}}
\footnote[$\ddagger$]{This approach is the same as the Class A approach in \cite{zhang2011order}, and should not 
be confused with the so-called dimensionally split approach. 
Our spatial integration scheme in this paper is directionally unsplit.} 
\cite{buchmuller2014improved,zhang2011order}.

In Section \ref{Sec:results} we test the PCM scheme on a wide spectrum of benchmark
problems in 1D, 2D and 3D, both for hydrodynamics and magnetohydrodynamics (MHD) applications.
We also compare the PCM solutions with PPM and WENO solutions in order to examine
numerical accuracy, capability and efficiency in both smooth and shock flow regimes.
We conclude our paper in Section \ref{Sec:conclusions} with a brief summary.
%
%

\section{The One-Dimensional Piecewise Cubic Method (PCM) Spatial Reconstruction}
\label{Sec:pcm}
In this section we describe a new PCM scheme in a one-dimensional 
finite volume formulation for  solving hyperbolic conservation laws of hydrodynamics and magnetohydrodynamics.
The new PCM scheme is a higher-order extension of Godunov's method \cite{godunov1959difference},
bearing its key components in the reconstruction algorithm on the relevant ideas of its high-order predecessors, 
the PPM scheme \cite{colella1984piecewise}, the WENO schemes 
\cite{jiang1996efficient,shi2002technique,borges2008improved,castro2011high},
and Hermite-WENO schemes \cite{qiu2004hermite, qiu2005hermite, zhu2009hermite, balsara2007sub}.

For the purpose of this section, 
we take the $3 \times 3$ hyperbolic system of conservation laws of the 1D Euler equations
\beq
\frac{\partial \bU}{\partial t} + \frac{\partial {F}(\bU)}{\partial x} = 0.
\label{Eq:Euler1D}
\eeq
The notations used are the vector of the conservative variables $\bU$ and fluxes $F(\bU)$, respectively, defined as
\beq
\bU = 
\left [
\begin{array}{c}
\rho \\ \rho u \\E
\end{array}
\right], \;\;\; 
%
F(\bU) = 
\left [
\begin{array}{c}
\rho u \\ \rho u^2 + p \\u(E+p)
\end{array}
\right].
\eeq
Here $\rho$ is the fluid density, $u$ is the fluid velocity in $x$-direction, 
and $E$ is the total energy as the sum of the internal energy $\epsilon={p}/({\gamma-1})$ and
the kinetic energy
obeying the ideal gas law,
\beq
E = \frac{p}{\gamma -1} + \frac{\rho u^2}{2},
\eeq 
where $p$ is the gas pressure, with the ratio of specific heats denoted as $\gamma$.
We denote the cells in $x$-direction by $I_i = [x_{i-1/2}, x_{i+1/2}]$. 
We assume our grid is configured on an equidistant uniform grid for simplicity.

In addition to the system of the Euler equations in the conserved variables $\bU$ as given in Eq. (\ref{Eq:Euler1D}),
we often use the two other equivalent system of equations each of which can be written either in
the primitive variables $\bV=[\rho,u,p]^T$ or in the characteristic variables $\bW$.
The characteristic variable $\bW$ is readily obtained from $\bU$ or $\bV$ by multiplying the
left eigenvectors corresponding to either $\bU$ or $\bV$, for instance, $\bW = \bL \bU$.
In the latter $\bL \; (\equiv \bR^{-1}$) represents the $3\times 3$ matrix obtained from diagonalizing
the coefficient matrix $\bA=\partial \mathcal{F}/\partial \bU = \bR \mathbf{\Lambda}\bR^{-1}$, 
whose rows are the $k$-th left eigenvectors $\bl^{(k)}$, $k=1, 2, 3$. The representation of the system
in $\bW$ furnishes a completely linearly decoupled 1D system of equations,
\beq
\frac{\partial \bW}{\partial t} + \mathbf{\Lambda}\frac{\partial \bW}{\partial x} = 0.
\label{Eq:Euler1D_char}
\eeq
The above system in the characteristic variables $\bW$ is therefore very handy for analyses,
and also is a preferred choice of variable in order to furnish numerical solutions 
more accurate than third-order especially with better non-oscillatory controls,
in particular when considering wave-by-wave propagations in a system of equations \cite{shu2009high}. 
In this reason the characteristic variable $\bW$ 
is taken as our default variable choice in the 1D PCM reconstruction steps via characteristic decompositions, albeit with
an increased computational cost, among the other two choices of the primitive $\bV$ or the conservative variables $\bU$.

The methodology presented below can be similarly applied to the 1D ideal MHD equations
(see for instance, \cite{brio1988upwind}).

\subsection{Piecewise Cubic Profile}
To begin with we first define a cubic polynomial $p_i(x)$ to approximate a 
$k$-th characteristic variable $q \in \bW$ on each interval $I_i$ by
\beq
p_i(x) = c_0 + c_1(x-x_i) + c_2(x-x_i)^2 + c_3(x-x_i)^3.
\label{Eq:pcm_poly}
\eeq
The goal is now to determine the four coefficients $c_i$, $i \in \Z$, $0 \le i \le 3$, which can be achieved 
by imposing the following four conditions:
\bea
\frac{1}{\dx}\int_{I_i}p_i(x) dx &=& \overline{q}_i,\label{Eq:pcm_cond1}\\
p_i(x_{i-1/2}) &=& q_{L,i},\label{Eq:pcm_cond2}\\
p_i(x_{i+1/2}) &=& q_{R,i},\label{Eq:pcm_cond3}\\
p_i'(x_{i}) &=& q'_{C,i},\label{Eq:pcm_cond4}
\eea
where 
\beq
\overline{q}_i = \frac{1}{\dx}\int_{I_i}q(x,t^n) dx
\eeq
is the cell-averaged quantity at $t^n$ on $I_i$ which is given as an initial condition;
\beq
q_{L,i} = q(x_{i-1/2},t^n) + \mcal{O}(\dx^p), \;\;\; q_{R,i} = q(x_{i+1/2},t^n) + \mcal{O}(\dx^p)
\eeq
are respectively the $p$-th order accurate pointwise left and the right Riemann states at $t^n$
on the cell $I_i$ that are unknown yet 
but are to be determined as described below; 
and lastly
\beq
q'_{C,i} = q'(x_i,t^n) + \mcal{O}(\dx^r)
\eeq
is the $r$-th order accurate approximation to the slope of $q$ at $t^n$ evaluated at $x_i$, 
which is again unknown at this point but is to be determined as below.

For the moment let us assume that all four quantities $\overline{q}_i, q_{L,i}, q_{R,i}$ and $q'_{C,i}$ are known.
It can be shown that the system of relations in Eqs. (\ref{Eq:pcm_cond1})  $\sim$ (\ref{Eq:pcm_cond4}) is equivalent to
a system given as:

\bea
c_0 + c_2 \frac{\dx^2}{12} &=& \overline{q}_i, \label{Eq:pcm_cond1a}\\
c_0 - c_1 \frac{\dx}{2} + c_2 \frac{\dx^2}{4} - c_3 \frac{\dx^3}{8} &=& q_{L,i}, \label{Eq:pcm_cond2b}\\
c_0 + c_1 \frac{\dx}{2} + c_2 \frac{\dx^2}{4} + c_3 \frac{\dx^3}{8} &=& q_{R,i}, \label{Eq:pcm_cond3b}\\
c_1 &=& q'_{C,i},\label{Eq:pcm_cond4b}
\eea
which, in turn, can be solved for all four $c_i$, $i=1, \dots, 4$. 
The final expressions of the coefficients in terms of $\overline{q}_i, q_{L,i}, q_{R,i}$, and $q'_{C,i}$ are given as:
\bea
c_0 &=& \frac{1}{4} \Big( - q_{R,i} - q_{L,i} + 6\overline{q}_i  \Big), \label{Eq:pcm_cond1c}\\
c_1 &=& q'_{C,i}, \label{Eq:pcm_cond2c}\\
c_2 &=& \frac{3}{\dx^2} \Big(q_{R,i} + q_{L,i} -2\overline{q}_i \Big),\label{Eq:pcm_cond3c}\\
c_3 &=& \frac{4}{\dx^3} \Big( q_{R,i} - q_{L,i} -\dx q'_{C,i} \Big).\label{Eq:pcm_cond4c}
\eea 

Therefore once we figure out the three unknowns, $q_{L,i}, q_{R,i}$, and $q'_{C,i}$,
the cubic profile $p_i(x)$ in Eq. (\ref{Eq:pcm_poly}) can be completely determined and
is ready to approximate $q$ on each $I_i$.

We now devote the following sections to describe how to determine $q_{L,i}, q_{R,i}$, and $q'_{C,i}$
so that the resulting PCM approximation to the variable $q$
lend its accuracy a fifth-order in space (Sections \ref{Sec:edges} and  \ref{Sec:qc_prime}) 
and a fourth-order in time (Section \ref{Sec:chartracing}).

\subsection{Reconstruction of the Riemann States $q_{L,i}$ and $q_{R,i}$}
\label{Sec:edges}
We follow the fifth-order finite volume WENO approach, 
either of the classical WENO-JS \cite{jiang1996efficient}
or WENO-Z \cite{borges2008improved,castro2011high},
in order to reconstruct the left and right Riemann states,
$q_{L,i}$ and $q_{R,i}$, on each cell $I_i$.
For the sake of providing a full self-contained description of the PCM scheme, we briefly present
the two WENO Riemann state reconstruction strategies here.

The main idea in WENO is to employ its reconstruction procedure
according to the nonlinear smoothness measurements
on three ENO sub-stencils, $S_\ell$, $\ell=1,2,3$, each of which consisting 
three cells $I_i$, $i=i_1,i_2,i_3$. Let us first define
\begin{eqnarray}
S_1&=&\{I_{i-2},I_{i-1},I_{i}\}, \\
S_2&=&\{I_{i-1},I_{i},I_{i+1}\}, \\
S_3&=&\{I_{i},I_{i+1},I_{i+2}\}.
\end{eqnarray}
Formulating the WENO reconstruction consists of the following three steps:
\paragraph{Step 1: ENO-Build}
We begin with building a second degree polynomial for each $\ell=1,2,3$,
\begin{equation}
\label{Eq:WENO_p(x)}
p_\ell(x)=\sum_{k=0}^2a_{\ell,k}(x-x_i)^k,
\end{equation}
each of which is defined on $S_\ell$, satisfying
\begin{equation}
\label{Eq:WENO_p(x)_constraints}
\frac{1}{\Delta x}\int_{I_k}p_\ell(x)\mathop{dx}=\bar{q}_k,
\end{equation}
for  $k=i+\ell-3, \dots, i+\ell-1$.
After a bit of algebra, we obtain the coefficients $a_{\ell,k}$ that determine $p_\ell(x)$ in
Eq. (\ref{Eq:WENO_p(x)}).

For $\ell=1$,
\begin{eqnarray}
\label{Eq:WENO_p(x)_coeffs_l=1}
&&a_{1,0}=\left(-\frac{1}{24}\bar q_{i-2} + \frac{1}{12}\bar q_{i-1}+\frac{23}{24}\bar q_{i}\right),\\
&&a_{1,1}=\left( \frac{1}{2}  \bar q_{i-2}                 -2 \bar q_{i-1} +\frac{3}{2}    \bar q_{i}\right)\frac{1}{\Delta x},\\
&&a_{1,2}=\left( \frac{1}{2}  \bar q_{i-2}                    -\bar q_{i-1}+\frac{1}{2}     \bar q_{i}\right)\frac{1}{\Delta x^2},
\end{eqnarray}

and for $\ell=2$,
\begin{eqnarray}
\label{Eq:WENO_p(x)_coeffs_l=2}
&&a_{2,0}=\left(-\frac{1}{24}\bar q_{i-1} + \frac{13}{12}\bar q_{i}-\frac{1}{24}\bar q_{i+1}\right),\\
&&a_{2,1}=\left(-\frac{1}{2}  \bar q_{i-1}                                      +\frac{1}{2}    \bar q_{i+1}\right)\frac{1}{\Delta x},\\
&&a_{2,2}=\left( \frac{1}{2}  \bar q_{i-1}                    -\bar q_{i}+\frac{1}{2}     \bar q_{i+1}\right)\frac{1}{\Delta x^2}.
\end{eqnarray}

Lastly, for $\ell=3$, we get
\begin{eqnarray}
\label{Eq:WENO_p(x)_coeffs_l=3}
&&a_{3,0}=\left( \frac{23}{24}\bar q_{i} + \frac{1}{12}\bar q_{i+1}  -\frac{1}{24}\bar q_{i+2}\right),\\
&&a_{3,1}=\left(-\frac{3}{2}   \bar q_{i}                   +2\bar q_{i+1} -\frac{1}{2}  \bar q_{i+2}\right)\frac{1}{\Delta x},\\
&&a_{3,2}=\left( \frac{1}{2}   \bar q_{i}                      -\bar q_{i+1} +\frac{1}{2} \bar q_{i+2}\right)\frac{1}{\Delta x^2}.
\end{eqnarray}
Then the three sets of left and right states follow as
\begin{equation}
\{p_1(x_{i-1/2}),p_2(x_{i-1/2}),p_3(x_{i-1/2})\},\mbox{ and } \{p_1(x_{i+1/2}),p_2(x_{i+1/2}),p_3(x_{i+1/2})\},
\end{equation}
where each of $p_\ell(x_{i\pm 1/2})$ is the ENO approximation and is given by, first for $p_1$,
\begin{eqnarray}
\label{Eq:WENO_p(x)_value_l=1}
&&p_{1}(x_{i-1/2})=-\frac{1}{6}\bar q_{i-2} + \frac{5}{6}\bar q_{i-1}  +\frac{1}{3}\bar q_{i},\\
&&p_{1}(x_{i+1/2})=\frac{1}{3}\bar q_{i-2} - \frac{7}{6}\bar q_{i-1}  +\frac{11}{6}\bar q_{i},
\end{eqnarray}
and for $p_2$,
\begin{eqnarray}
\label{Eq:WENO_p(x)_value_l=2}
&&p_{2}(x_{i-1/2})= \frac{1}{3}\bar q_{i-1} + \frac{5}{6}\bar q_{i}  -\frac{1}{6}\bar q_{i+1},\\
&&p_{2}(x_{i+1/2})=-\frac{1}{6}\bar q_{i-1} + \frac{5}{6}\bar q_{i}  +\frac{1}{3}\bar q_{i+1},
\end{eqnarray}
and finally for $p_3$,
\begin{eqnarray}
\label{Eq:WENO_p(x)_value_l=3}
&&p_{3}(x_{i-1/2})=\frac{11}{6}\bar q_{i} - \frac{7}{6}\bar q_{i+1}  +\frac{1}{3}\bar q_{i+2},\\
&&p_{3}(x_{i+1/2})=\frac{1}{3}\bar q_{i} + \frac{5}{6}\bar q_{i+1}  -\frac{1}{6}\bar q_{i+2}.
\end{eqnarray}

These left and right states respectively approximate the pointwise values at the interfaces $q(x_{i\pm 1/2})$
with third-order accuracy, i.e., $p_\ell(x_{i\pm 1/2})-q({x_{i\pm 1/2}})=O(\Delta x^3)$ (see \cite{jiang1996efficient})
by using the given cell-averaged quantities $\bar q_k$.
%
%
%
\paragraph{Step 2: Linear Constant Weights}
The next step is to construct a fourth-degree polynomial 
\begin{equation}
\label{Eq:WENO_phi}
\phi(x) = \sum_{k=0}^4 b_k(x-x_i)^k
\end{equation}
over the entire stencil $S=\cup_{\ell=1}^3 S_\ell$ which satisfies
\begin{equation}
\label{Eq:WENO_phi(x)_constraints}
\frac{1}{\Delta x}\int_{I_k}\phi(x)dx=\bar{q}_k,
\end{equation}
for $k=i-2,\dots,i+2$.
We can show that the coefficients $b_k$ are given as
\begin{eqnarray}
\label{Eq:WENO_phi_coeffs}
&&b_0=\frac{3}{640}\bar q_{i-2} -\frac{29}{480}\bar q_{i-1}+ \frac{1067}{960}\bar q_{i} -\frac{29}{480}\bar q_{i+1} +\frac{3}{640}\bar q_{i+2}, \label{Eq:b0}\\
&&b_1=\Big(\frac{5}{48}\bar q_{i-2} -\frac{17}{24}\bar q_{i-1}+ \frac{17}{24}\bar q_{i+1} -\frac{5}{48}\bar q_{i+2}\Big)\frac{1}{\dx}, \label{Eq:b1}\\
&&b_2=\Big(-\frac{1}{16}\bar q_{i-2} +\frac{3}{4}\bar q_{i-1} -\frac{11}{8}\bar q_{i} +\frac{3}{4}\bar q_{i+1}- \frac{1}{16}\bar q_{i+2}\Big)\frac{1}{\dx^2},\label{Eq:b2}\\
&&b_3=\Big(-\frac{1}{12}\bar q_{i-2} +\frac{1}{6}\bar q_{i-1} - \frac{1}{6}\bar q_{i+1} +\frac{1}{12}\bar q_{i+2}\Big)\frac{1}{\dx^3}, \label{Eq:b3}\\
&&b_4=\Big(\frac{1}{24}\bar q_{i-2} -\frac{1}{6}\bar q_{i-1}+ \frac{1}{4}\bar q_{i} -\frac{1}{6}\bar q_{i+1} +\frac{1}{24}\bar q_{i+2}\Big)\frac{1}{\dx^4}. \label{Eq:b4}
\end{eqnarray}
%
%
WENO uses $\phi(x)$ to determine three linear constant weights $\gamma_{\ell}^{\pm}$, 
$\ell=1,2,3$, with $\sum_\ell \gamma_\ell^{\pm}=1$, such that
\begin{equation}
\label{Eq:WENO_gammas}
\phi(x_{i\pm 1/2})=\sum_{\ell=1}^3 \gamma_\ell^{\pm} p_\ell(x_{i\pm 1/2}).
\end{equation}
The values on the left-hand side become
%
\begin{equation}
\label{Eq:WENO_phi_values_left}
\phi(x_{i-1/2})=-\frac{1}{20}\bar q_{i-2} +\frac{9}{20}\bar q_{i-1} + \frac{47}{60}\bar q_{i} -\frac{13}{60}\bar q_{i+1} +\frac{1}{30}\bar q_{i+2},
\end{equation}
and
\begin{equation}
\label{Eq:WENO_phi_values_right}
\phi(x_{i+1/2})=\frac{1}{30}\bar q_{i-2} -\frac{13}{60}\bar q_{i-1} + \frac{47}{60}\bar q_{i} +\frac{9}{20}\bar q_{i+1} -\frac{1}{20}\bar q_{i+2}.
\end{equation}
Now, by inspection, we obtain a set of linear weights for the left state,
\begin{equation}
\label{Eq:WENO_gamma_left}
\gamma_1^-=\frac{3}{10},\gamma_2^-=\frac{6}{10},\gamma_3^-=\frac{1}{10},
\end{equation}
and for the right state,
\begin{equation}
\label{Eq:WENO_gamma_right}
\gamma_1^+=\frac{1}{10},\gamma_2^+=\frac{6}{10},\gamma_3^+=\frac{3}{10}.
\end{equation}
\paragraph{Step 3: Nonlinear Weights}
The last step that imposes the non-oscillatory feature in the WENO approximations
is to measure how smoothly the three polynomials 
$p_\ell(x)$ vary on $I_i$. This is done by determining non-constant, nonlinear weights
$\omega_\ell^{\pm}$ (three of them for each $\pm$ state) 
that rely on the so-called smoothness indicator $\beta_\ell$, defined by
%
\begin{equation}
\label{Eq:WENO_beta}
\beta_\ell=\sum_{s=1}^{2}\left(\Delta x^{2s-1} 
\int_{I_i} \Big[\frac{d^s }{dx^s} p_\ell(x) \Big]^2 dx \right).
\end{equation}
With this definition $\beta_\ell$ becomes small for smooth flows,
and large for discontinuous flows. 


For explicit expressions, we attain
\begin{eqnarray}
\label{Eq:WENO_beta_explicit}
&&\beta_1=\frac{13}{12}\left(\bar q_{i-2}-2\bar q_{i-1}+\bar q_{i} \right)^2+\frac{1}{4}\left(\bar q_{i-2}-4\bar q_{i-1}+3\bar q_{i}\right)^2,\\
&&\beta_2=\frac{13}{12}\left(\bar q_{i-1}-2\bar q_{i}+\bar q_{i+1} \right)^2+\frac{1}{4}\left(\bar q_{i-1}-\bar q_{i+1}\right)^2,\\
&&\beta_3=\frac{13}{12}\left(\bar q_{i}-2\bar q_{i+1}+\bar q_{i+2} \right)^2+\frac{1}{4}\left(3\bar q_{i}-4\bar q_{i+1}+\bar q_{i+2}\right)^2.
\end{eqnarray}
Equipped with these $\beta_\ell$, the nonlinear weights $\omega_\ell^{\pm}\ge 0$ are defined as:
\begin{itemize}
\item For WENO-JS:
\begin{equation}
\label{Eq:WENO5_omega}
\omega_\ell^{\pm} = \frac{\tilde{\omega}_\ell^{\pm}}{ \sum_{s}\tilde{\omega}_s^{\pm}}, \mbox{ where }
\tilde{\omega}_\ell^{\pm} = \frac{\gamma_\ell^{\pm}}{(\epsilon + \beta_\ell)^m},
\end{equation}

\item For WENO-Z:
\begin{equation}
\label{Eq:WENOZ_omega}
\omega_\ell^{\pm} = \frac{\tilde{\omega}_\ell^{\pm}}{ \sum_{s}\tilde{\omega}_s^{\pm}}, \mbox{ where }
\tilde{\omega}_\ell^{\pm} = {\gamma_\ell^{\pm}}\Biggl(1+\Bigl(\frac{|\beta_0-\beta_2|}{\epsilon + \beta_\ell}\Bigr)^m\Biggr).
\end{equation}
\end{itemize}
Here $\epsilon$ is any arbitrarily small positive number that prevents division by zero, for which
we choose $\epsilon=10^{-36}$. One of the classical choice of $\epsilon$ in many WENO literatures
is found to be $\epsilon=10^{-6}$ \cite{jiang1996efficient, shi2002technique}; however, it was
suggested in \cite{borges2008improved} that $\epsilon$ should be chosen to be much smaller
in order to force this parameter to play only its original role of avoiding division by zero
in the definitions of the weights, Eqs. (\ref{Eq:WENO5_omega}) and (\ref{Eq:WENOZ_omega}).


Another closely related point of discussion is with the value of $m$, the power in the denominators in Eqs. 
(\ref{Eq:WENO5_omega}) and (\ref{Eq:WENOZ_omega}).
The parameter $m$ determines the rate of
changes in $\beta_\ell$, and most of the WENO
literatures use $m=2$. However, we observe that using $m=1$ resolves discontinuities
sharper in most of our numerical simulations without exhibiting any numerical instability, 
so became the default value in our implementation.
For more detailed discussions on the choices of $\epsilon$ and $m$, 
see \cite{borges2008improved,castro2011high}.

Using these nonlinear weights, we complete the WENO reconstruction procedure of producing the
fifth-order spatially accurate, non-oscillatorily reconstructed values at each cell interface at each time step $t^n$
\cite{jiang1996efficient,shu2009high},
\begin{equation}
\label{Eq:WENO_final_states}
q_{L;R,i} = \sum_{\ell=1}^{3}\omega_\ell^{\pm} p_\ell(x_{i\pm 1/2}).
\end{equation}

\subsection{Reconstruction of the Derivative $q'_{C,i}$}
\label{Sec:qc_prime}

The spatial reconstruction part of PCM proceeds to the next final step to obtain the derivative $q'_{C,i}$
in Eq. (\ref{Eq:pcm_cond4}). The approach is again to take the WENO-type reconstruction as before, 
but this time, to approximate a first derivative of a function \cite{shu2009high}, i.e., $q'(x_i,t^n)$.

For this, we might consider using the same ENO-build strategy in Section \ref{Sec:edges} in which the three 
second degree ENO polynomials in Eq. (\ref{Eq:WENO_p(x)}) are constructed over the five-point stencil 
$S=\cup_{\ell=1}^3 S_\ell$. However, this setup will provide only a third-order accurate approximation $q'_{C,i}$
to the exact derivative $q'(x_i)$. 
To see this, we first observe that the smoothness indicators $\beta_\ell$ with 
this setup will be including only a single term,
\begin{equation}
\label{Eq:WENO_beta_1}
\beta_\ell=\Delta x^{3} 
\int_{I_i} \Big[p''_\ell(x) \Big]^2 dx, \;\;\; \ell=1, 2, 3.
\end{equation}
Through a Taylor expansion analysis on Eq. (\ref{Eq:WENO_beta_1}) we see
\begin{equation}
\label{Eq:WENO_beta_order}
\beta_\ell=D(1+\mcal{O}(\dx)),
\end{equation}
where $D=(q''\dx^2)^2$ is a nonzero quantity independent of $\ell$ but may depend on $\dx$, assuming
$q'' \ne 0$ on $S$.
This results in a set of three nonlinear weights $\omega_\ell$, $\ell=1,2,3$, obtained either by 
Eq. (\ref{Eq:WENO5_omega}) or Eq. (\ref{Eq:WENOZ_omega}),
satisfying
\beq
\label{Eq:WENO_omega_order}
\omega_\ell = \gamma_\ell + \mcal{O}(\dx),
\eeq
where the linear constant weights $\gamma_\ell$ are assumed to exist, when $q'(x,t^n)$ is smooth in $S$, such that
\beq
q'_{C,i} =  \sum_{\ell=1}^3\gamma_\ell p_\ell'(x_{i}) = q'(x_i,t^n) + \mcal{O}(\dx^3).
\eeq
This finally implies
the accuracy of $q'_{C,i}$ is found out to be third-order,
\begin{equation}
\label{Eq:qC_prime_third_order}
q'_{C,i} = \sum_{\ell=1}^3\omega_\ell p_\ell'(x_{i}) = q'(x_i,t^n) + \mcal{O}(\dx^3),
\end{equation}
because
\bea
\label{Eq:qC_prime_third_order_derivation}
 \sum_{\ell=1}^3\omega_\ell p_\ell'(x_{i}) -  \sum_{\ell=1}^3\gamma_\ell p_\ell'(x_{i}) 
 &=&\sum_{\ell=1}^3\big(\omega_\ell- \gamma_\ell\big) \Big(p_\ell'(x_{i}) -q'(x_i,t^n)\Big)\nonumber\\
 &=&\sum_{\ell=1}^3\mcal{O}(\dx) \mcal{O}(\dx^2) = \mcal{O}(\dx^3).
\eea
In the last equality, we used the fact that, for each $\ell$, $p'_\ell(x)$ is only a first degree polynomial
which is accurate up to second-order when approximating $q'(x,t^n)$.

For this reason, we want a better strategy to obtain
an approximation $q'_{C,i}$ at least fourth-order accurate in order that
the overall nominal accuracy of the 1D PCM scheme achieves {\it{at least}} fourth-order accurate 
in both space and time. 

\paragraph{Step 1: PPM-Build}
An alternate strategy for this goal therefore would be to use 
a set of third degree polynomials instead.
This can be designed using the two third degree polynomials, 
$\phi_{\pm}(x)$, from the PPM algorithm \cite{colella1984piecewise},
\beq
\label{Eq:PPM_phi_all}
\phi_{\pm}(x)=\sum_{k=0}^3 a_k^{\pm}(x-x_{i\pm 1/2})^k.
\eeq
Following the description of PPM, we carry out to determine the coefficients $a_k^{\pm}$ by
imposing the following constraints on $\phi_{\pm}(x)$ 
that are essential to keeping the volume averages on each cell $I_i$:
\begin{equation}
\label{Eq:PPM_phi_minum}
\frac{1}{\Delta x}\int_{I_k} \phi_{-}(x)\mathop{dx}= \bar q_k^n, \mbox{ for } i-2\le k \le i+1,
\end{equation}
and
\begin{equation}
\label{Eq:PPM_phi_plus}
\frac{1}{\Delta x}\int_{I_k} \phi_{+}(x)\mathop{dx}= \bar q_k^n, \mbox{ for } i-1\le k \le i+2.
\end{equation}
After a bit of algebra we obtain the coefficients $a_k^{\pm}$, with $s=1$ for $a_k^{+}$, while
$s=0$ for  $a_k^{-}$:
\begin{equation}
\label{Eq:PPM_a0}
a_0^{\pm}=\frac{1}{12}\Big(-\bar q_{i-2+s} +7\bar q_{i-1+s} + 7\bar q_{i+s} -\bar q_{i+1+s}  \Big),
\end{equation}
\begin{equation}
\label{Eq:PPM_a1}
a_1^{\pm}=\frac{1}{12\Delta x}\Big(\bar q_{i-2+s} -15\bar q_{i-1+s} +15\bar q_{i+s} -\bar q_{i+1+s}  \Big),
\end{equation}
\begin{equation}
\label{Eq:PPM_a2}
a_2^{\pm}=\frac{1}{4\Delta x^2}\Big(\bar q_{i-2+s} -\bar q_{i-1+s} -\bar q_{i+s} +\bar q_{i+1+s}  \Big),
\end{equation}
\begin{equation}
\label{Eq:PPM_a3}
a_3^{\pm}=\frac{1}{6\Delta x^3}\Big(-\bar q_{i-2+s} +3\bar q_{i-1+s} -3\bar q_{i+s} +\bar q_{i+1+s}  \Big).
\end{equation}

\paragraph{Step 2: Linear Constant Weights}
Now that the polynomials are determined over the stencil $S = \cup_{\ell=1}^3 S_\ell$, we use their first derivatives
$\phi'_{\pm}$
to obtain a convex combination with two linear weights $\gamma_-$ and $\gamma_+$,
\beq
\label{Eq:PCM_ucPrime_linear}
q'_{C,i} = \gamma_- \phi'_{-}(x_i) + \gamma_+ \phi'_{+}(x_i).
\eeq
The two linear weights can be determined by comparing Eq. (\ref{Eq:PCM_ucPrime_linear})
with $\phi'(x_i)$ in Eq. (\ref{Eq:WENO_phi}), 
\beq
\gamma_- \phi'_{-}(x_i) + \gamma_+ \phi'_{+}(x_i) = \phi'(x_i)
\eeq
This gives us 
\beq
\gamma_- \Big(a_1^- + a_2^- \dx + 3 a_3^- \frac{\dx^2}{4} \Big) + 
\gamma_+ \Big(a_1^+ - a_2^+ \dx + 3 a_3^+ \frac{\dx^2}{4} \Big) = b_1
\eeq
where $b_1$ is defined in Eq. (\ref{Eq:b1}).
By inspection, we obtain
\beq
\gamma_- = \gamma_+ = \frac{1}{2}.
\eeq

\paragraph{Step 3: Nonlinear Weights}
The smoothness indicators $\beta_\pm$ are now constructed using $\phi_{\pm}(x)$ as
\beq
\label{Eq:PCM_beta}
\beta_\pm=\sum_{s=2}^{3}\left(\Delta x^{2s-1} 
\int_{I_i} \Big[\frac{d^s }{dx^s} \phi_\pm(x) \Big]^2 dx \right).
\eeq
They can be written explicitly as
\bea
\beta_- 
&=& 4 (a_2^-)^2 \dx^4 + 12 (a_2^-) (a_3^-) \dx^5 + 48 (a_3^-)^2 \dx^6\\
&=& \frac{1}{4}\Big( \bar q_{i-2} -\bar q_{i-1} -\bar q_{i} +\bar q_{i+1} \Big)^2\nonumber\\
&+& \frac{1}{2}\Big( \bar q_{i-2} -\bar q_{i-1} -\bar q_{i} +\bar q_{i+1} \Big)\Big( -\bar q_{i-2} +3\bar q_{i-1} -3\bar q_{i} +\bar q_{i+1} \Big)\nonumber\\
&+&\frac{4}{3}\Big( -\bar q_{i-2} +3\bar q_{i-1} -3\bar q_{i} +\bar q_{i+1} \Big)^2,
\eea
and
\bea
\beta_+ 
&=& 4 (a_2^+)^2 \dx^4 - 12 (a_2^+) (a_3^+) \dx^5 + 48 (a_3^+)^2 \dx^6\\
&=& \frac{1}{4}\Big( \bar q_{i-1} -\bar q_{i} -\bar q_{i+1} +\bar q_{i+2} \Big)^2\nonumber\\
&+& \frac{1}{2}\Big( \bar q_{i-1} -\bar q_{i} -\bar q_{i+1} +\bar q_{i+2} \Big)\Big( -\bar q_{i-1} +3\bar q_{i} -3\bar q_{i+1} +\bar q_{i+2} \Big)\nonumber\\
&+&\frac{4}{3}\Big( -\bar q_{i-1} +3\bar q_{i} -3\bar q_{i+1} +\bar q_{i+2} \Big)^2.
\eea
Upon conducting Taylor series expansion analysis on $\beta_\pm$, we can see that
\beq
\label{Eq:beta_pcm}
\beta_\pm = D (1+\mcal{O}(\dx)),
\eeq
where $D=(q''\dx^2)^2$ is a nonzero quantity independent of $\pm$ but might depend on $\dx$, assuming
$q'' \ne 0$ on $S$.

The remaining procedure is to obtain the two nonlinear weights $\omega_\pm$
in the similar way done in the edge reconstructions
in Eqs. (\ref{Eq:WENO5_omega}) -- (\ref{Eq:WENOZ_omega}),
\begin{itemize}
\item For WENO-JS:
\begin{equation}
\label{Eq:PCM_WENO5_omega}
\omega_{\pm} = \frac{\tilde{\omega}_{\pm}}{\tilde{\omega}_{-} + \tilde{\omega}_{+}}, \mbox{ where }
\tilde{\omega}_{\pm} = \frac{\gamma_{\pm}}{(\epsilon + \beta_\pm)^m},
\end{equation}

\item For WENO-Z:
\begin{equation}
\label{Eq:PCM_WENOZ_omega}
\omega_{\pm} = \frac{\tilde{\omega}_{\pm}}{\tilde{\omega}_{-} + \tilde{\omega}_{+}}, \mbox{ where }
\tilde{\omega}_{\pm} = {\gamma_{\pm}}\Biggl(1+\Bigl(\frac{|\beta_+-\beta_-|}{\epsilon + \beta_\pm}\Bigr)^m\Biggr).
\end{equation}
\end{itemize}

The final representation of the approximation $q'_{C,i}$ becomes
\beq
q'_{C,i} = \omega_- \phi'_-(x_i) + \omega_+ \phi'_+(x_i).
\eeq
Let us now verify that this approximation is fourth-order accurate after all, that is,
\beq
\label{Eq:PCM_ucPrime_final}
q'_{C,i} = \omega_- \phi_-(x_i) + \omega_+ \phi_+(x_i) = q'(x_i,t^n) + \mcal{O}(\dx^4).
\eeq
Similarly as before, using Eqs. (\ref{Eq:beta_pcm}),  (\ref{Eq:PCM_WENO5_omega}), and (\ref{Eq:PCM_WENOZ_omega}), 
we can see that, with the help of the binomial series expansion,
\beq
\label{Eq:PCM_omega_order}
\omega_\pm = \gamma_\pm + \mcal{O}(\dx).
\eeq
Therefore, the desired accuracy claimed in Eq. (\ref{Eq:PCM_ucPrime_final}) is readily verified by
repeating the similar relationship in Eq. (\ref{Eq:qC_prime_third_order_derivation}):
\bea
\label{Eq:qC_prime_fourth_order_derivation}
 \sum_{\ell=-,+}\omega_\ell\phi_\ell'(x_{i}) -  \sum_{\ell=-,+}\gamma_\ell \phi_\ell'(x_{i}) 
 &=&\sum_{\ell=-,+}\big(\omega_\ell- \gamma_\ell\big) \Big(\phi_\ell'(x_{i}) -q'(x_i,t^n)\Big)\nonumber\\
 &=&\sum_{\ell=-,+}\mcal{O}(\dx) \mcal{O}(\dx^3) = \mcal{O}(\dx^4).
\eea
Comparing Eq. (\ref{Eq:qC_prime_fourth_order_derivation}) with Eq. (\ref{Eq:qC_prime_third_order_derivation}), 
we now see that it is fourth-order accurate due to the improved third-order accuracy
in calculating $\phi_\ell'(x_{i}) -q'(x_i,t^n)$.
This is a result of using
the third degree PPM polynomials $\phi_\pm(x)$, with which $q'(x,t^n)$ can be accurately
approximated by the second degree polynomials $\phi'_\pm(x)$ up to third-order.

We notice that there are some cases when 
$q'_{C,i}$ in Eq. (\ref{Eq:PCM_ucPrime_final}) 
differs from $({q_{R,i}-q_{L,i}})/{\dx}$ by an order of magnitude.
This may happen in two different cases:
(i) they both are very small, approximating zero slopes, or
(ii) one is larger (or smaller) than the other in regions where $q'(x,t^n)$ becomes singular at 
discontinuities or kinks at which the derivatives $q'(x_i,t^n)$ are not well defined.
The first is simply due to the level of machine accuracy (e.g., one being $10^{-16}$ and the other
being $10^{-15}$) and does not affect the overall spatial approximation of PCM.
However, the latter needs to be taken with some spacial care because,
at those singular points, any over/under predictions of $q'_{C,i}$ will result in undesirable
oscillations, which can yield negative states in approximating density or pressure.
To prevent this situation, we limit both $q'_{C,i}$ and $({q_{R,i}-q_{L,i}})/{\dx}$ using
the MC slope limiter when it is detected there is an order of magnitude difference between the two, 
that is,
%
\beq
\label{Eq:qPrime_flattening}
c_1 = \mbox{MC\_limiter}\left(q'_{C,i}, \frac{q_{R,i}-q_{L,i}}{\dx}\right)
\eeq
if $\Big|\frac{\dx q'_{C,i}}{q_{R,i} - q_{L,i}}\Big|> 10$ or $\Big|\frac{\dx q'_{C,i}}{q_{R,i} - q_{L,i}}\Big| < 0.1$.
This limiting does not get activated on smooth flows in general and does not 
affect the overall fifth-order accuracy of PCM (see Section \ref{Sec:convegence_performance}).
However, in case the limiting is fully turned on and is
activated on smooth flows, the accuracy is reduced to
third-order because the solution accuracy is limited by
the third-order dissipation of the MC limiter in smooth regions (see \cite{toth2008hall}).
In what follows we call Eq. (\ref{Eq:qPrime_flattening}) the PCM flattening.

On a separate note, the PCM scheme reduces to
a PPM-like algorithm when setting
\beq
\label{Eq:qPrime_PPM}
q'_{C,i} = \frac{q_{R,i}-q_{L,i}}{\dx},
\eeq
because, in this case, we have $c_3 = 0$ in Eq. (\ref{Eq:pcm_cond4c}) 
so that $p_i(x)$ in Eq.  (\ref{Eq:pcm_poly}) loses
its highest term, becoming  a piecewise parabolic polynomial at most.
The solution accuracy becomes third-order accurate, similar to the solution
accuracy of PPM on 1D smooth flows.

This completes the PCM spatial reconstruction steps that provide
the fifth-order accurate Riemann states $q_{L;R,i}$, and the fourth-order accurate
derivative $q'_{C,i}$ in space.

The remaining task includes conducting a temporal updating step via
tracing the characteristic lines using the piecewise cubic polynomials in Eq. (\ref{Eq:pcm_poly}).
This step produces the Riemann states 
$(q_L,q_R)=(q_{R,i}^{n+1/2},q_{L,i+1}^{n+1/2})$
as predictor.
We will show in the next section that these predictors are at least fourth-order accurate in time, and they 
are provided as the initial value problems for the Godunov fluxes at each interface $x_{i+1/2}$.
%
%

\section{The PCM Characteristic Tracing for Temporal Updates}
\label{Sec:chartracing}
The PCM proceeds to the last step which advances the pointwise Riemann interface states at $t^n$
\beq
q_{L;R,i} = p_i(x_{i\pm 1/2}),
\eeq
where $p_i(x)$ is the piecewise cubic polynomial in Eq. (\ref{Eq:pcm_poly}),
to the half-time updated predictor states
\beq
q_{L;R,i}^{n+1/2}
\eeq
by tracing
characteristics. The idea is same as how the PPM characteristic tracing is performed \cite{colella1984piecewise},
in which we seek a time averaged state. For instance, at the interface $x_{i+1/2}$, we consider
\beq
q^{n+1/2}_{x+1/2}=\frac{1}{\dt} \int_{t^n}^{t^{n+1}} q(x_{i+1/2},t) dt.
\eeq
The initial condition at $t^n$ of a generalized Riemann problem is given as
\beq
q(x_{i+1/2},t^n) =
\left \{
\begin{array}{ll}
p_i(x_{i+1/2}),       & x \in I_i \\
p_{i+1}(x_{i+1/2}), & x \in I_{i+1}.
\end{array}
\right.
\eeq
Given a linear characteristic equation as in
Eq.~(\ref{Eq:Euler1D_char}), and for $t>t^n$ we then have
\beq
\label{Eq:char_tracing_IC}
q(x_{i+1/2},t) = 
\left \{
\begin{array}{ll}
p_i(x_{i+1/2}-\lambda_i (t-t^n)),       & x \in I_i, \;\; \lambda_i > 0, \\
p_{i+1}(x_{i+1/2} - \lambda_{i+1}(t-t^n)), & x \in I_{i+1}, \;\; \lambda_{i+1} < 0.
\end{array}
\right.
\eeq
%
Here $\xi(t) = x_{i+1/2} - \lambda (t-t^n)$ is a characteristic line for an eigenvalue $\lambda$, assuming $t-t^n < \dt$.

We argue that the characteristically traced solution in Eq. (\ref{Eq:char_tracing_IC}) is {\it{almost}} exact, provided the 
stability condition $t-t^n < \dt$ is satisfied (which is always true),
inheriting all the desirable high-order accurate properties built in to the initial conditions which are, 
in this case, given by the piecewise cubic polynomial $p_i(x)$ (see \cite{leveque2007finite}).
Therefore, the spatial accuracy designed in $p_i(x)$ naturally gets transferred to the evaluation of
the time averaged state in Eq. (\ref{Eq:char_tracing_IC}).
In particular, for our case, the expected accuracy of the characteristic tracing using our cubic polynomial $p_i(x)$ 
for predicting a future state at $t>t^n$, satisfying $t-t^n < \dt$, is to be at least fourth-order accurate.

We now illustrate, for exposition purpose, the case with $x\in I_i$ with $\lambda_i > 0$ first. Using $\uparrow$ to denote
the state from the left of $x_{i+1/2}$,
\bea
q^{n+1/2}_{R,i} = q^{n+1/2}_{x+1/2,\uparrow}
&=&\frac{1}{\dt} \int_{t^n}^{t^{n+1}} q(x_{i+1/2},t) dt \label{Eq:ERP_pcm_1}\\
&=&\frac{1}{\lambda_i \dt} \int_{x_{i+1/2 - \lambda_i \dt}}^{x_{i+1/2}} p_i(x) dx. \label{Eq:ERP_pcm_2}
\eea
Again, as seen in Eqs. (\ref{Eq:ERP_pcm_1}) and (\ref{Eq:ERP_pcm_2}), 
the half-time advancement of the spatially reconstructed state
is given by the average of the reconstructed variable $p_i(x)$ over 
the domain of dependence $[x_{i+1/2 - \lambda_i \dt}, x_{i+1/2}]$ of the interface $x_{i+1/2}$.
Therefore the accuracy of $q^{n+1/2}_{R,i}$ is inherited from that of the reconstruction algorithm of $p_i(x)$.

The outcome of the integration yields
\bea
q^{n+1/2}_{x+1/2,\uparrow} &=& 
c_0 + 
\frac{c_1}{2}\left(1-\frac{\lambda_i \dt}{\dx} \right) \dx +
\frac{c_2}{4}\left(1-2\frac{\lambda_i \dt}{\dx} + \frac{4}{3}\Big(\frac{\lambda_i \dt}{\dx} \Big)^2\right) \dx^2\nonumber\\
&+&\frac{c_3}{8}\left( 1-3\frac{\lambda_i \dt}{\dx} + 4\Big(\frac{\lambda_i \dt}{\dx}\Big)^2 - 2\Big(\frac{\lambda_i \dt}{\dx}\Big)^3 \right) \dx^3.
\eea

The case for $x\in I_{i+1}$ with $\lambda_{i+1} < 0$ can be obtained similarly,
\bea
q^{n+1/2}_{L,i+1} &=& q^{n+1/2}_{x+1/2, \downarrow} \nonumber \\
&=& 
c_0 + 
\frac{c_1}{2}\left(-1-\frac{\lambda_{i+1} \dt}{\dx} \right) \dx +
\frac{c_2}{4}\left(1+2\frac{\lambda_{i+1} \dt}{\dx} + \frac{4}{3}\Big(\frac{\lambda_{i+1} \dt}{\dx}\Big)^2 \right) \dx^2\nonumber\\
&+&\frac{c_3}{8}\left( -1-3\frac{\lambda_{i+1} \dt}{\dx} - 4\Big(\frac{\lambda_{i+1} \dt}{\dx}\Big)^2 - 2\Big(\frac{\lambda_{i+1} \dt}{\dx}\Big)^3 \right) \dx^3.
\eea

In the general case of a system of Euler equations, the above treatment is to be extended to include multiple characteristic waves
correspondingly depending on the sign of each $k$-th eigenvalue $\lambda^{(k)}_i$. 
This gives us, for the two predictor states $\bV_{L;R,i}^{n+1/2}$ on each cell $I_i$ in primitive form,
\bea
\label{Eqn:PCM_right_state_final}
&&\bV_{R,i}^{n+1/2} = {\mbf C}_0+
\frac{1}{2}\sum_{k;\lambda^{(k)}_i>0}\Big(1-\frac{\lambda^{(k)}_i\Delta t}{\Delta x} \Big)r^{(k)} 
\Delta \mbf C^{(k)}_1\nonumber\\
&&+\frac{1}{4}\sum_{k;\lambda^{(k)}_i>0}\left(1-2\frac{\lambda^{(k)}_i\Delta t}{\Delta x} 
+\frac{4}{3}\Big(\frac{\lambda_i^{(k)}\Delta t}{\Delta x}\Big)^2\right)r^{(k)}
\Delta \mbf C^{(k)}_2,\nonumber\\
&&+\frac{1}{8}\sum_{k;\lambda^{(k)}_i>0}\left(1-3\frac{\lambda^{(k)}_i\Delta t}{\Delta x} 
+{4}\Big(\frac{\lambda^{(k)}_i\Delta t}{\Delta x}\Big)^2
-{2}\Big(\frac{\lambda^{(k)}_i\Delta t}{\Delta x} \Big)^3\right)r^{(k)} \Delta \mbf C^{(k)}_3,\nonumber\\
\eea
and
\bea
\label{Eqn:PCM_left_state_final}
&&\bV_{L,i}^{n+1/2} = {\mbf C}_0+
\frac{1}{2}\sum_{k;\lambda^{(k)}_i<0}\Big(-1-\frac{\lambda^{(k)}_i\Delta t}{\Delta x} \Big)r^{(k)} 
\Delta \mbf C^{(k)}_1\nonumber\\
&&+\frac{1}{4}\sum_{k;\lambda^{(k)}_i<0}\left(1+2\frac{\lambda^{(k)}_i\Delta t}{\Delta x} 
+\frac{4}{3}\Big(\frac{\lambda_i^{(k)}\Delta t}{\Delta x}\Big)^2\right)r^{(k)}
\Delta \mbf C^{(k)}_2,\nonumber\\
&&+\frac{1}{8}\sum_{k;\lambda^{(k)}_i<0}\left(-1-3\frac{\lambda^{(k)}_i\Delta t}{\Delta x} 
-{4}\Big(\frac{\lambda^{(k)}_i\Delta t}{\Delta x}\Big)^2
-{2}\Big(\frac{\lambda^{(k)}_i\Delta t}{\Delta x} \Big)^3\right)r^{(k)} \Delta \mbf C^{(k)}_3. \nonumber\\
\eea
Those new notations introduced in Eqs. (\ref{Eqn:PCM_right_state_final}) and (\ref{Eqn:PCM_left_state_final}) 
represent the $k$-th right eigenvector $r^{(k)}$, $k=1,2,3$, 
which is the $k$-th column vector of the $3\times 3$ matrix $\bR$ evaluated at $I_i$,
\beq
\bR = \Big[ r^{(1)} | r^{(2)} | r^{(3)} \Big],
\eeq
and the $k$-th characteristic variable vector $\Delta \mbf C^{(k)}_m$ given as
\beq
\Delta \mbf C^{(k)}_m = \dx^m \ell^{(k)} \cdot \mathbf{C}_m,
\eeq
where, for $m=0,\dots,3$,
\beq
\mathbf{C}_m =  \Big[ c_m^{(1)} | c_m^{(2)} | c_m^{(3)} \Big]^T,
\eeq
in which $c^{(k)}_m$ is the $m$-th coefficient in Eqs. (\ref{Eq:pcm_cond1c}) $\sim$ (\ref{Eq:pcm_cond4c}) 
of the piecewise cubic polynomial in Eq. (\ref{Eq:pcm_poly})
applied to each of the $k$-th characteristic variable $\bar {q}_i$.

It is worth mentioning that, unlike the characteristic tracing of PPM (see \cite{colella1984piecewise}), 
the PCM scheme does not necessarily require any extra
monotonicity enforcements on $\bV^{n+1/2}_{L;R,i}$.
First of all, this is because the use of the WENO reconstruction algorithms provides
$q_{L;R,i}$ and $q'_{C,i}$, all of which are, by design, non-oscillatory.
Secondly, such monotonicity enforcements on the PPM's parabolic polynomials 
are now redundant in PCM, since our building block polynomials are piecewise cubic. 
Compared to the parabolic polynomials, 
the cubic polynomials can easily adapt to fit  $q_{L;R,i}$, $q'_{C,i}$, and $\bar{q}_i$ uniquely on each $I_i$, 
without needing to preserve such monotonicity constraints as in PPM, by readily
varying its rate of change $p'_i(x)$ at an inflection point if needed, taking an advantage of
an extra degree of freedom by being cubic.
\section{Final Update Step in 1D}
The only remaining task at this point is the final update to evolve $\avg{\bU}_i^n$ to $\avg{\bU}^{n+1}_i$.
We proceed this using the high-order Godunov fluxes 
$\bF_{i+1/2}^{n+1/2}=\mathcal{RP}(\bU_L,\bU_R)=\mathcal{RP}(\bU_{R,i}^{n+1/2},\bU_{L,i+1}^{n+1/2})$ as corrector,
where $\mathcal{RP}$ implies a solution of the Riemann problem. Note that the Riemann states in conservative variables
$\bU_{R,i}^{n+1/2},\bU_{L,i+1}^{n+1/2}$ are obtained either 
by conversions from $\bV_{R,i}^{n+1/2},\bV_{L,i+1}^{n+1/2}$ in 
Eq. (\ref{Eqn:PCM_right_state_final}) and Eq. (\ref{Eqn:PCM_left_state_final}),
or projecting the characteristic variables directly to the conservative variables in 
Eq. (\ref{Eqn:PCM_right_state_final}) and Eq. (\ref{Eqn:PCM_left_state_final}).
We note that the first needs to be processed using high-order approximation 
\cite{mccorquodale2011high}, in particular for multidimensonal problems, while
such a high-order conversion is not required in 1D. In this regards the latter could be a better choice
in multi spatial dimensions, because there is no need for any high-order conversion
from the primitive Riemann states to the conservative Riemann states,
knowing the fact that the {\it{conservative}} states variables are the type of inputs
for the Riemann problems.

\section{Multidimensional Extension of the 1D PCM Scheme}
\label{Sec:pcmMultiD}
Our primary purpose in the current paper is to focus on laying down the key algorithmic components of PCM in 1D.
As described, the 1D PCM algorithm is formally fifth-order in space and fourth-order in time. 
Our test problems of one-dimensional smooth flows in Section \ref{Sec:convegence_performance} show 
that the algorithm delivers nominally a fifth-order accurate convergence rate, 
particularly with smaller $L_1$ errors than WENO-JS with RK4.

Although possible, extending such a high-order 1D algorithm to multiple spatial dimensions 
in a way to preserve the same order of convergence in 1D is 
an attentive task that requires some extra cares and attentions 
\cite{shu2009high,buchmuller2014improved,zhang2011order,mccorquodale2011high} in the finite volume formulation.
On the other hand, one of the simplest and easiest multidimensional extensions that has been widely adopted in many algorithmic choices
\cite{mignone2007pluto, mignone2011pluto, mignone2010second, stone2008athena, fryxell2000flash, dubey2009extensible, lee2009unsplit, lee2013solution, bryan1995piecewise, bryan2014enzo, teyssier2002cosmological}
is to use the dimension-by-dimension formalism in which the baseline 1D algorithm is extended in each normal sweep direction, 
requiring a very minimal effort for extension.
However, the order of convergence from the resulting multidimensional extension is limited to be at most second-order 
due to the lack of accuracies that may arise in a couple of places in code implementations: 
(i) mis-using averaged quantities in place of pointwise quantities for Riemann states, 
(ii) using low-order approximations in converting between primitive and conservative variables, (iii)
and applying low-order quadrature rules in flux function estimations 
\cite{shu2009high,buchmuller2014improved,zhang2011order,mccorquodale2011high}.
In our case, it takes more coding efforts, practically because 
the multidimensional PCM results we demonstrate in this paper
have been obtained by integrating the PCM algorithm in the FLASH code framework 
\cite{fryxell2000flash, dubey2009extensible, dlee_flash};
hence carrying out the above-mentioned code changes 
in a large code such as FLASH requires extra efforts that are
not the main points of the current paper.
We leave such a high-order, multidimensional extension in our future work,
and instead, we adopt the simple dimension-by-dimension formalism 
for our multidimensional extension of the 1D PCM algorithm.

Additionally, for our choice of multidimensional extension we use 
the computationally efficient unsplit corner transport upwind (CTU) formulation in FLASH \cite{lee2009unsplit, lee2013solution}, 
which requires smaller number of Riemann solves in both 2D and 3D than the conventional
CTU approaches \cite{colella1990multidimensional, gardiner2008unsplit, saltzman1994unsplit}, while 
achieving the maximum Courant condition of CFL $\approx$ 1 \cite{lee2009unsplit, lee2013solution}.
%
%

\section{Results}
\label{Sec:results}

In this section we present numerical results of PCM in 1D, 2D and 3D for hydrodynamics and magnetohydrodynamics.
The PCM results are compared with numerical solutions of other popular choices of reconstruction schemes including 
the second-order PLM \cite{colella1985direct}, 
the third-order PPM \cite{colella1984piecewise} and 
the fifth-order WENO methods \cite{jiang1996efficient, borges2008improved,castro2011high}.
As mentioned, the second-order accurate dimension-by-dimension approach has been
adopted to extend all of the above baseline 1D reconstruction algorithms to multiple spatial dimensions.
As this is the case, for multidimensional problems
we have chosen the predictor-corrector type of characteristic tracing methods (charTr)
for PLM, PPM, and WENO\footnote[$\dagger$]{To implement a characteristic tracing for WENO we first reconstruct the fifth-order
Riemann states, $q_{S,i}^{weno,n}$, $S=L,R$, using WENO. 
They are then temporally evolved by $\dt/2$ to get
$q_{S,i}^{weno,n+1/2}$ by integrating over each corresponding domain of dependence the piecewise parabolic polynomials 
defined by $q_{L,i}^{weno,n}$, $q_{R,i}^{weno,n}$ and $\bar{q}_i$  \cite{dlee_flash}.},
not to mention PCM by design. In 1D problems, however, we
treat WENO differently and integrate its spatial reconstruction with RK4 
in consideration of fully demonstrating its orders of accuracy due from
both space (i.e., $\mathcal{O}(\dx^5)$) and time (i.e., $\mathcal{O}(\dt^4)$).
It should also be noted that the orders of WENO + RK4 in 1D are to be well comparable to
those of PCM. Hence the choice 
provides a set of good informative comparisons between PCM and WENO + RK4 in particular, 
in which we will illuminate the advantages of PCM.
In what follows, unless otherwise mentioned, we set the WENO-JS approach in Eq. (\ref{Eq:PCM_WENO5_omega})
as the default choice for $q'_{C,i}$ in our PCM results. 
This default setting will be referred to as  {\it{PCM-JS}} (or simply {\it{PCM}}), while the 
choice with the WENO-Z approach in Eq. (\ref{Eq:PCM_WENOZ_omega}) will be referred to as {\it{PCM-Z}}.

\subsection{1D Tests}

\subsubsection{1D Convergence and Performance Tests}
\label{Sec:convegence_performance}
\paragraph{\underline{Gaussian and Sinusoidal Wave Advections}}
In our first test we consider two configurations of 1D passive advection of smooth flows, 
involving initial density profiles of Gaussian and sinusoidal waves.
We initialize the both problems on a computational box 
on [0,1] with periodic boundary conditions.
The initial density profile of the Gaussian advection is defined by $\rho(x) = 1 + e^{-100(x-x_0)^2}$, 
with $x_0=0.5$, whereas for the sinusoidal advection the density is initialized by $\rho(x) = 1.5 - 0.5 \sin(2 \pi x)$.
In both cases, we set constant velocity, $u=1$, and pressure, $P=1/\gamma$, and the specific heat ratio, $\gamma=5/3$. 

The resulting profiles are propagated for one period through the boundaries, reaching $t=1$. At this point,
both profiles return to its initial positions at which we conduct the $L_1$ error convergence tests compared with
the initial conditions on the grid resolutions of $N_x=16, 32,64,128,256, 512$ and $1024$.
Since the nature of the problem is a pure advection in both, any deformation of the initial profile is due to either phase errors or numerical diffusion. 
For stability we use a fixed Courant number, $C_{\mbox{cfl}}=0.8$ for both tests. 
We choose the HLLC Riemann solver \cite{toro1994restoration} in all cases.

\begin{figure}
\begin{tabular}{c}
\hspace{-0.5in}
\subfigure[][]{
\includegraphics[width=3.0in, trim= -0.5in 0in 0.in 0.in,clip=true]{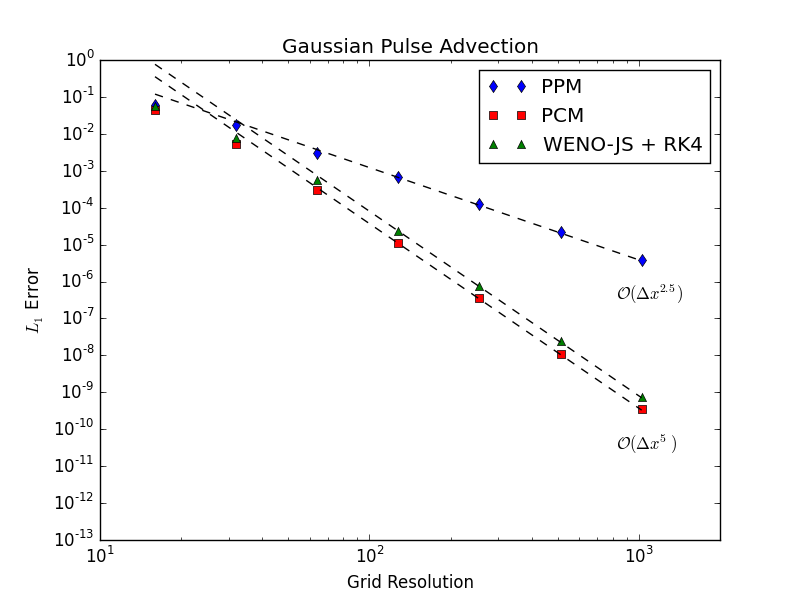}}\hspace{-0.5in}
\subfigure[][]{
\includegraphics[width=3.0in, trim= -0.5in 0in 0.in 0.in,clip=true]{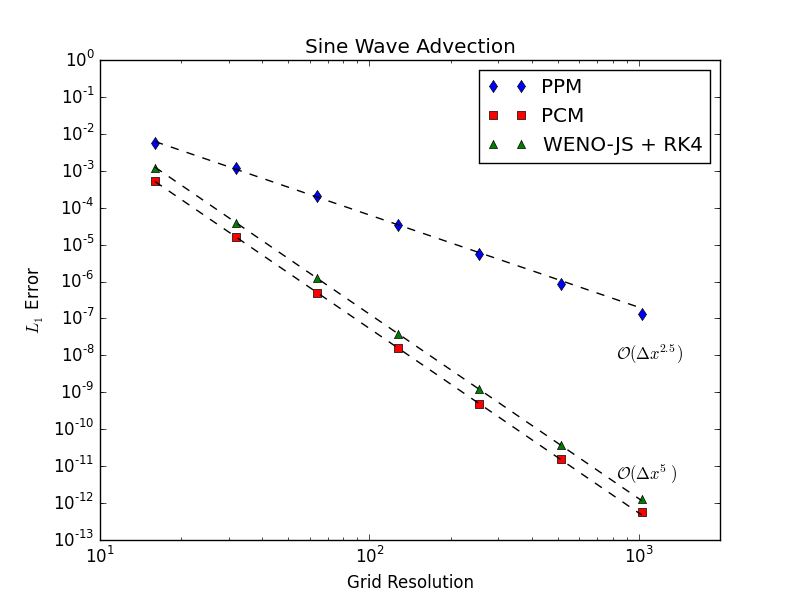}}
\end{tabular}
\caption{Convergence test of (a) 1D Gaussian pulse advection, and (b) 1D sinusoidal wave advection.}
\label{Fig:1DAdvections}
\end{figure}

The results of this study are shown in Fig.~\ref{Fig:1DAdvections}. 
From these numerical experiments, the PCM reconstruction shows the fifth-order convergence rates in both tests.
Although both PCM and WENO-JS + RK4 demonstrate the same fifth-order of convergence rate,
the $L_1$ errors of PCM are more than twice smaller than those of WENO-JS + RK4.
The solutions of PPM converge with the rate of 2.5 which is the slowest among the three.
Parameter choices for the PPM runs include the use of the MC slope limiter applied to characteristic variables, no flattening,
no contact discontinuity steepening, and no artificial viscosity (this setting for PPM remains the same in what follows). 
\begin{table}[ht!]
  \centering
  \begin{tabular}{|| l | c ||}
    \hline
    Scheme & Speedup \\ \hline
    PPM      & 0.65 \\
    PCM     & 1.00 \\
    WENO-JS + RK4  & 1.71 \\
    \hline
  \end{tabular}
  \caption{Relative speedup of the PPM and WENO schemes compared to
    the PCM scheme for the 1D Gaussian and sine advection problems. 
    The comparisons have been obtained from a serial calculation on a single CPU.}
  \label{tab:performance}
\end{table}

In Table \ref{tab:performance}  we compare the relative performance speedups of PCM, PPM and WENO-JS + RK4,
all testing the Gaussian and sinusoidal advection problems. We can clearly see that
there is a big performance advantage in PCM over WENO-JS + RK4 in delivering the target fifth-order accuracy.
The major gain in PCM lies in its predictor-corrector type of characteristic tracing
which affords not only the accuracy but also the computational efficiency.
Such a relative computational efficiency of PCM in 1D is expected to grow much larger in multidimensional problems,
considering that there have to be added algorithmic complexities in 
achieving high-order accurate solutions in multidimensional finite volume reconstruction 
\cite{shu2009high,buchmuller2014improved,zhang2011order,mccorquodale2011high}
from the perspectives of balancing optimal numerical stability and accuracy.
We will report our strategies of multidimensional extension of PCM in our future work.

%

\subsubsection{1D Discontinuous Tests}
In this section we test PCM on a series of well-benchmarked shock-tube problems of one dimensional hydrodynamics and MHD 
that involve discontinuities and shocks. As all the tests here have already been well discussed in various literatures, 
we will describe their setups only briefly and put our emphasis more on discussing the code performance of PCM.
Readers are encouraged to refer to the cited references in the texts for more detailed descriptions on each setup.

\paragraph{\underline{(a) Sod Shock Tube}}

\begin{figure}[htpb!]
\centering
\begin{tabular}{cccc}
\subfigure[][]{
\includegraphics[width=2.5in, trim= 0.7in 0.3in 0.in 0.35in,clip=true]{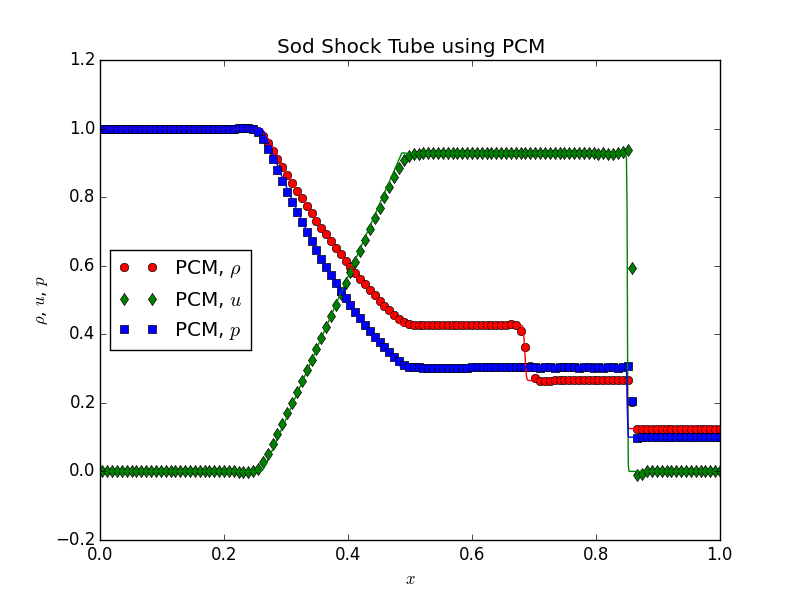}}
\subfigure[][]{
\includegraphics[width=2.5in, trim= 0.7in 0.3in 0.in 0.35in,clip=true]{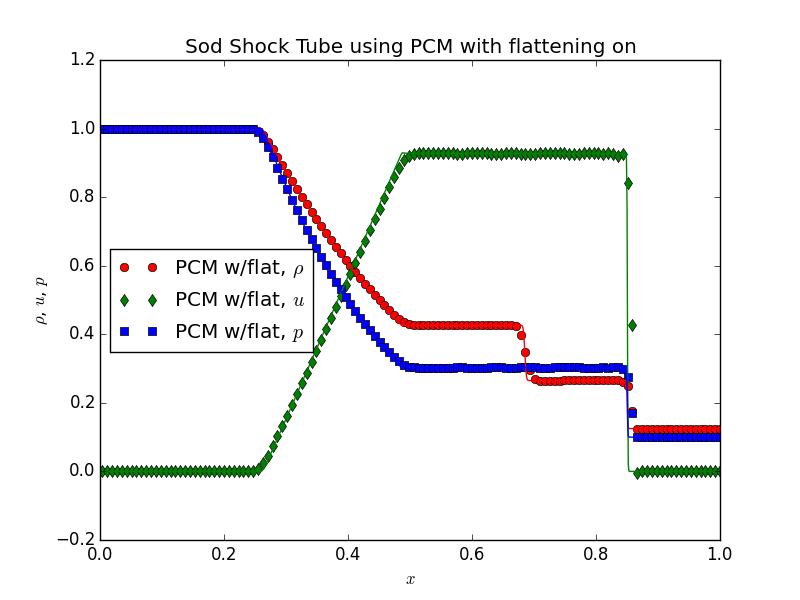}}\\
\subfigure[][]{
\includegraphics[width=2.5in, trim= 0.7in 0.3in 0.in 0.35in,clip=true]{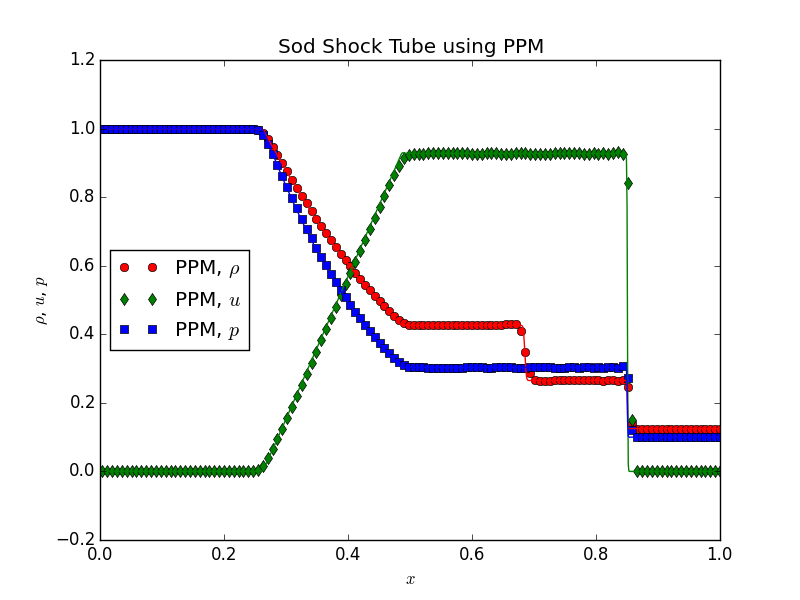}}
\subfigure[][]{
\includegraphics[width=2.5in, trim= 0.7in 0.3in 0.in 0.35in,clip=true]{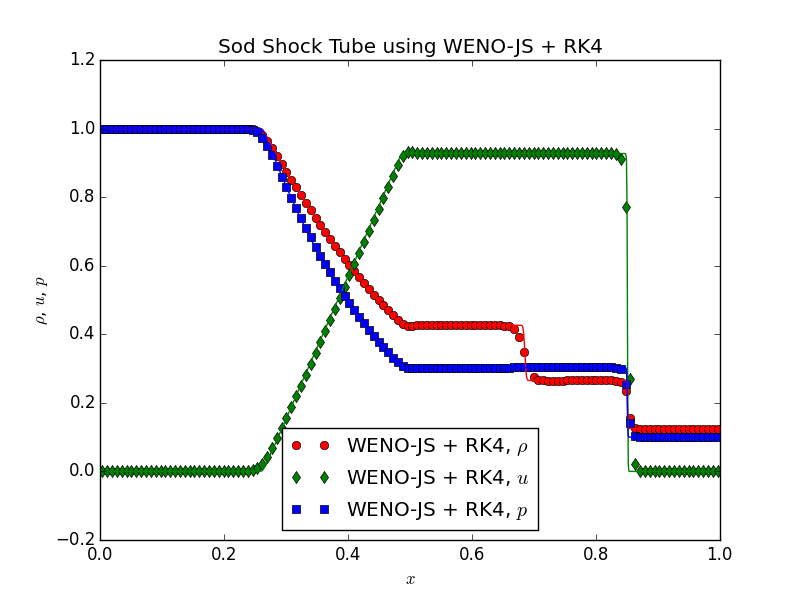}}\\
\end{tabular}
\caption{The Sod shock tube problem at $t=0.2$.
(a) PMC without flattening, (b) PCM with flattening, (c) PPM with MC slope limiter,
and (d) WENO-JS + RK4.}
\label{Fig:Sod}
\end{figure}

The Sod's problem \cite{sod1978survey} has been one of the most widely chosen popular
tests in 1D  to assess a code's capability to handle shocks and contact discontinuities. 
The initial condition is consist of the left and the right states given as
\beq
\label{Eq:Sod}
(\rho, u, p) = 
\left \{
\begin{array}{ll}
(1, 0, 1)             & x < 0.5, \\
(0.125, 0, 0.1) & x > 0.5,
\end{array}
\right.
\eeq
with the ratio of specific heats $\gamma = 1.4$ on the entire domain $[0,1]$. The outflow 
boundary conditions are imposed at $x=0$ and $x=1$.
Shown in Fig. \ref{Fig:Sod} include two numerical solutions of PCM, with and without 
the use of the PCM flattening given by Eq. (\ref{Eq:qPrime_flattening}); and two solutions of
using PPM and WENO-JS + RK4. The Roe Riemann solver \cite{roe1981approximate} 
was used in all cases. 
The test cases (denoted in symbols) are resolved on the grid size of $N_x=128$, 
and are compared with the reference solutions (denoted in solid curves) 
computed using WENO-JS + RK4 on the grid resolution of $N_x=1024$.
A fixed value of $C_{\mbox{cfl}}=0.8$ was used for all tests.

The result in Fig. \ref{Fig:Sod}(a) shows that the solutions of PCM without using the flattening 
well predict all nonlinear flow characteristics of the rarefaction wave, the contact discontinuity, and the shock.
A notable thing in PCM is the number of points at the shock. 
We see in Fig. \ref{Fig:Sod}(a) that at the shock there is only one single point in all flow variables,
whereas in all other cases, there are two points spread over the shock width.

We also tested the PCM flattening in in Fig. \ref{Fig:Sod}(b). We observe that the switch 
introduces some level of noisy oscillations, easily seen in the region between 
the rarefaction tail and the shock. As anticipated, the solution looks very similar
to that of PPM in  Fig. \ref{Fig:Sod}(c) because, in the limit of Eq. (\ref{Eq:qPrime_PPM}),
the PCM flattening reduces the PCM scheme to a PPM-like algorithm.

\paragraph{\underline{(b) The Shu-Osher Test}}

\begin{figure}\centering
\begin{tabular}{cccc}
\subfigure[][]{
\includegraphics[width=4.5in]{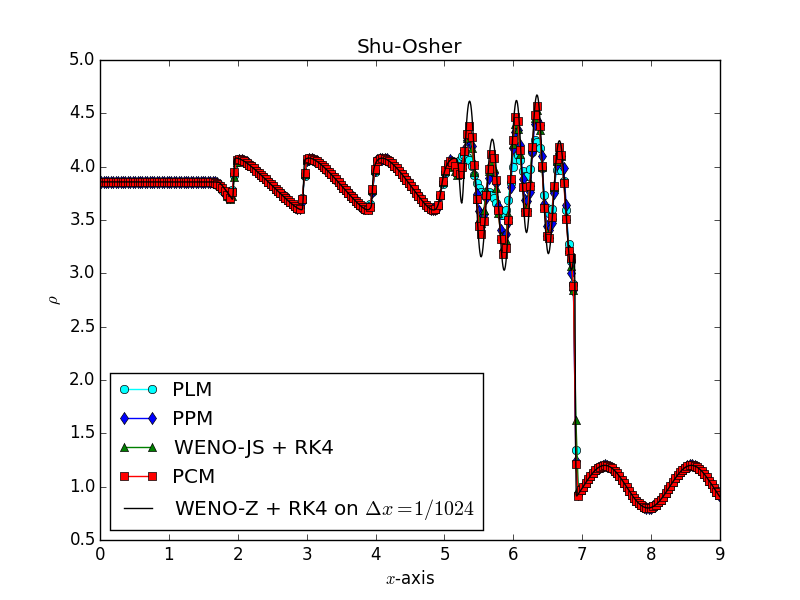}}\\
\subfigure[][]{
\includegraphics[width=4.in]{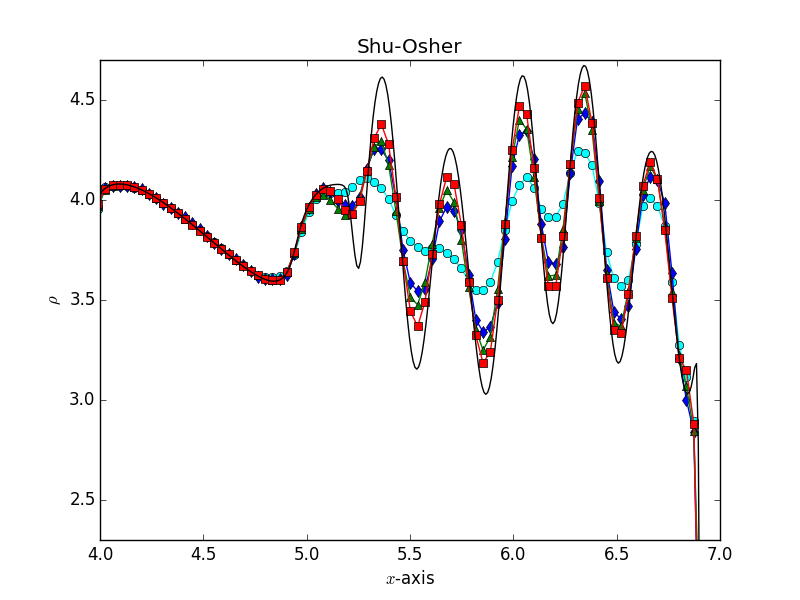}}
\end{tabular}
\caption{The Shu-Osher Riemann problem at $t=1.8$. (a) All four reconstruction schemes on $N_x=256$ are
compared with WENO-JS + RK4 on $N_x=1024$. For PPM, the MC slope limiter
is used. (b) A close-up view to demonstrate the schemes' numerical diffusivity.}
\label{Fig:ShuOsher}
\end{figure}

The second test is the Shu-Osher problem~\cite{Shu1989}. 
In this problem we test PCM's ability to resolve both small-scale smooth flow features and the shock. 
On [-4.5, 4.5], the initial condition launches a nominally Mach 3 shock wave at $x=-4.0$
propagating into a region ($x>-4.0$) of a constant density field with 
sinusoidal perturbations. As the shock advances, two sets of density features appear behind the shock.
The first set has the same spatial frequency as the un-shocked perturbations, whereas
and the second set behind the shock involves the frequency that is doubled. 
The important point of the test is to see how well a code can accurately resolve 
strengths of the oscillations behind the shock, as well as the shock itself.

The results of this test are shown for PLM, PPM, WENO-JS + RK4, and PCM in Fig.~\ref{Fig:ShuOsher}. 
The solutions are calculated at $t=1.8$ using a 
resolution of $N_x=256$ and are compared to a reference solution resolved on $N_x=1024$.
All methods were solved using the Roe Riemann solver, with $C_{\mbox{cfl}}=0.8$.
It is evident in  Fig.~\ref{Fig:ShuOsher}(b) that the PCM solution exhibits the least diffusive solution among 
the tested methods, producing  
a very-high order accurate solution that is more quickly approaching to the high resolution reference solution.

%
%

\paragraph{\underline{(c) The Einfeldt Strong Rarefaction Test}}

\begin{figure}[ht!]
\centering
\begin{tabular}{cccc}
\subfigure[][]{
\includegraphics[width=2.5in,  trim= 0.7in 0.3in 0.in 0.35in,clip=true]{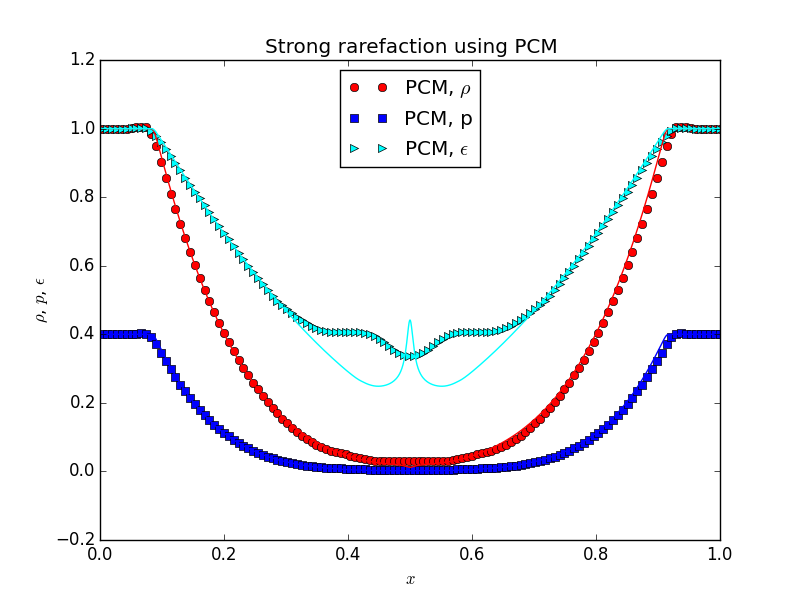}}
\subfigure[][]{
\includegraphics[width=2.5in,  trim= 0.7in 0.3in 0.in 0.35in,clip=true]{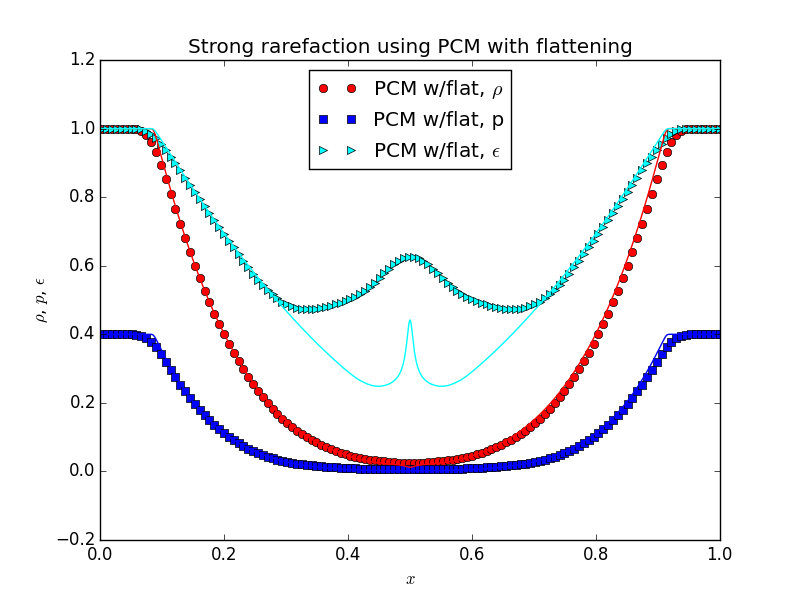}}\\
\subfigure[][]{
\includegraphics[width=2.5in,  trim= 0.7in 0.3in 0.in 0.35in,clip=true]{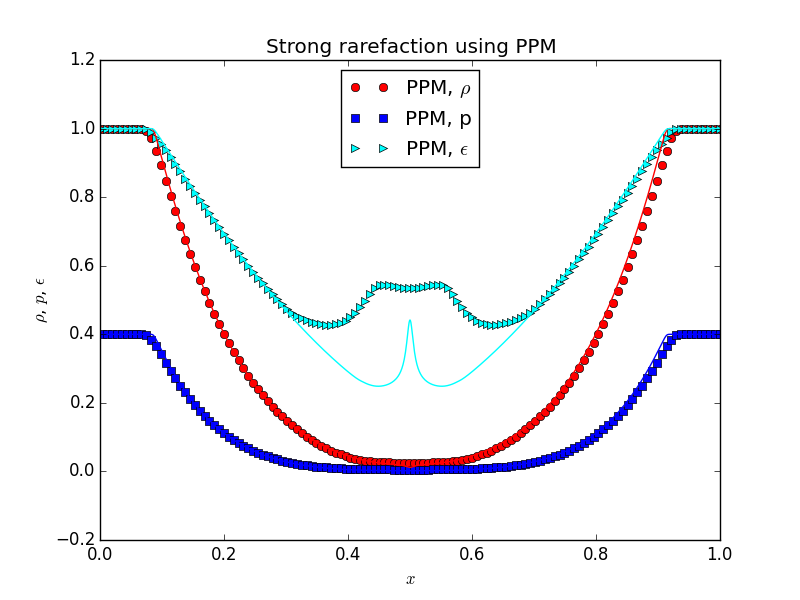}}
\subfigure[][]{
\includegraphics[width=2.5in,  trim= 0.7in 0.3in 0.in 0.35in,clip=true]{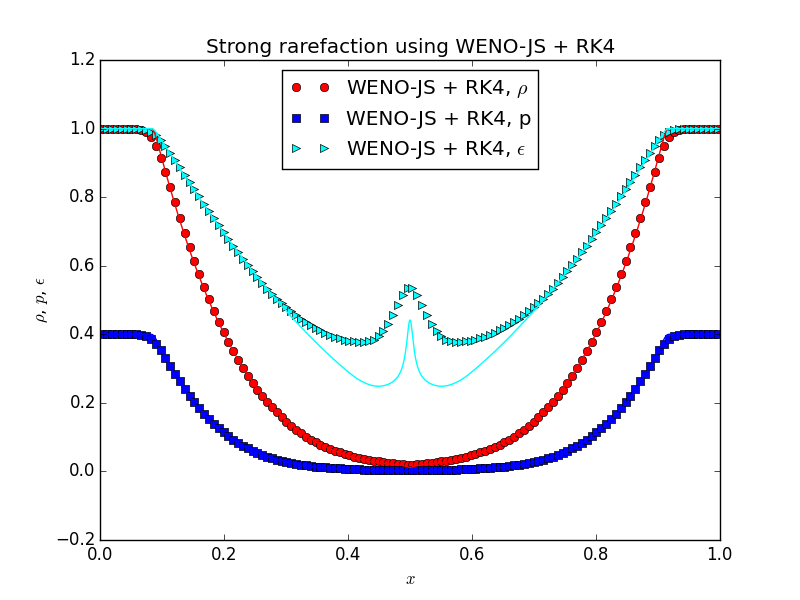}}\\
\end{tabular}
\caption{The Einfeldt strong rarefaction test at $t=0.15$. All tests used the HLLC Riemann solver
on $N_x=128$ with $C_{\mbox{cfl}}=0.8$. The solid curves represent the reference solution 
computed using WENO-JS + RK4 on $N_x = 1024$.}
\label{Fig:Rarefaction}
\end{figure}

First described by Einfeldt et al. \cite{einfeldt1991godunov} the main test point in this problem is to
see how satisfactorily a code can compute physical variables, $p, u, \rho, \epsilon$, etc. in the low density region.
Among the variables the internal energy $\epsilon=p/(\rho(\gamma-1))$, where $\gamma=1.4$, is the hardest to get it right 
due to the ratio of the pressure and density that are both close to zero. The ratio of the two small
quantities will amplify any small errors in each, hence making the error in $\epsilon$ appear to be
the largest in general \cite{Toro2009}. 

The large errors in $\epsilon$ are indeed observed in Fig. \ref{Fig:Rarefaction}(b) $\sim$ Fig. \ref{Fig:Rarefaction}(d) 
in that the error
is the largest at or around $x=0.5$ in the presence of sudden increase of its peak values. On the
contrary, the internal energy computed using PCM shown in Fig. \ref{Fig:Rarefaction}(a) behaves in a
uniquely different way such that the value continues to drop when approaching $x=0.5$.
From this viewpoint, and with the help of the exact solution available in \cite{Toro2009}, it's fair to say that
the PCM solution in Fig. \ref{Fig:Rarefaction}(a) appears to predict the internal energy most accurately.
It is seen that there are two slight bumps produced in Fig. \ref{Fig:Rarefaction}(a), at $x\approx 0.08$
and $x \approx 0.92$ in $\rho$, which disappear by turning on the PCM flattening 
as shown in Fig. \ref{Fig:Rarefaction}(b).

\paragraph{\underline{(d) Two-Blast}}

\begin{figure}[pbht!]
\centering
\begin{tabular}{cccc}
\subfigure[][]{
\includegraphics[width=4.5in]{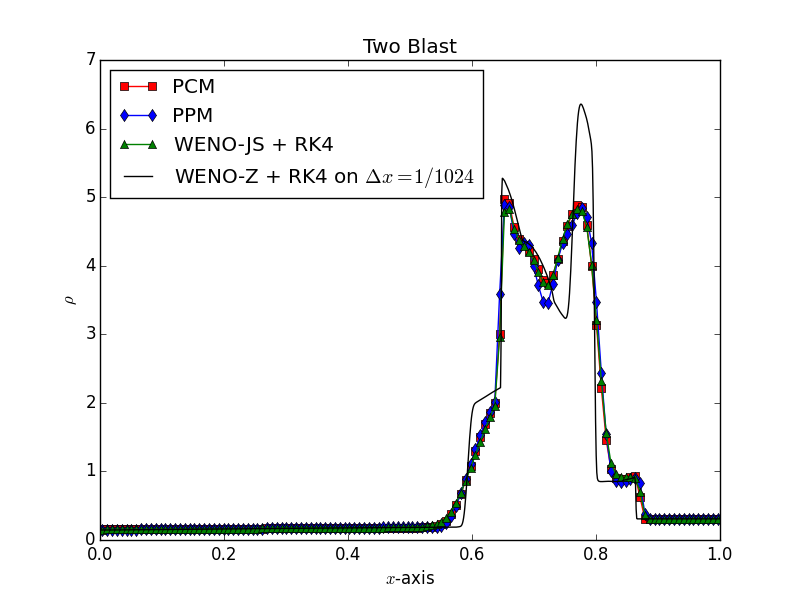}}\\
\subfigure[][]{
\includegraphics[width=4.in]{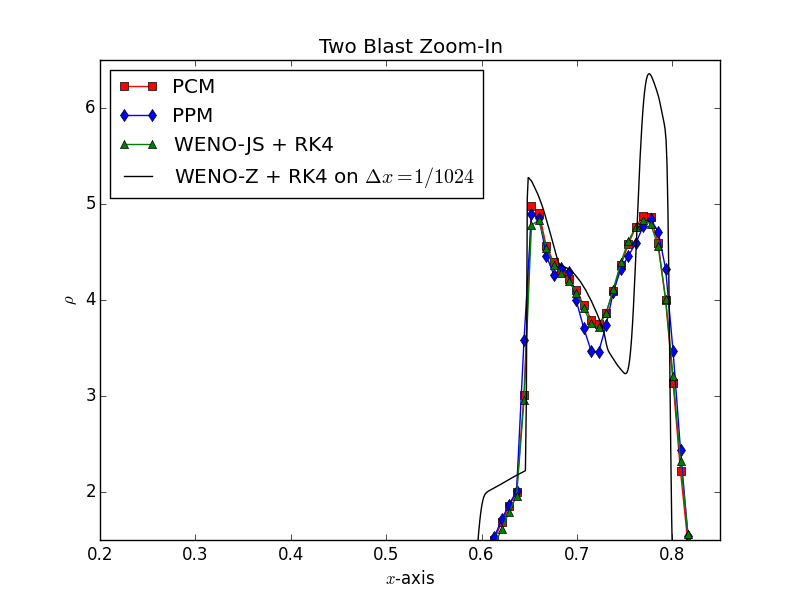}}
\end{tabular}
\caption{Two blast problem at $t=0.038$. (a) All three reconstruction schemes on $N_x=128$ are
compared with WENO-JS + RK4 on $N_x=1024$. For PPM, the MC slope limiter
is used. (b) A close-up view to demonstrate the schemes' numerical accuracy. 
In all tests we used the HLLC Riemann solver with $C_{\mbox{cfl}}=0.8$.}
\label{Fig:2blast}
\end{figure}
%

This problem was introduced by Woodward and Colella \cite{woodward1984numerical} and was designed to
test a code performance particularly on interactions of strong shocks and discontinuities. We follow the original
setup to test PCM, and compare its solution with those of PPM and WENO-JS + RK4, using 128 grid points to
resolve the domain $[0,1]$. In Fig. \ref{Fig:2blast} the three density profiles at $t=0.038$ are plotted against
the high-resolution solution of WENO-JS + RK4 on 1024 grid points.
Overall, all methods we tested here produce an acceptable quality of solutions as illustrated in Fig. \ref{Fig:2blast}(a).
Note however that, among the three methods, the PCM solution in Fig. \ref{Fig:2blast}(b) demonstrates the highest peak heights, 
following more closely the high-resolution solutions.
As reported in \cite{stone2008athena} we see that all methods also smear out the contact discontinuity at $x\approx 0.6$ 
pretty much the same amount.

\paragraph{\underline{(e) Brio-Wu MHD Shock Tube}}

\begin{figure}[htb!]
\centering
\begin{tabular}{cccc}
\subfigure[][]{
\includegraphics[width=2.5in, trim= 0.7in 0.3in 0.in 0.35in,clip=true]{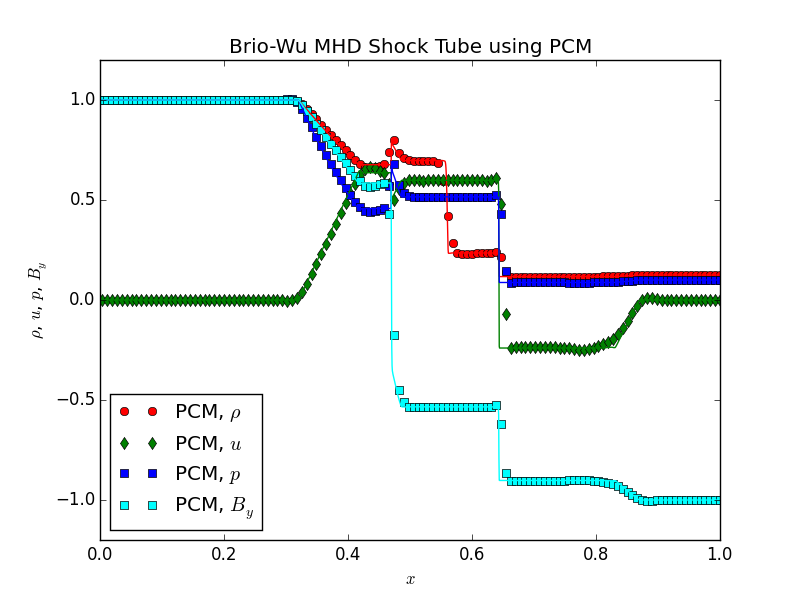}}
\subfigure[][]{
\includegraphics[width=2.5in, trim= 0.7in 0.3in 0.in 0.35in,clip=true]{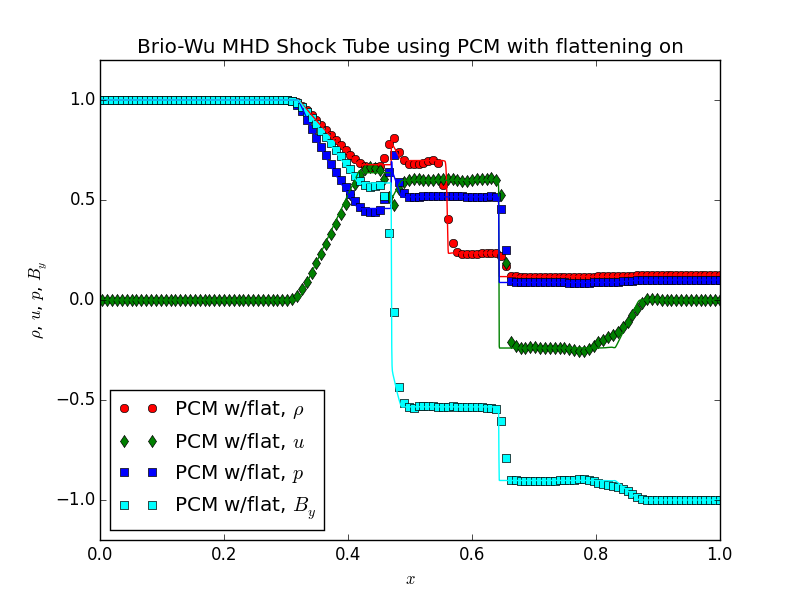}}\\
\subfigure[][]{
\includegraphics[width=2.5in, trim= 0.7in 0.3in 0.in 0.35in,clip=true]{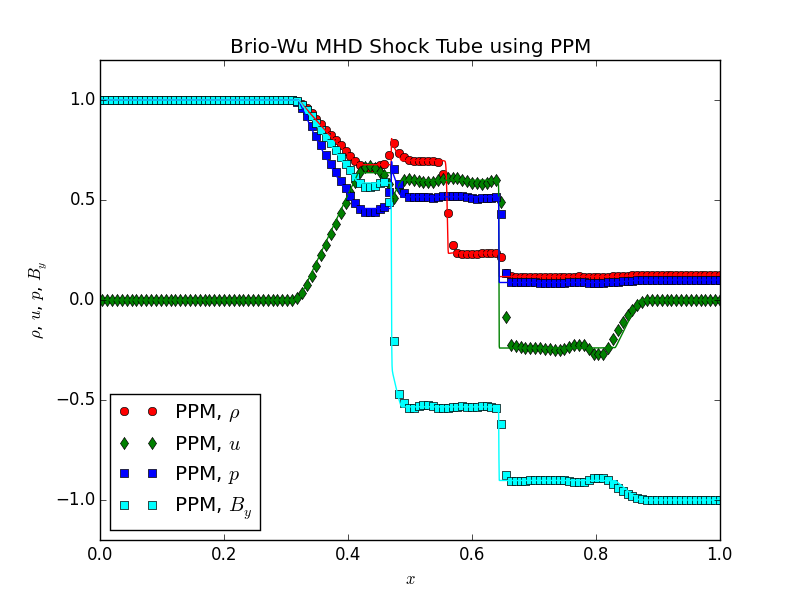}}
\subfigure[][]{
\includegraphics[width=2.5in, trim= 0.7in 0.3in 0.in 0.35in,clip=true]{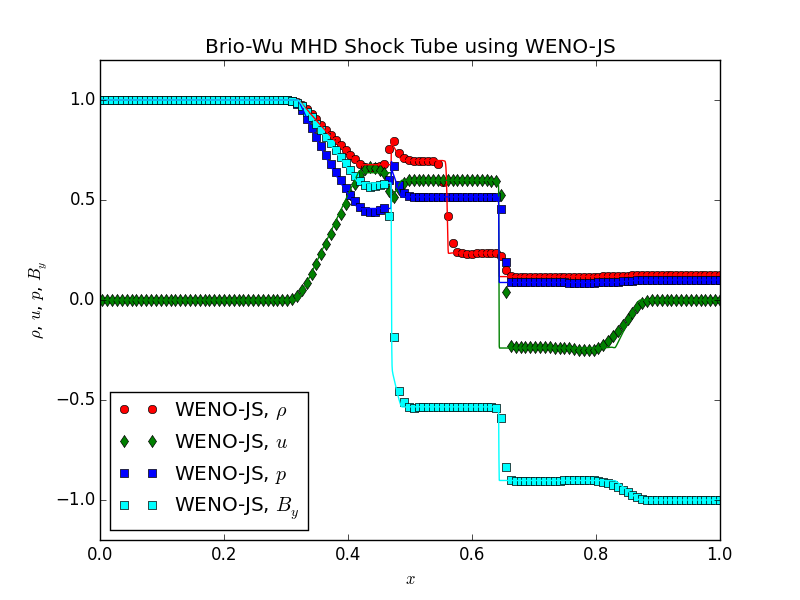}}\\
\end{tabular}
\caption{The Brio-Wu MHD shock tube problem. The Roe Riemann solver is consistently used for all methods. 
The test problems are computed on $N_x=128$, while the high-resolution reference solution is 
obtained using PLM on $N_x=2048$ with the MC slope limiter. The same slope limiter was used in PPM too.
The Courant number is fixed as $0.8$ in all runs.}
\label{Fig:BrioWu}
\end{figure}

An MHD version of the Sod's shock tube problem was first studied by
Brio and Wu \cite{brio1988upwind}, and it has become a must-to-do test
for MHD codes. Since then, the problem has revealed a couple of interesting
findings including not only the discovery of the compound wave \cite{brio1988upwind},
but also the existence of non-unique solutions \cite{torrilhon2003uniqueness,torrilhon2003non}.
More recently, Lee \cite{lee2011upwind} realized that there are unphysical numerical oscillations
in using PPM and studied an approach to suppress the level of oscillations based on the upwind
slope limiter. The presence of such oscillations in PPM has been also briefly reported in \cite{stone2008athena}.
The study reported in  \cite{lee2011upwind} shows that the origin of the oscillations arise from 
the numerical nature of a slowly moving shock as a function of
the magnetic strength of tangential component. 
The slowly moving shock was first identified in \cite{woodward1984numerical}, 
and the oscillatory behaviors have been studied  by many researchers 
for more than 30 years,
yet there is no ultimate resolution 
\cite{karni1997computations,stiriba2003numerical,
arora1997postshock,jin1996effects,roberts1990behavior,johnsen2008numerical}.

In this test, as just mentioned, 
there are observable numerical oscillations found in all methods, PPM, PCM and WENO-JS + charTr.
The results in Fig. \ref{Fig:BrioWu} show that the oscillations are the largest in PPM, consistent with the findings in \cite{lee2011upwind},
and there are less amount in PCM and WENO-JS + charTr. The PPM solutions are suffering from significant amount of
spurious oscillations in all four variables, $\rho, u, p$ and $B_y$, as shown in Fig. \ref{Fig:BrioWu}(c).
Such behaviors are less significant in PCM and WENO-JS + charTr, respectively illustrated in 
Fig. \ref{Fig:BrioWu}(a) and Fig. \ref{Fig:BrioWu}(d), in that, the oscillations in $\rho, p, B_y$ 
are much more controlled now,
while the most outstanding oscillations are found in $u$ near $x\approx0.8$.
Again, Fig. \ref{Fig:BrioWu}(b) shows that the PCM flattening makes PCM to perform very similar to PPM.
It is worth mentioning that the oscillatory behaviors remain to be consistent regardless 
of the choice of Riemann solvers such as 
HLL \cite{harten1997upstream},
HLLC \cite{li2005hllc}, 
HLLD \cite{miyoshi2005multi}, or 
Roe \cite{roe1981approximate} (tested here).

\paragraph{\underline{(f) RJ2a MHD Shock Tube}}

\begin{figure}[ht!]
\centering
\begin{tabular}{cccc}
\subfigure[][]{
\includegraphics[width=2.5in,  trim= 0.7in 0.3in 0.in 0.35in,clip=true]{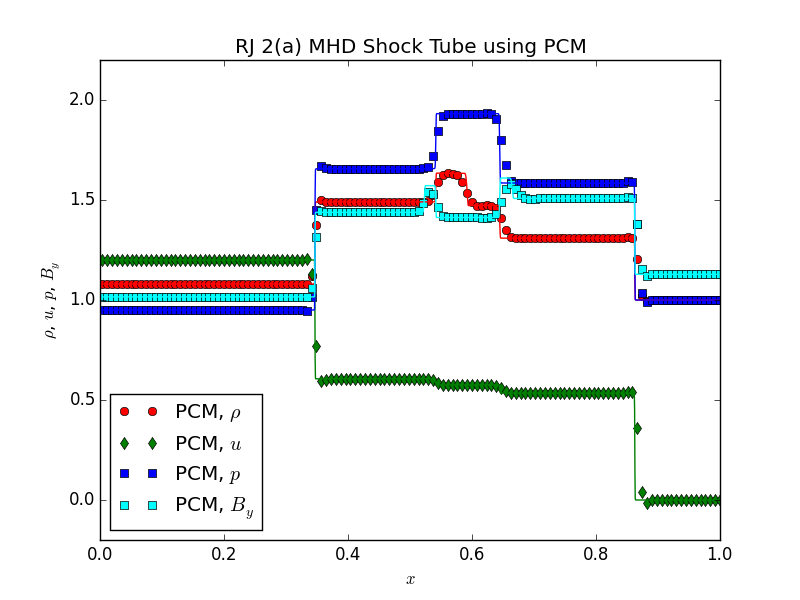}}
\subfigure[][]{
\includegraphics[width=2.5in,  trim= 0.7in 0.3in 0.in 0.35in,clip=true]{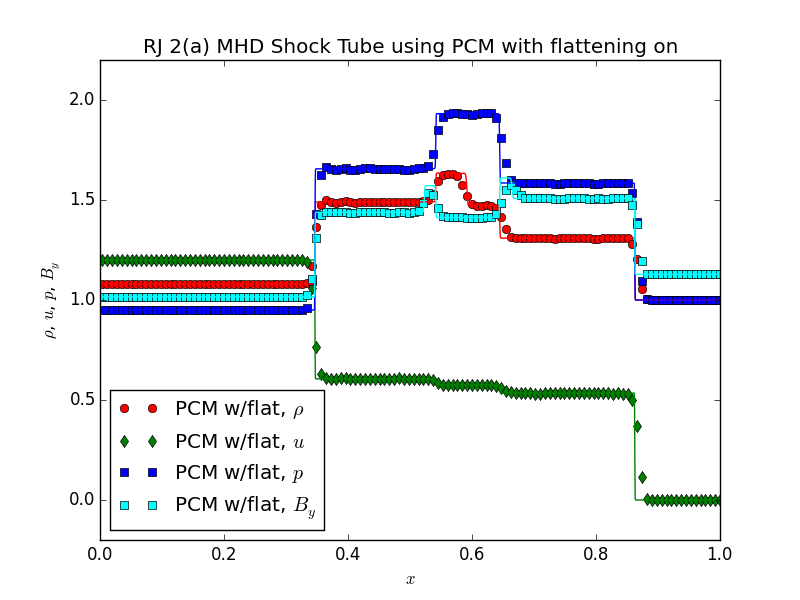}}\\
\subfigure[][]{
\includegraphics[width=2.5in,  trim= 0.7in 0.3in 0.in 0.35in,clip=true]{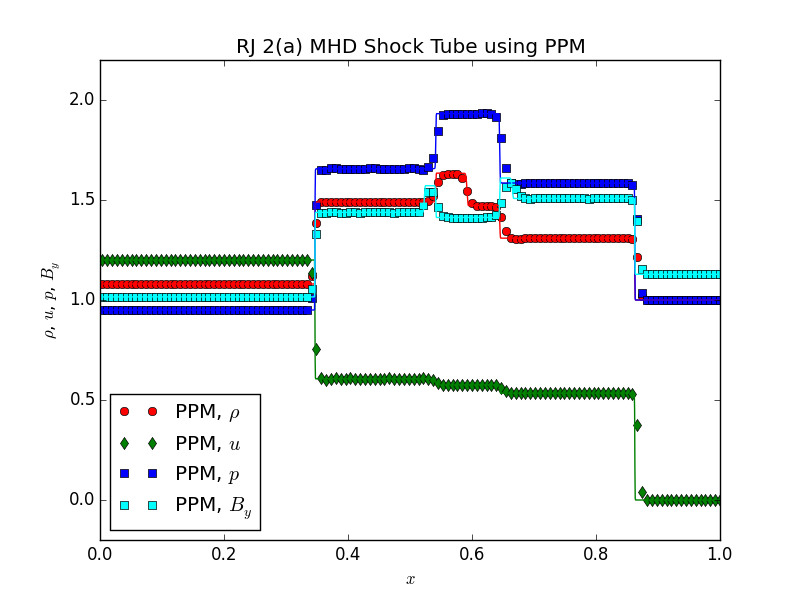}}
\subfigure[][]{
\includegraphics[width=2.5in,  trim= 0.7in 0.3in 0.in 0.35in,clip=true]{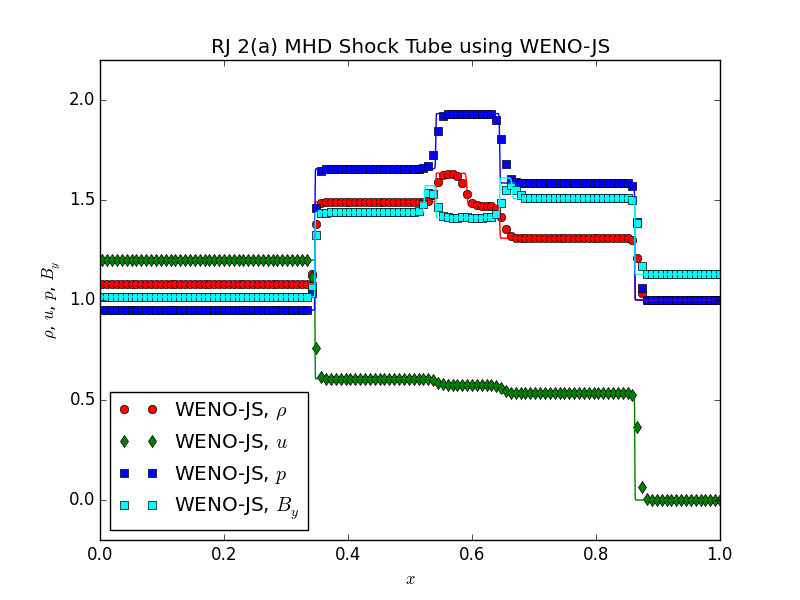}}\\
\end{tabular}
\caption{The RJ2a MHD shock tube problem. All runs used the HLLD Riemann solver, computed on $N_x=128$
with $C_{\mbox{cfl}}=0.8$.
The reference solution was obtained using PLM on $N_x=2048$. The PLM and PPM methods used the MC slope
limiter.}
\label{Fig:RJ2a}
\end{figure}

Ryu et al. \cite{ryu1994numerical} studied a class of one dimensional MHD shock tube problems
that are informative to run as a code verification test. We have chosen one of their setups, introduced
in their figure 2a. In what follows the problem is referred to be as the RJ2a test.
The viewpoint of this test is to monitor if all three dimensional MHD waves are successfully captured.
We see that in Fig. \ref{Fig:RJ2a} all structures of left- and right-going fast shocks, 
left- and right-going slow shocks, and a contact discontinuity are well captured in all methods tested,
including PCM.

\subsection{2D Tests}
We present two dimensional tests of hydrodynamics and MHD in this section.
All test cases are computed using the second-order dimension-by-dimension
extension of the baseline 1D algorithms, including PCM.

\subsubsection{2D Convergence Test of the Isentropic Vortex Advection}

\begin{figure}
\centering
\includegraphics[width=3.4in]{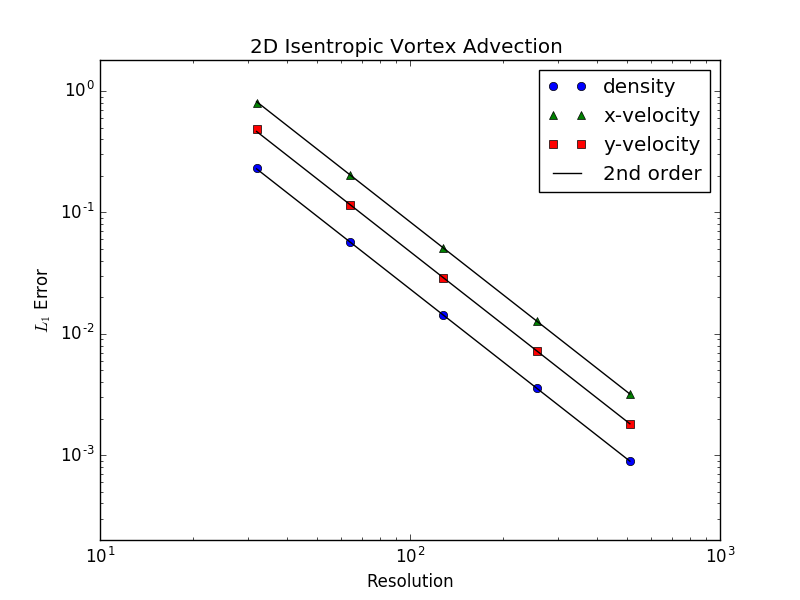}
\caption{Convergence test of the 2D isentropic vortex advection problem.
The errors in $\rho, u$, and $v$ are calculated in $L_1$ sense against the initial conditions. 
The tested PCM solutions are solved on $N_x \times N_y$, where $N_x=N_y= 32, 64, 128, 256$ 
and $512$. All runs reached to $t=10$ using the HLLC Riemann solver with $C_{\mbox{cfl}}=0.8$.}
\label{Fig:2dIsentropicVortex}
\end{figure}

The first 2D test problem, considered in \cite{yee2000entropy}, consists of the advection of an
isentropic vortex along the diagonal of a cartesian computational box.
The dynamics of the problem allows to quantify a code's dissipative properties and the correct
discretization balance of multidimensional terms through monitoring the preservation
of the initial circular shape of the vortex. At $t=10$ the vortex finishes one periodic advection over
the domain and returns to the initial position, where we can measure the solution accuracy against
the initial condition. 
As such we have chosen this problem particularly to access the PCM's order of convergence rate in 2D.
We omit the details of the initial problem setup which can be found in \cite{yee2000entropy}.

As expected, the results presented in Fig. \ref{Fig:2dIsentropicVortex} clearly confirm that, when PCM is
extended to 2D using the simple dimension-by-dimension formulation, the overall numerical
solution accuracy converges in second-order, regardless of its inherent fifth-order property in 1D.

\subsubsection{2D Discontinuous Tests}

\paragraph{\underline{(a) Sedov}}

\begin{figure}[ht!]
\centering
\begin{tabular}{cccc}
\subfigure[][]{
\includegraphics[width=2.5in, trim= 2.2in 1.6in 0.in 0.5in,clip=true]{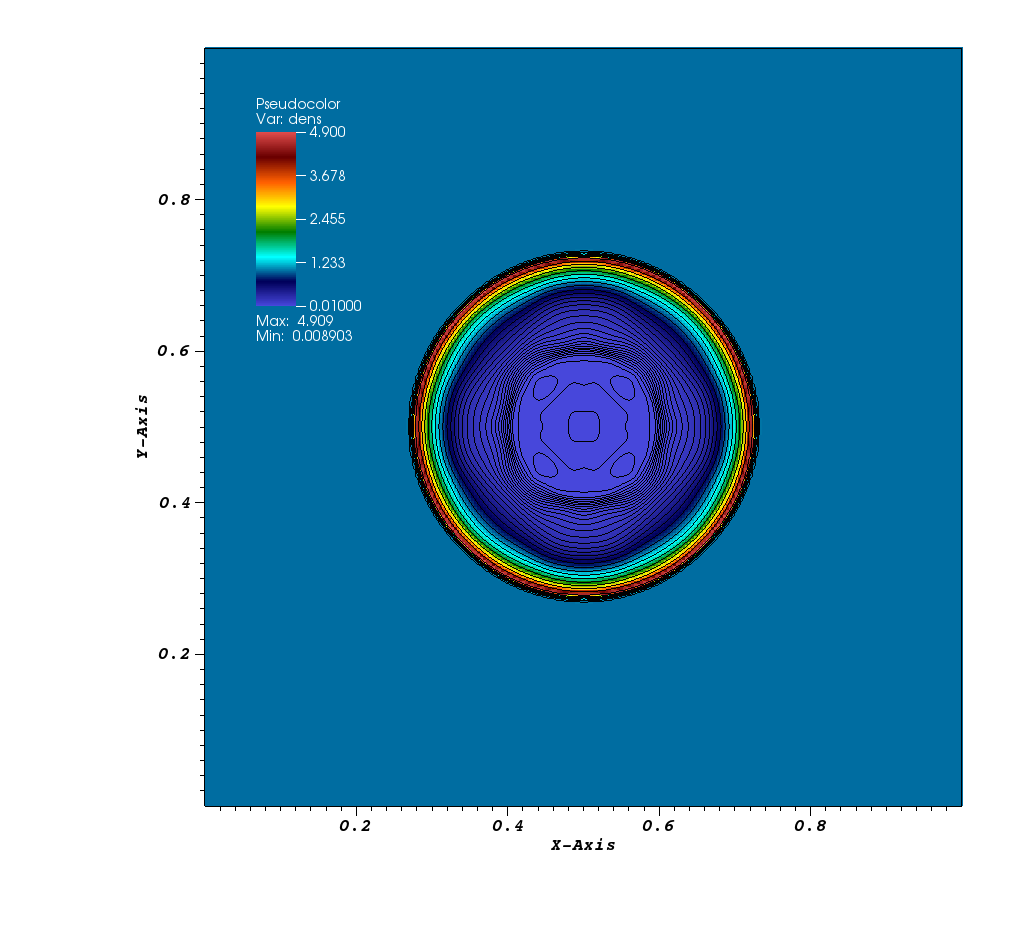}}
\subfigure[][]{
\includegraphics[width=2.5in, trim= 2.2in 1.6in 0.in 0.5in,clip=true]{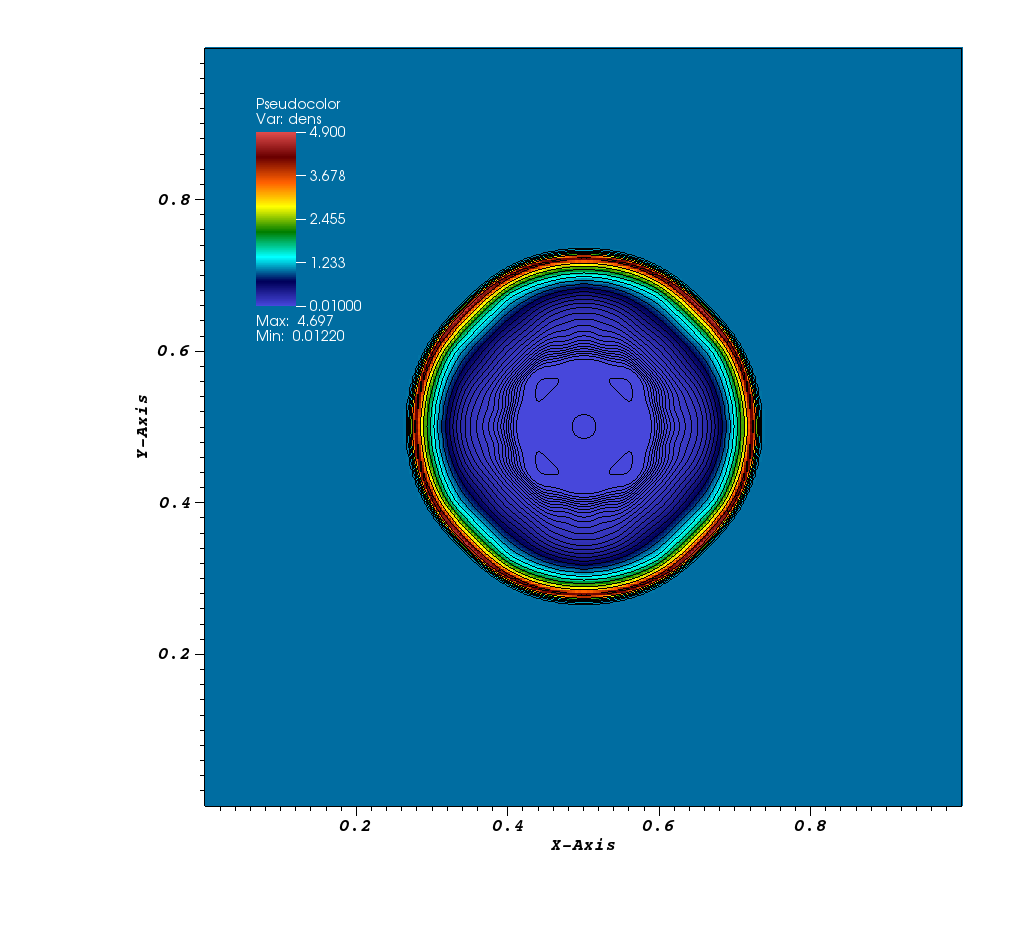}}\\
\subfigure[][]{
\includegraphics[width=2.5in, trim= 2.2in 1.6in 0.in 0.5in,clip=true]{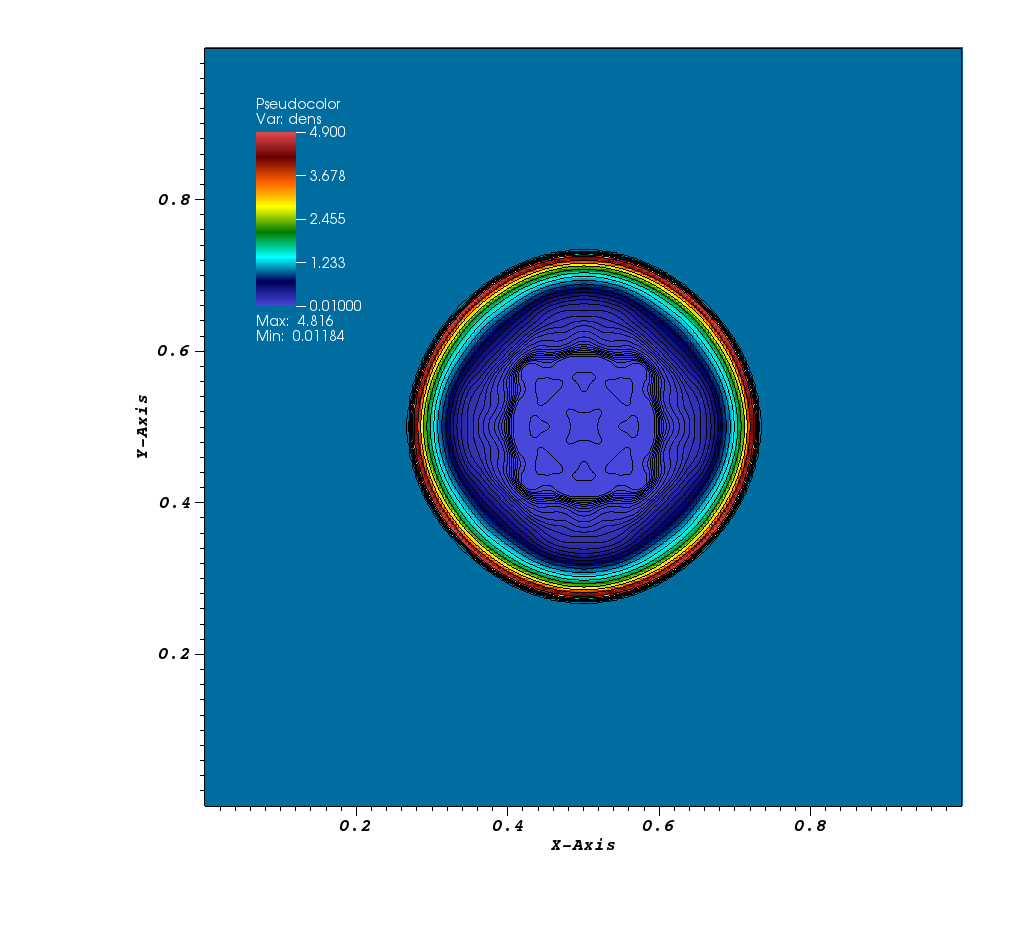}}
\subfigure[][]{
\includegraphics[width=2.5in, trim= 2.2in 1.6in 0.in 0.5in,clip=true]{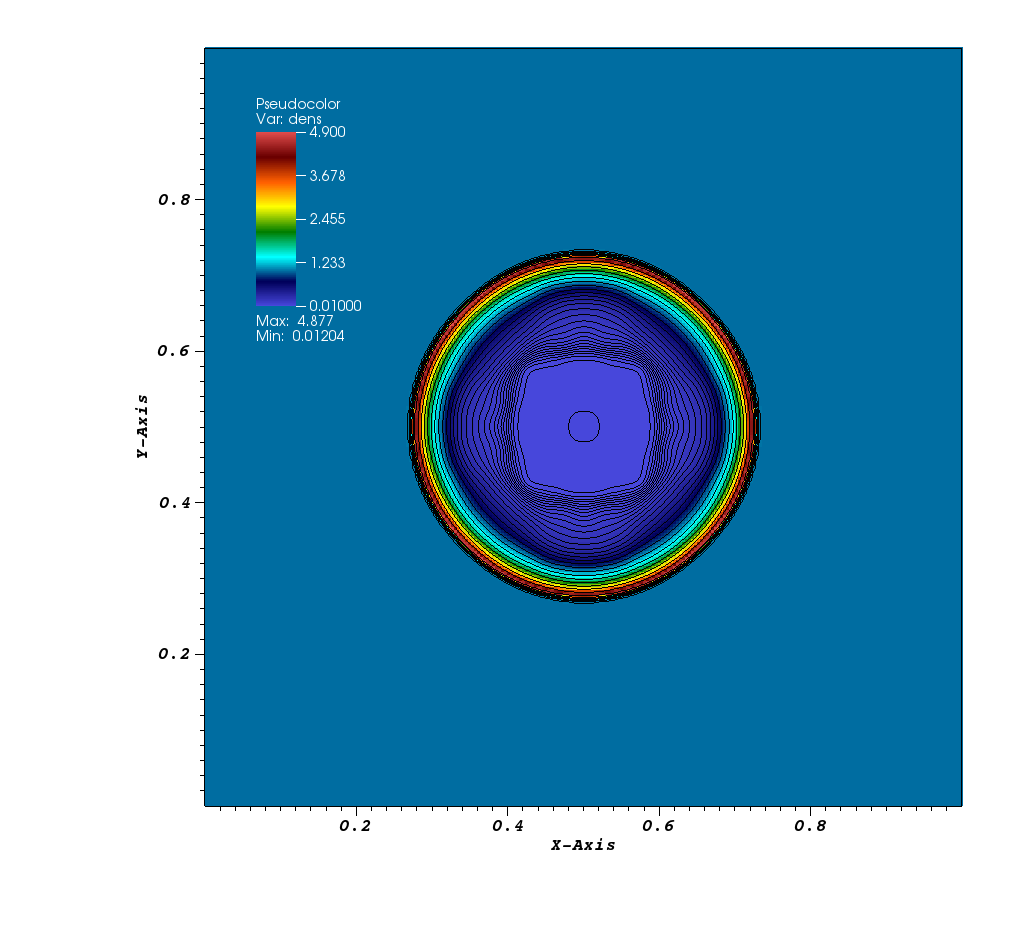}}
\end{tabular}
\caption{The Sedov explosion test. (a) PCM without flattening, (b) PCM with flattening on, (c) PPM, and (d) WENO-JS + charTr.
All calculations are done on $256 \times256$ cells with the HLLC Riemann solver and with $C_{\mbox{cfl}}=0.8$.
PPM is computed using the MC slope limiter. 30 equally spaced levels of density contour lines are shown in logarithmic scale.}
\label{Fig:Sedov_PCM_a}
\end{figure}

We consider the Sedov blast test \cite{sedov1993similarity} to check PCM's ability 
to handle a spherical symmetry of the strong hydrodynamical shock explosion. 
The problem studies a self-similar evolution of a spherical shock wave propagation
due to an initial point-source of a highly pressurized perturbation. The test has been 
used widely in various literatures, and we follow the same setup found in \cite{fryxell2000flash}.
Panels in Fig. \ref{Fig:Sedov_PCM_a} show the density field in linear scale
at $t=0.05$ resolved on a grid size of
$256\times 256$ for the domain $[0,1]\times [0,1]$. 
The HLLC Riemann solver was used in all runs with $C_{\mbox{cfl}}=0.8$.
The range of the plotted densities in colors in all four panels is $0.01 \le \rho \le 4.9$,
the same is also used for the 30 levels of density contour lines that are plotted in logarithmic scale.

\begin{figure}[th!]
\centering
\begin{tabular}{cccc}
\subfigure[][]{
\includegraphics[width=2.5in, trim= 2.2in 1.6in 0.in 0.5in,clip=true]{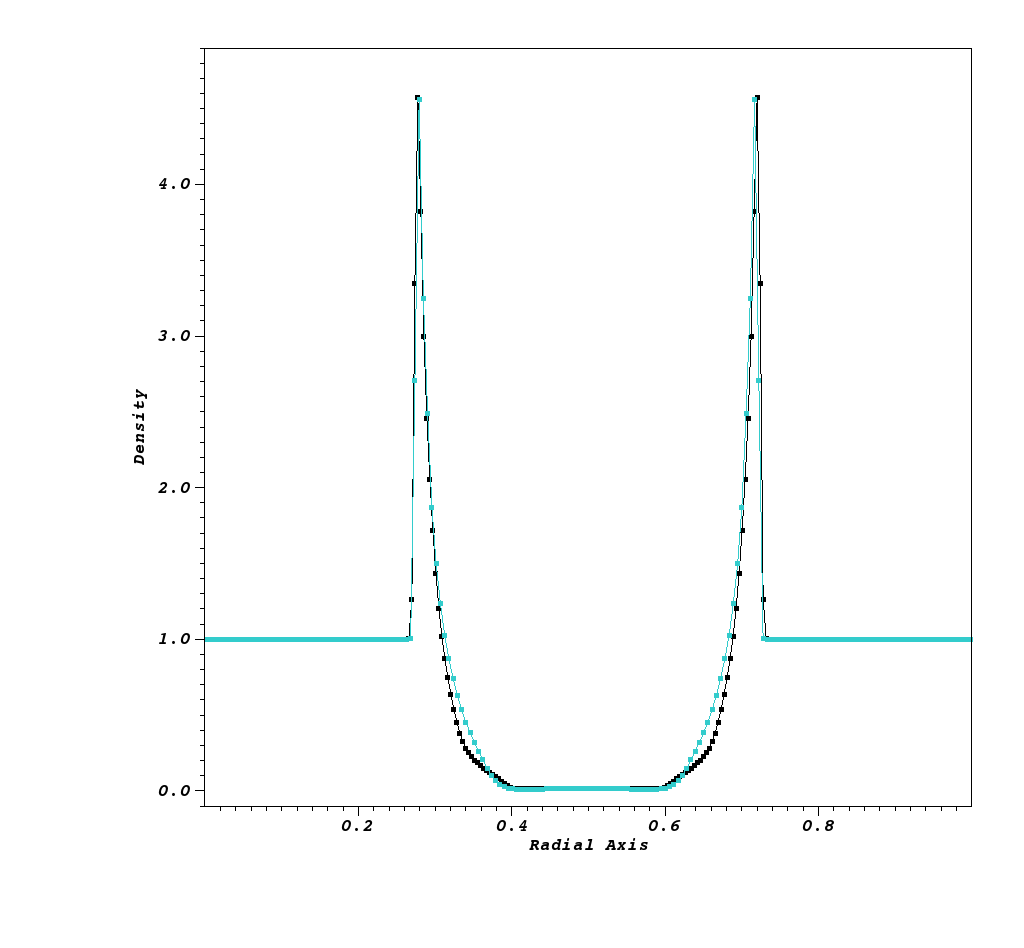}}
\subfigure[][]{
\includegraphics[width=2.5in, trim= 2.2in 1.6in 0.in 0.5in,clip=true]{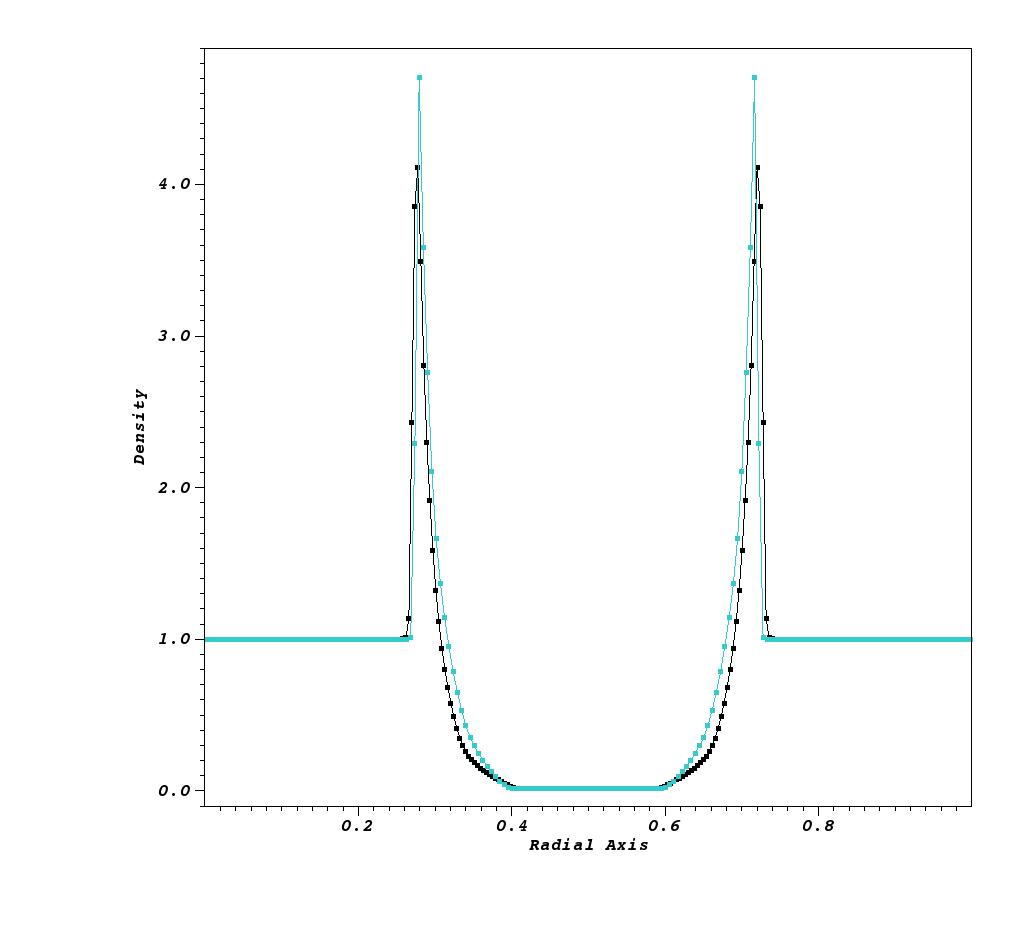}}\\
\subfigure[][]{
\includegraphics[width=2.5in, trim= 2.2in 1.6in 0.in 0.5in,clip=true]{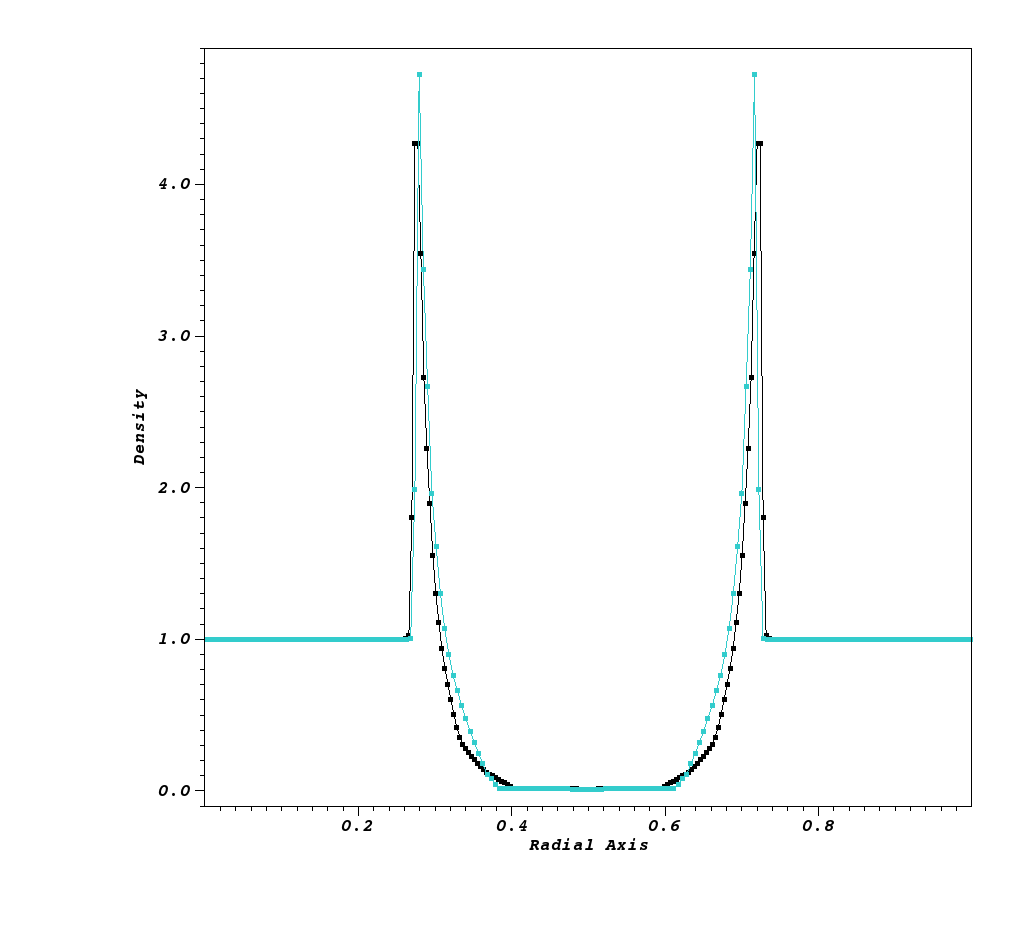}}
\subfigure[][]{
\includegraphics[width=2.5in, trim= 2.2in 1.6in 0.in 0.5in,clip=true]{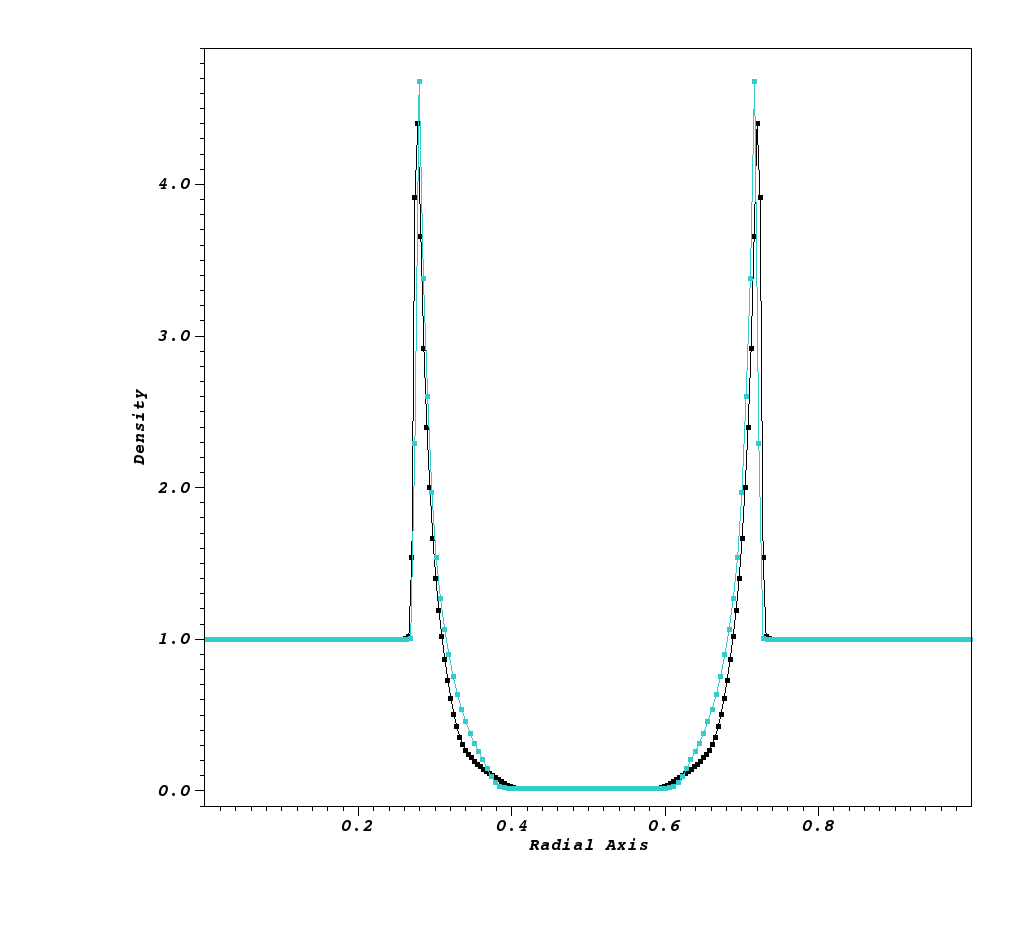}}

\end{tabular}
\caption{Two section cuts of $\rho$ from the Sedov test in Fig. \ref{Fig:Sedov_PCM_a}. 
The two lines represent section cuts of densities along $y=0.5$ (black curves) and $y=x$ (cyan curves).
(a) PCM without flattening, (b) PCM with flattening on, (c) PPM, and (d) WENO-JS + charTr.}
\label{Fig:Sedov_PCM_b}
\end{figure}

As can be seen, the PCM solution in Fig. \ref{Fig:Sedov_PCM_a}(a) is superior 
not only in preserving a great deal of the spherical symmetry at the outermost shock front, 
but also in revealing more flow structures in the central low density region, again 
in the most spherical manner. This great ability of preserving the spherical symmetry in PCM
is also found in Fig. \ref{Fig:Sedov_PCM_b}(a) where the two curves are the two section cuts
of density fields along $y=0.5$ (black) and $y=x$ (cyan), respectively. 
We see that the two peak values of each section cut are matching each other very closely
in terms of both their locations and their magnitudes.
In the other schemes there are clearly much larger disagreements in the magnitude of the peak density values.

It is also interesting to note that the PCM flattening makes the symmetry worse, as illustrated in
Fig. \ref{Fig:Sedov_PCM_a}(a) and Fig. \ref{Fig:Sedov_PCM_a}(b). 
For this reason, as well as for the observations we have collected in our 1D results, the default choice in PCM
is to keep the flattening off, unless otherwise stated in what follows.

%
%

\paragraph{\underline{(b) 2D Riemann Problems}}
Next we test PCM for a family of well-known benchmarked Riemann problems whose mathematical
classification was originally put forward by Zhang and Zheng \cite{zhang1990conjecture},
in which the original 16 of admissible configurations were conjectured on polytropic gas.
This claim was corrected by Schultz-Rinne \cite{schulz1993classification} that one of them was impossible,
and the numerical testings for such 15 configurations were studied in \cite{schulz1993numerical}.
Later, Lax and Liu showed  that there are total of 19 genuinely different configurations
available, providing numerical solutions of all 19 cases too \cite{lax1998solution}. See also \cite{chang19952}.
Until today, this family of 2D Riemann problems has been chosen by many people to demonstrate that
their numerical algorithms can predict these 19 configurations successfully in pursuance of code verification purposes
\cite{buchmuller2014improved,kurganov2002solution,don2016hybrid,balsara2010multidimensional}.

We follow the setup as described in \cite{don2016hybrid} in the following two
verification tests, Configuration 3 and Configuration 5. In both cases the calculations show
numerical solutions on $[0,1]\times [0,1]$ using outflow boundary conditions.

\paragraph{(b) -- Configuration 3}
\begin{figure}[pbht!]
\centering
\begin{tabular}{cccc}
\subfigure[][]{
\includegraphics[width=2.4in, trim= 0.7in 1.6in 0.in 0.5in,clip=true]{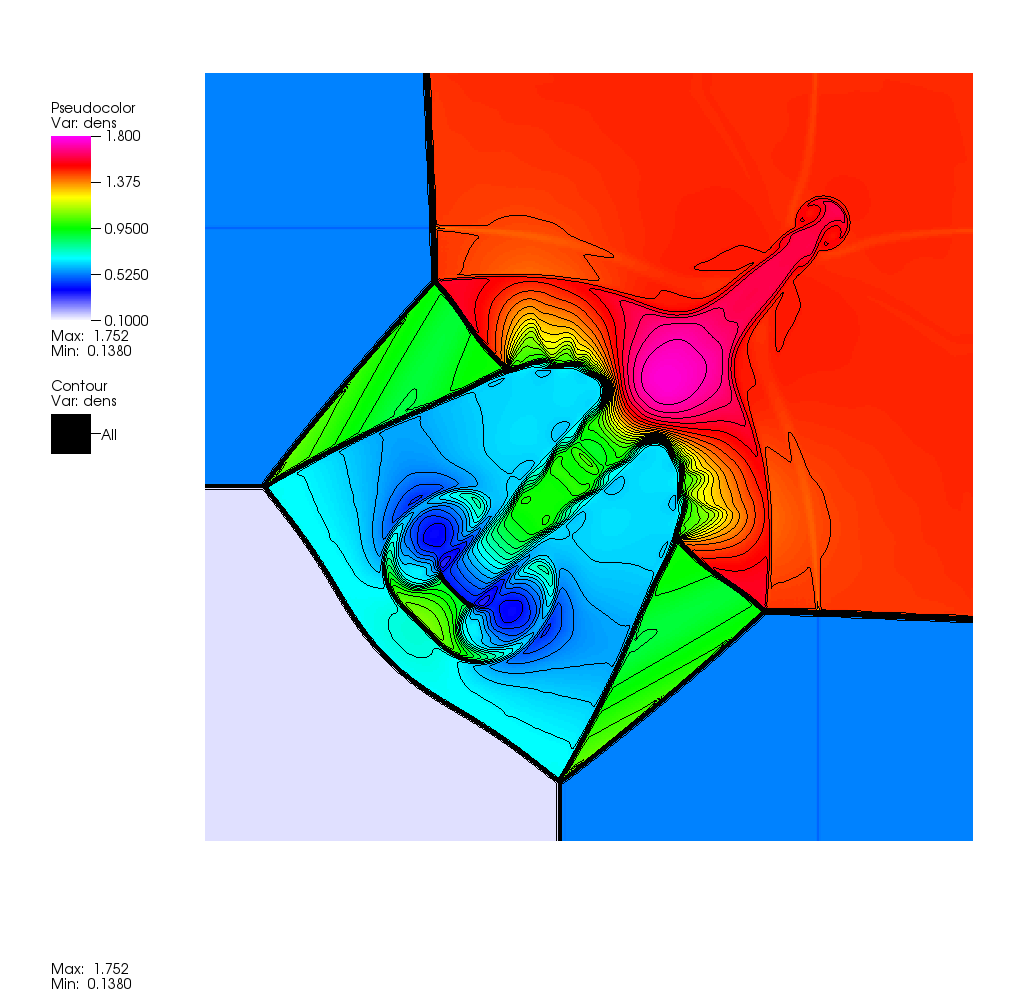}}
\subfigure[][]{
\hspace{-0.1in}
\includegraphics[width=2.4in, trim= 0.7in 1.6in 0.in 0.5in,clip=true]{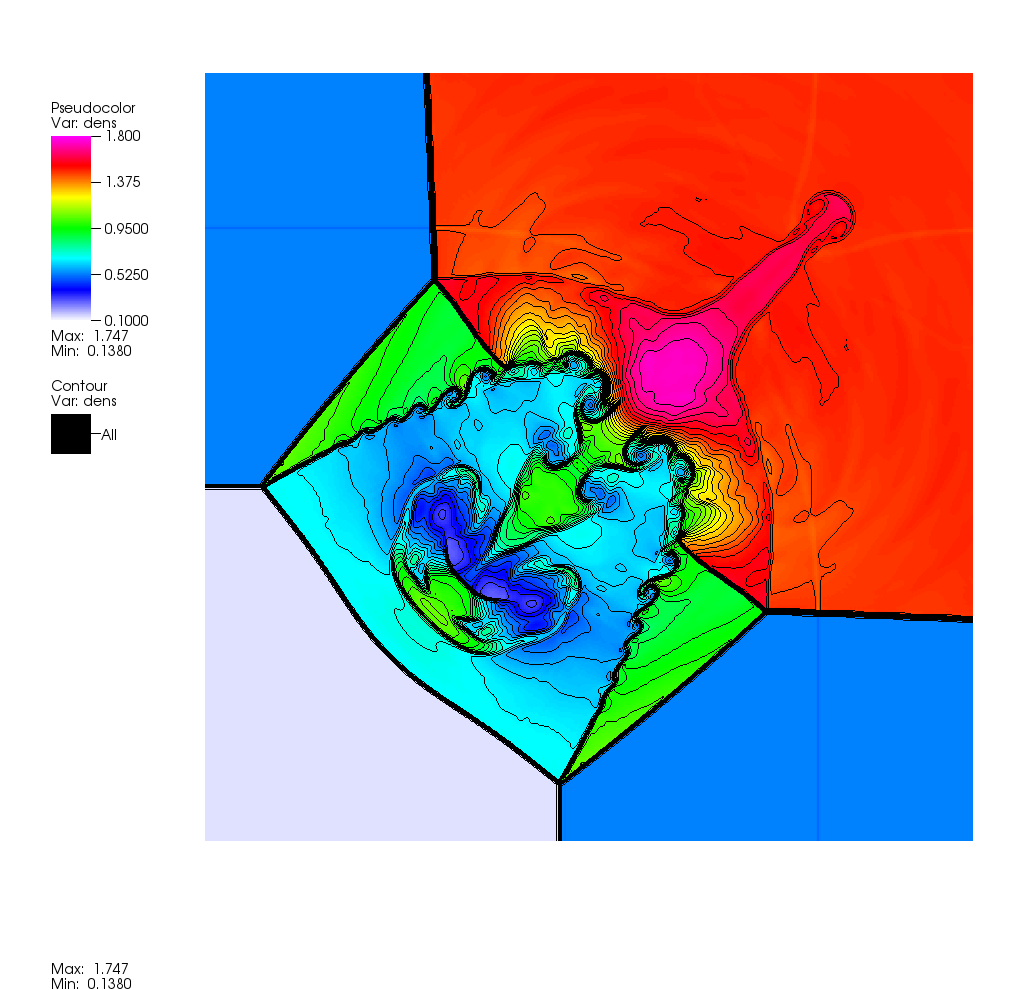}}\\
\subfigure[][]{
\includegraphics[width=2.4in, trim= 0.7in 1.6in 0.in 0.5in,clip=true]{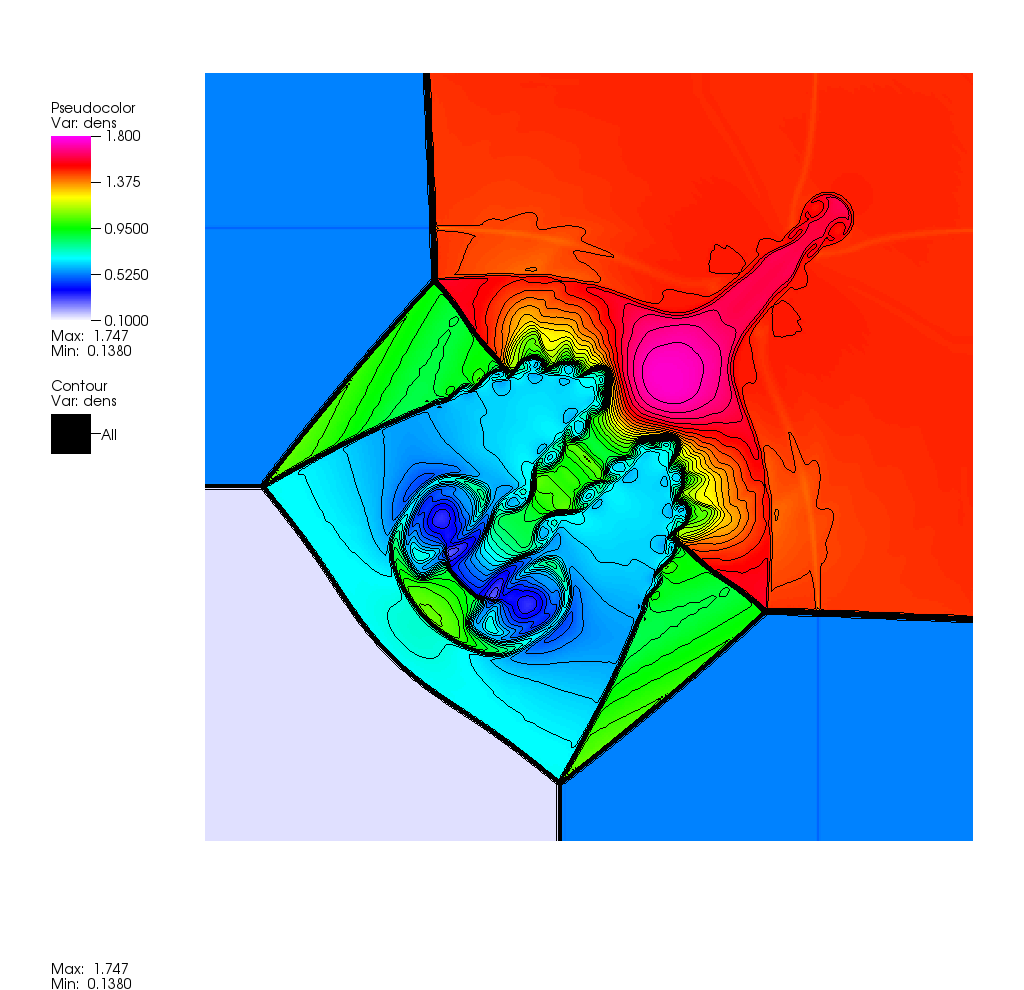}}
\subfigure[][]{
\hspace{-0.1in}
\includegraphics[width=2.4in, trim= 0.7in 1.6in 0.in 0.5in,clip=true]{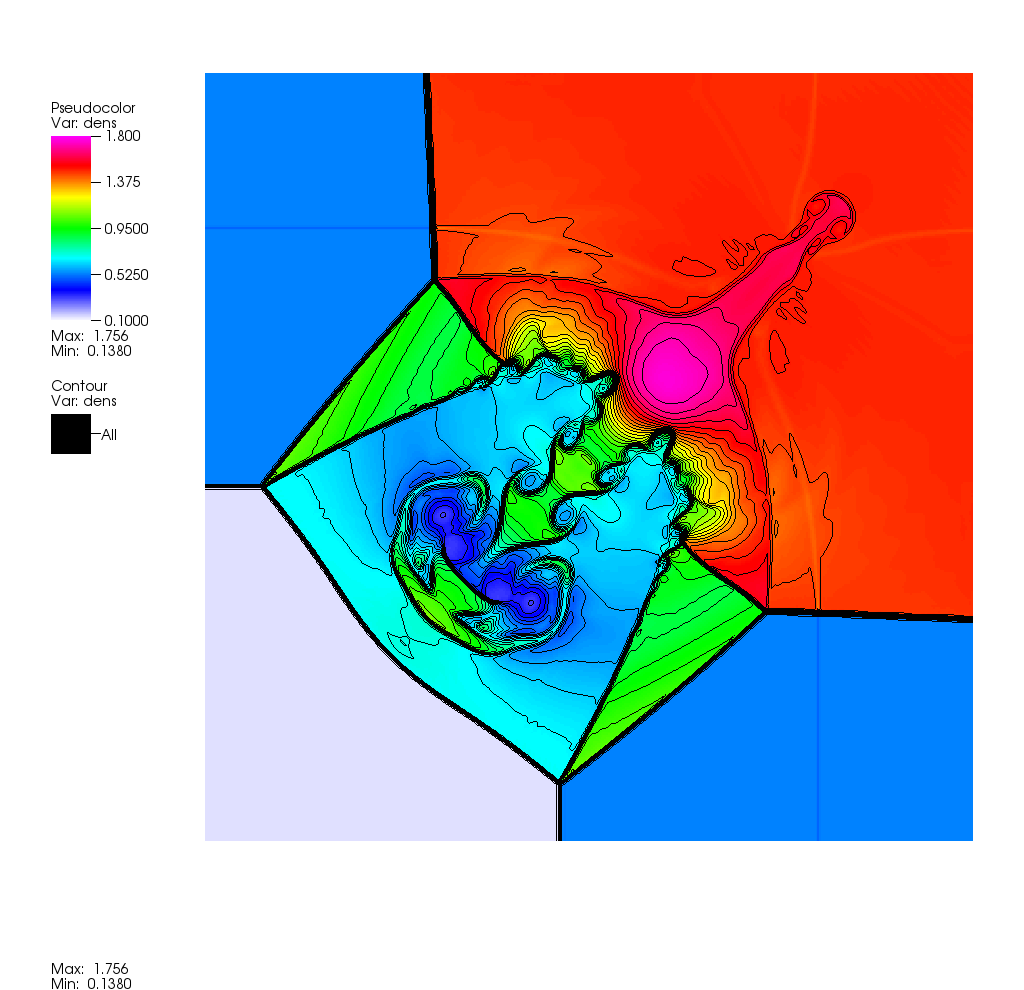}}\\
\subfigure[][]{
\includegraphics[width=2.4in, trim= 0.7in 1.6in 0.in 0.5in,clip=true]{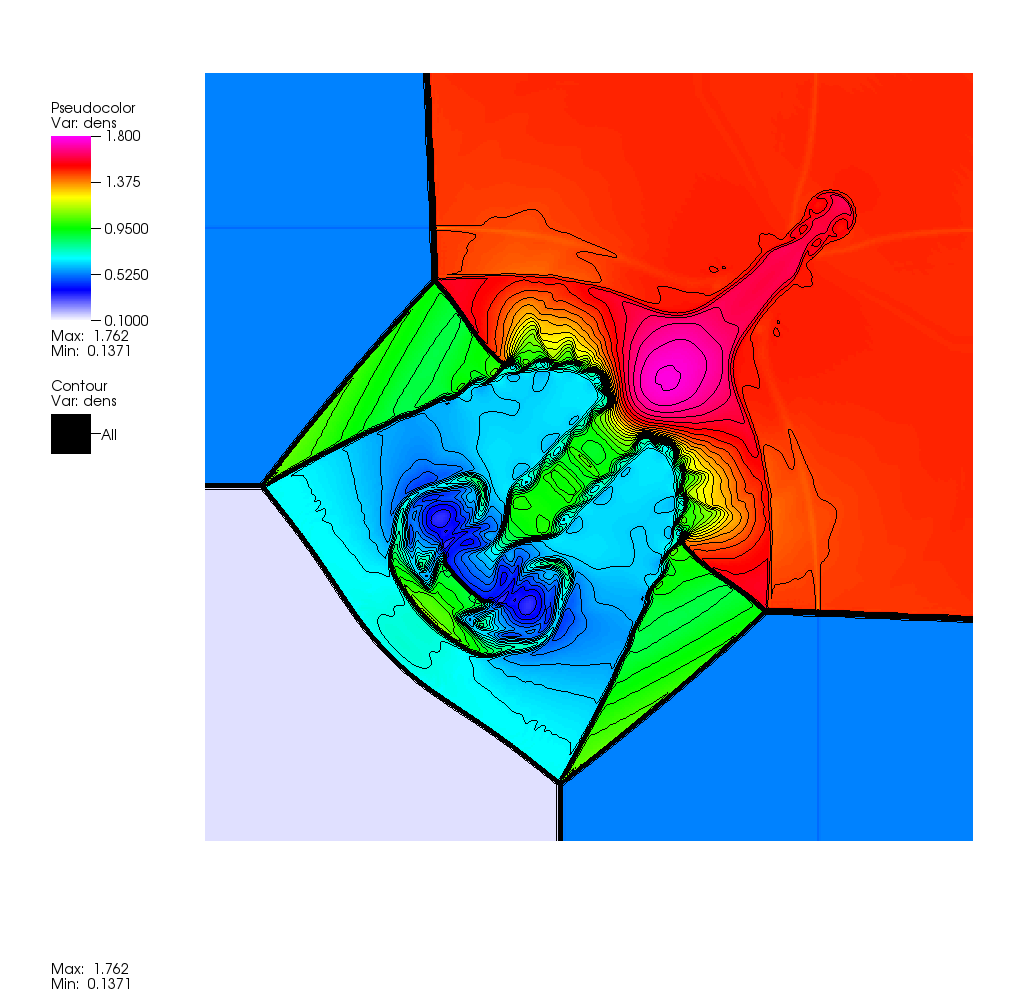}}
\subfigure[][]{
\hspace{-0.1in}
\includegraphics[width=2.4in, trim= 0.7in 1.6in 0.in 0.5in,clip=true]{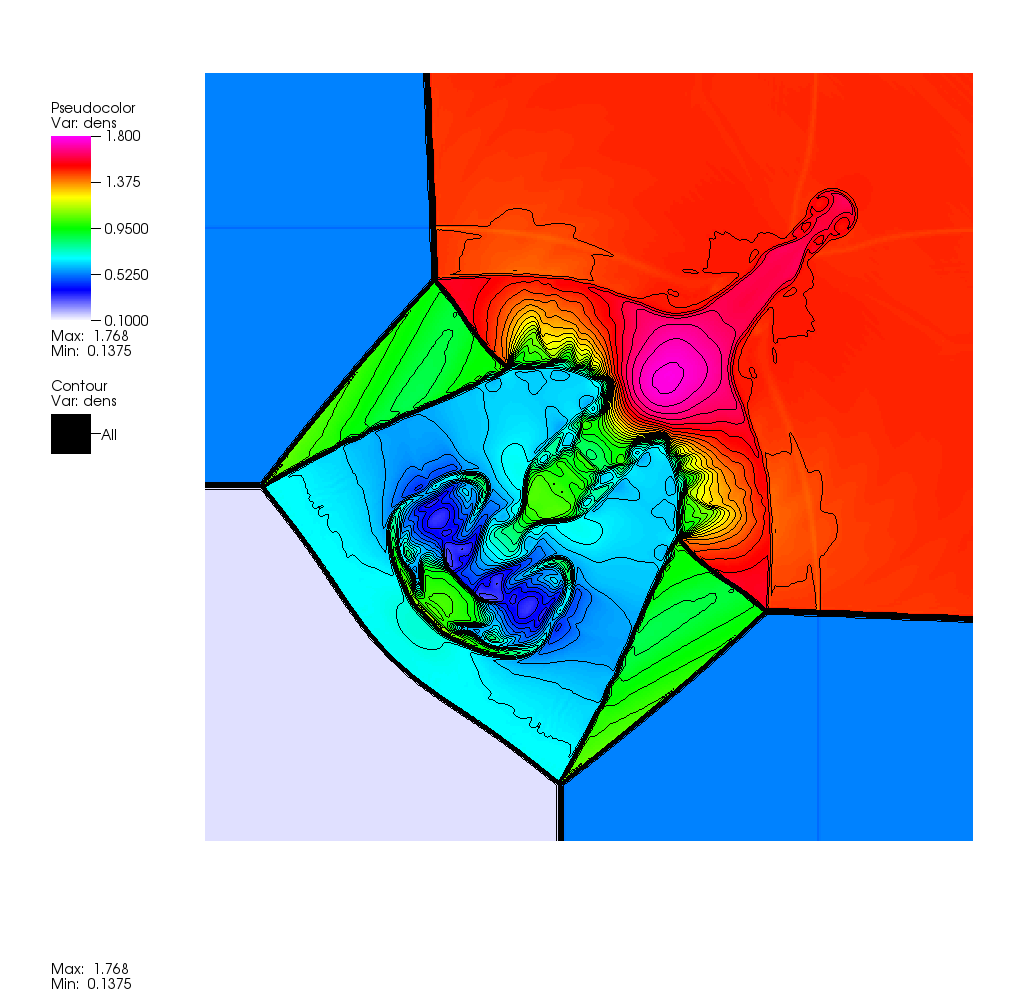}}
\end{tabular}
\caption{2D Riemann Problem -- Configuration 3.
(a) PLM, (b) PPM, 
(c) WENO-JS + CharTr, (d) WENO-Z + CharTr,
(e) PCM-JS, and (f) PCM-Z.
Each panel shows the density values at $t=0.8$ between $[0.1, 1.8]$ in linear scale, calculated using $400\times 400$ grid cells.
The total of 40 contour lines are over-plotted. The MC slope limiter is used in (a) and (b).}
\label{Fig:2dRiemann_conf3}
\end{figure}

Panels in Fig. \ref{Fig:2dRiemann_conf3} show numerical solutions of density at $t=0.8$ resolved on $400\times 400$
using the HLLC Riemann solvers with $C_{\mbox{cfl}}=0.8$. Also shown are the 40 contour lines of $\rho$.
The range of $\rho$ is fixed as $0.1 \le \rho \le 1.8$ in both the pseudo-color figures and the contour lines.
For the PCM method we employed both approaches of 
WENO-JS (see Eq. (\ref{Eq:PCM_WENO5_omega})) and
WENO-Z (see Eq. (\ref{Eq:PCM_WENOZ_omega}))
for the calculations of the smoothness stencils. The figures can be directly compared with Fig. 6 in \cite{don2016hybrid}
where they used the same grid resolution for their hybrid compact-WENO scheme.
First of all, including the two PCM solutions, 
we see that all calculations have produced their solutions successfully, in particular, without
suffering unphysical oscillations near shocks and contact discontinuities. 
This test confirms that the PCM results are well comparable to the other solutions, 
except for some expected minor discrepancies.

One thing to notice is that the PPM solution in Fig. \ref{Fig:2dRiemann_conf3}(b)
has interestingly much more formations of Kelvin-Helmholtz instabilities, identified as vortical rollups
along the slip lines (shown as
the interface boundaries between the green triangular regions and 
the sky blue areas surrounding the mushroom-shaped jet). 
This feature is also often found in a test known as ``Double Mach reflection" 
\cite{woodward1984numerical}, where the similar pattern of rollups are detected 
along the slip line.
As found in various studies \cite{zhang2011order,qiu2005hermite,mignone2007pluto}
it is conventional to say that the amount of such vortical rollups at slip lines
is one of the key factors to measure inherent numerical dissipations in a code.
If we follow this approach, it then leads us to say that the PPM method 
is the least dissipative method among the six methods we tested.
However, we find that this conclusion is somewhat arguable considering the nominal order of
accuracy of PPM is lower than those of WENO-JS + CharTr,  WENO-Z + CharTr and PCM.
We think that there is to be more accurate assessment regarding this type of conclusion.
Readers can find a very similar numerical comparison between PPM and WENO-JS 
in \cite{mignone2007pluto}.

\paragraph{(b) -- Configuration 5}

\begin{figure}[pbht!]
\centering
\begin{tabular}{cccc}
\subfigure[][]{
\includegraphics[width=2.4in, trim= 0.7in 1.6in 0.in 0.5in,clip=true]{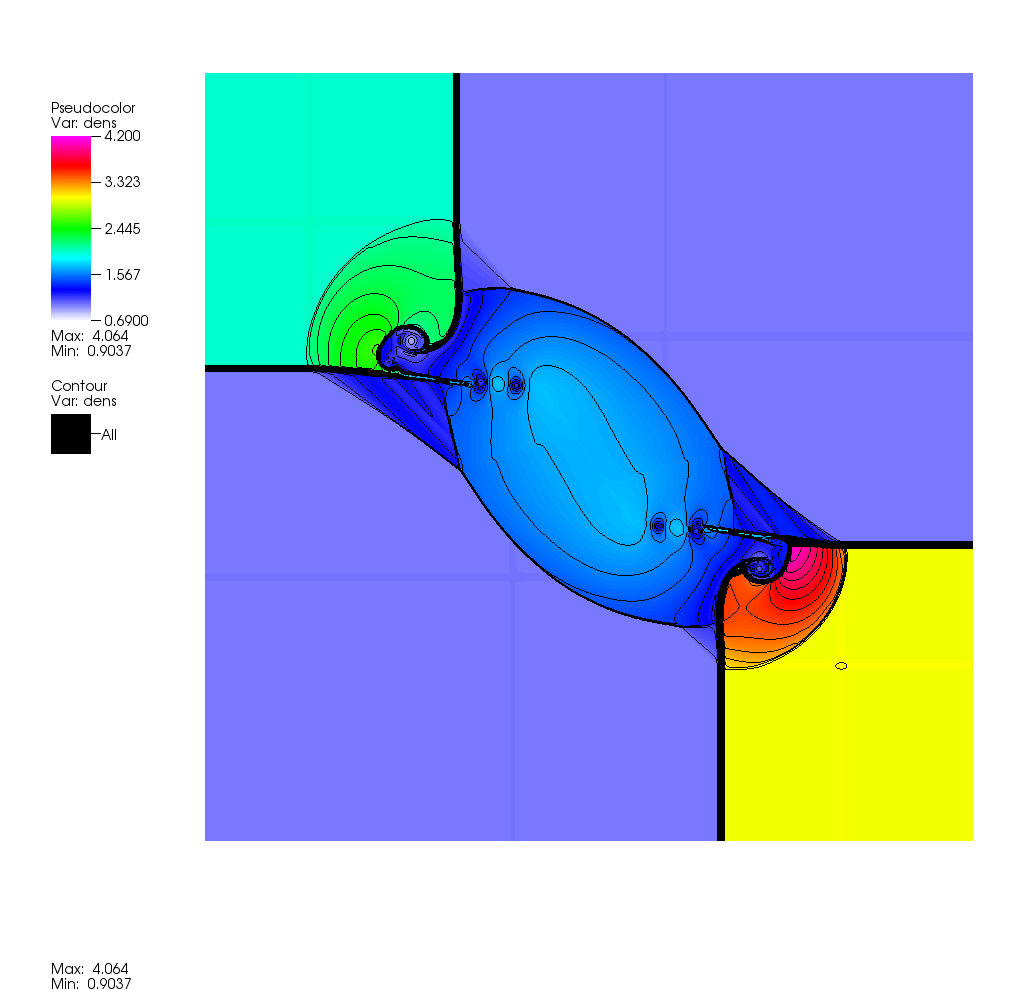}}
\subfigure[][]{
\includegraphics[width=2.4in, trim= 0.7in 1.6in 0.in 0.5in,clip=true]{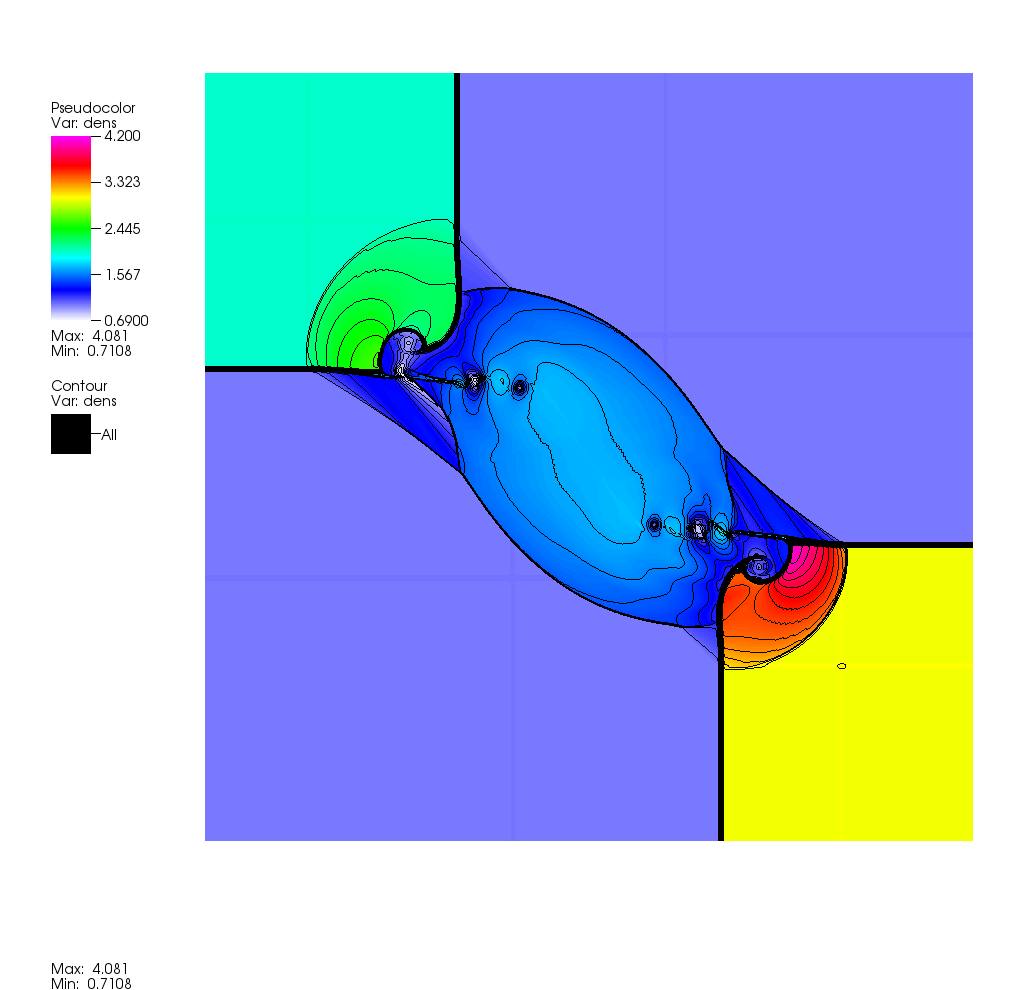}}\\
\subfigure[][]{
\includegraphics[width=2.4in, trim= 0.7in 1.6in 0.in 0.5in,clip=true]{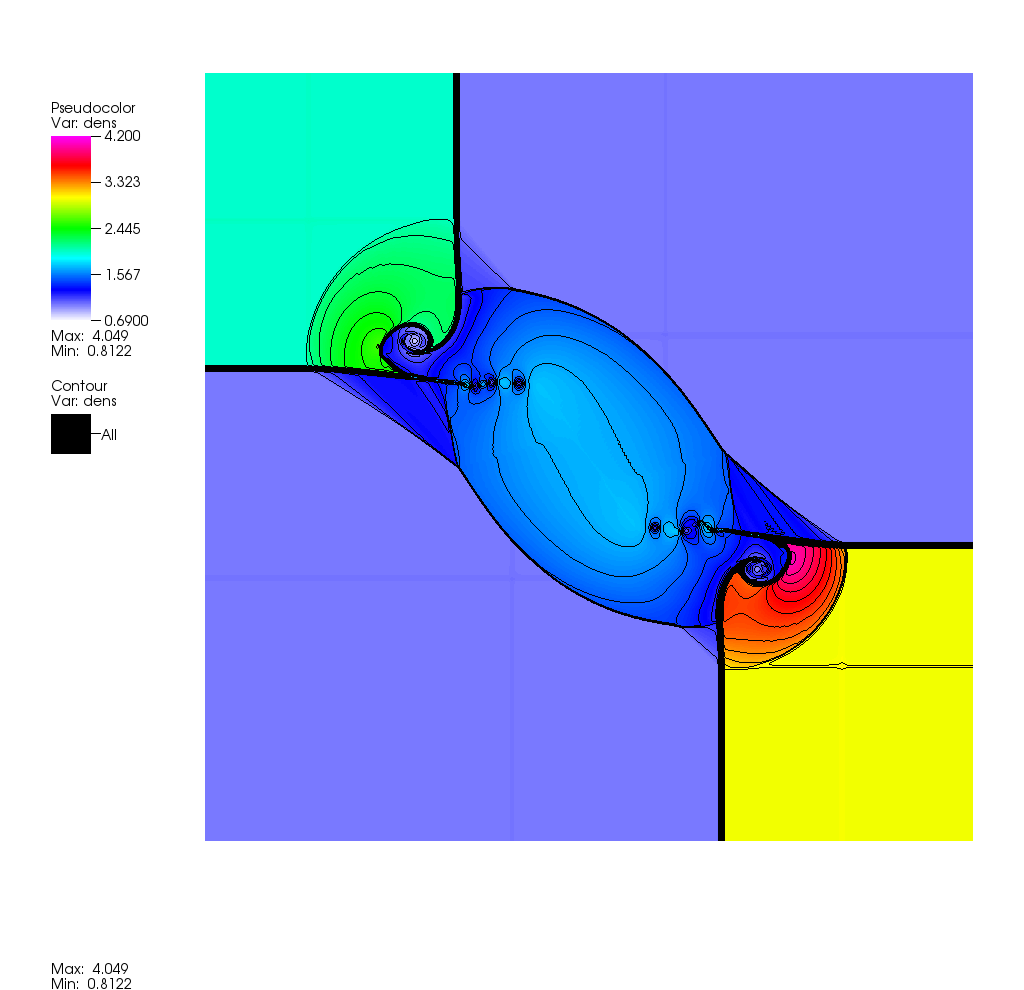}}
\subfigure[][]{
\includegraphics[width=2.4in, trim= 0.7in 1.6in 0.in 0.5in,clip=true]{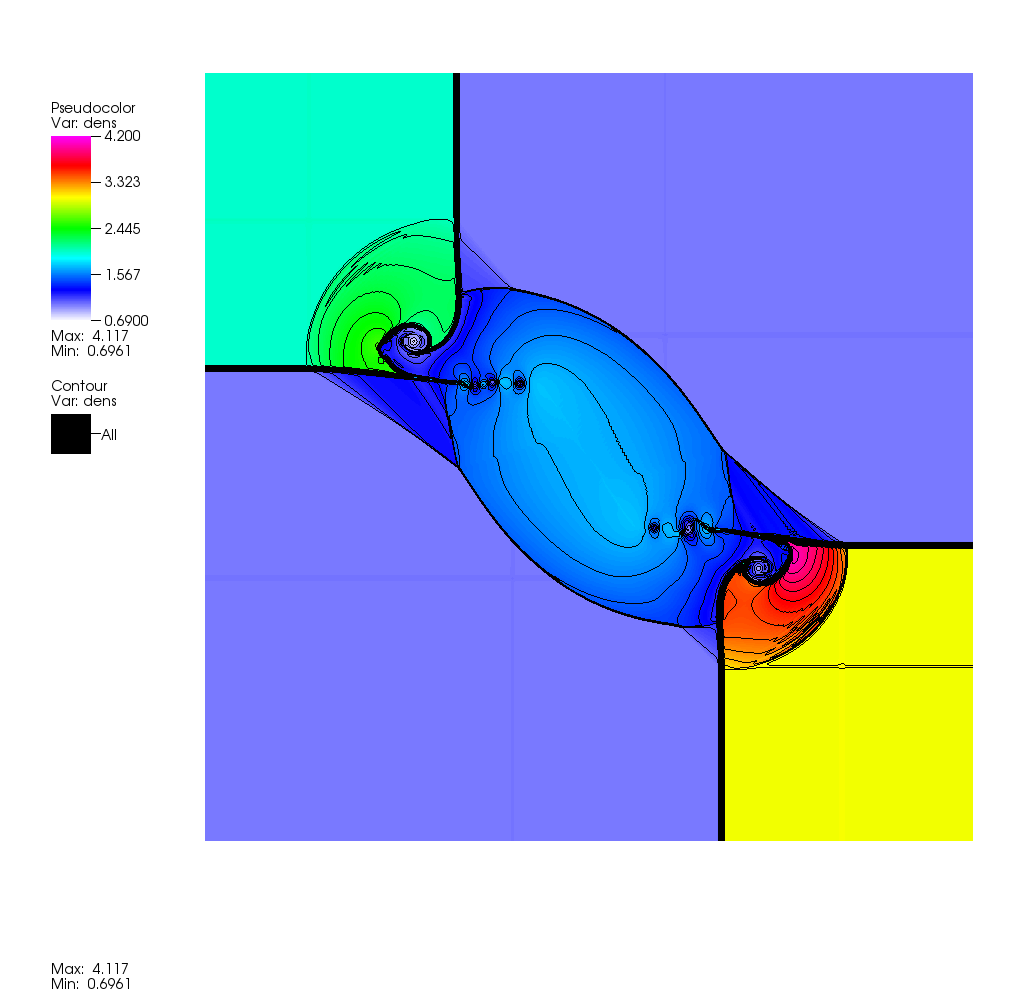}}
\end{tabular}
\caption{2D Riemann Problem -- Configuration 5. 
(a) PLM, (b) PPM,
(c) WENO-JS + CharTr., and (d) PCM-JS.
Each panel shows the density values at $t=0.23$ between $[0.69, 4.2]$ in linear scale, resolved on $1024\times 1024$ grid cells.
The total of 40 contour lines are over-plotted. The MC slope limiter is used in (a) and (b). All runs used the Roe Riemann solver
with  $C_{\mbox{cfl}}=0.8$.}
\label{Fig:2dRiemann_conf5}
\end{figure}
As a second 2D Riemann problem we consider Configuration 5 
to test PCM and compare its solution with three other solutions of PLM, PPM and WENO-JS + CharTr.
The WENO-JS approaches in 
Eqs. (\ref{Eq:WENO5_omega}) and (\ref{Eq:PCM_WENO5_omega}) are adopted for PCM
for the calculations of the smoothness stencils.
The choice of grid resolution is $1024\times 1024$ in this test in order to directly compare our results
with the results reported in Fig. 4 and Fig. 5 of \cite{buchmuller2014improved}.
As obtained in Fig. \ref{Fig:2dRiemann_conf5}, the PCM solution satisfactorily compares well
with the solutions of PPM and WENO-JS + CharTr, as well as with the 
high-resolution results in \cite{buchmuller2014improved}.
As also reported in \cite{buchmuller2014improved,zhang2011order},
when considering discontinuous flows in multiple space dimensions, the dimension-by-dimension approach
works just as fine in terms of producing comparably accurate solutions. 
Likewise, the results in Fig. \ref{Fig:2dRiemann_conf5} 
show that we observe the same qualitative performances in all the methods we tested here, including PCM.

Although all the solutions are comparably admissible, there are few distinctive features in the PCM solution, displayed in
Fig. \ref{Fig:2dRiemann_conf5} (d). We note that the minimum and maximum values of the computed
density are respectively the smallest and the largest among the four results. This trend is consistent 
with the increasing order of accuracy in the four panels here. 
The same observation can be found also in Fig. \ref{Fig:2dRiemann_conf3} too. 
We consider this as an indication that the small scale features
are better resolved in PCM with less amount of numerical diffusion.

%
%
\paragraph{\underline{(c) MHD Rotor}}
\begin{figure}[pbht!]
\centering
\begin{tabular}{cccc}
\vspace{-0.05in}
\subfigure[][]{
\includegraphics[width=3.0in, trim= 0.7in 1.1in 2.in 0.4in,clip=true]{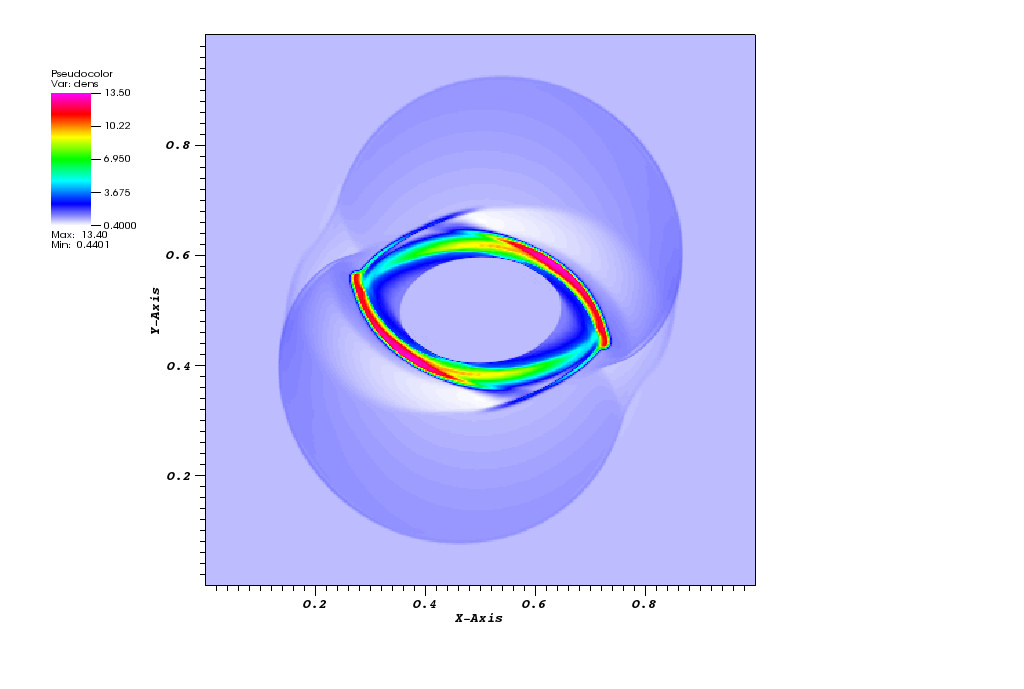}}
\hspace{-0.5in}
\subfigure[][]{
\includegraphics[width=2.85in, trim= 2.3in 1.1in 1.in 0.4in,clip=true]{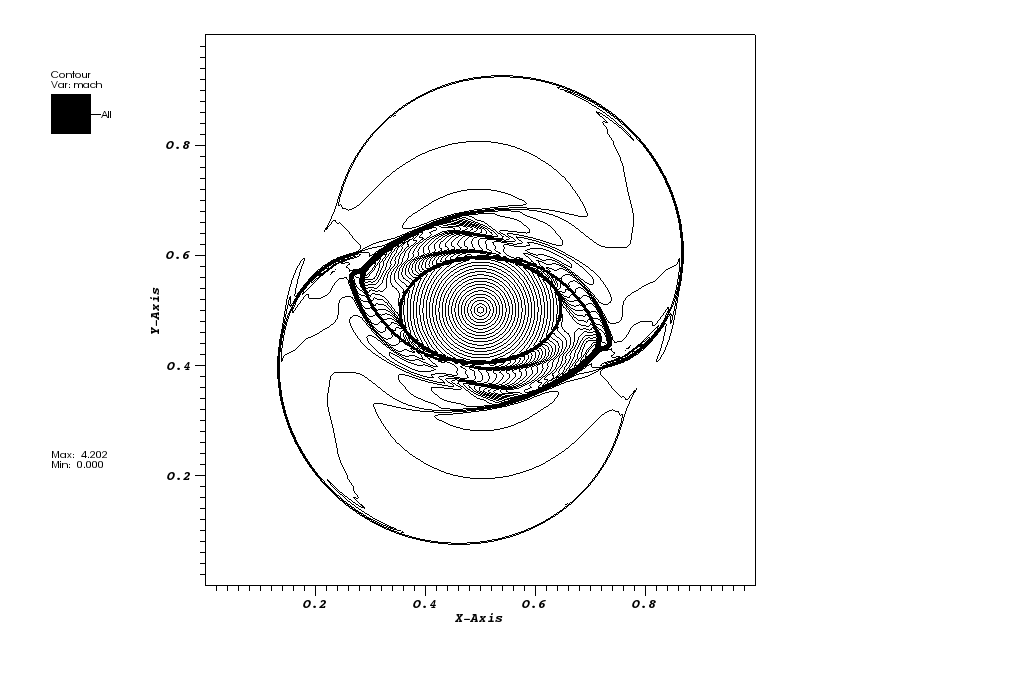}}\\
\vspace{-0.05in}
\subfigure[][]{
\includegraphics[width=3.0in, trim= 0.7in 1.1in 2.in 0.4in,clip=true]{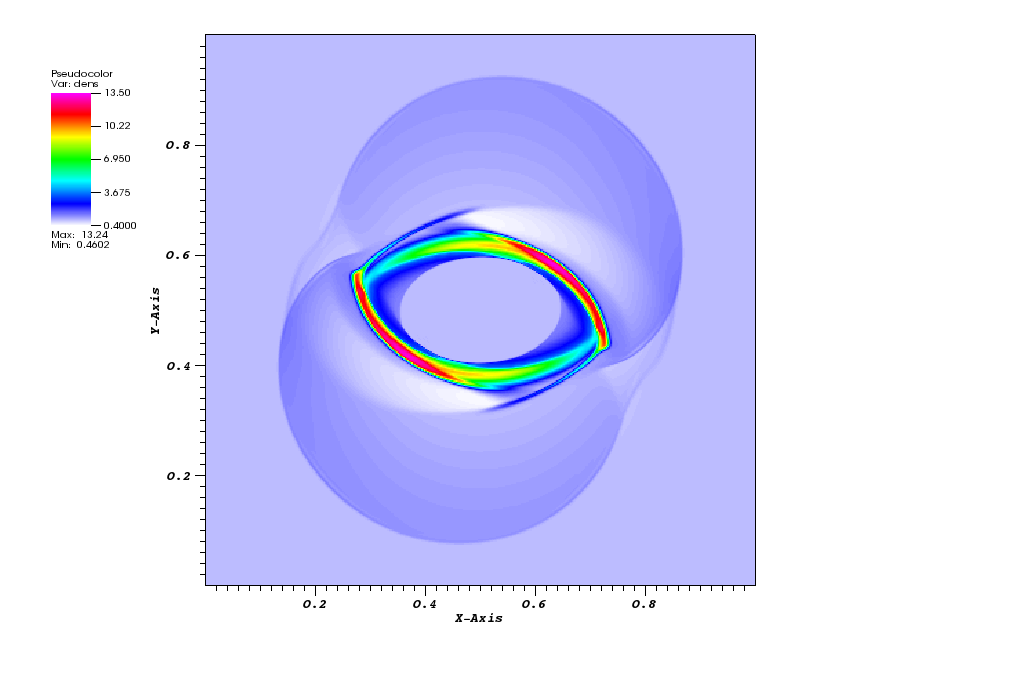}}
\hspace{-0.5in}
\subfigure[][]{
\includegraphics[width=2.85in, trim= 2.3in 1.1in 1.in 0.4in,clip=true]{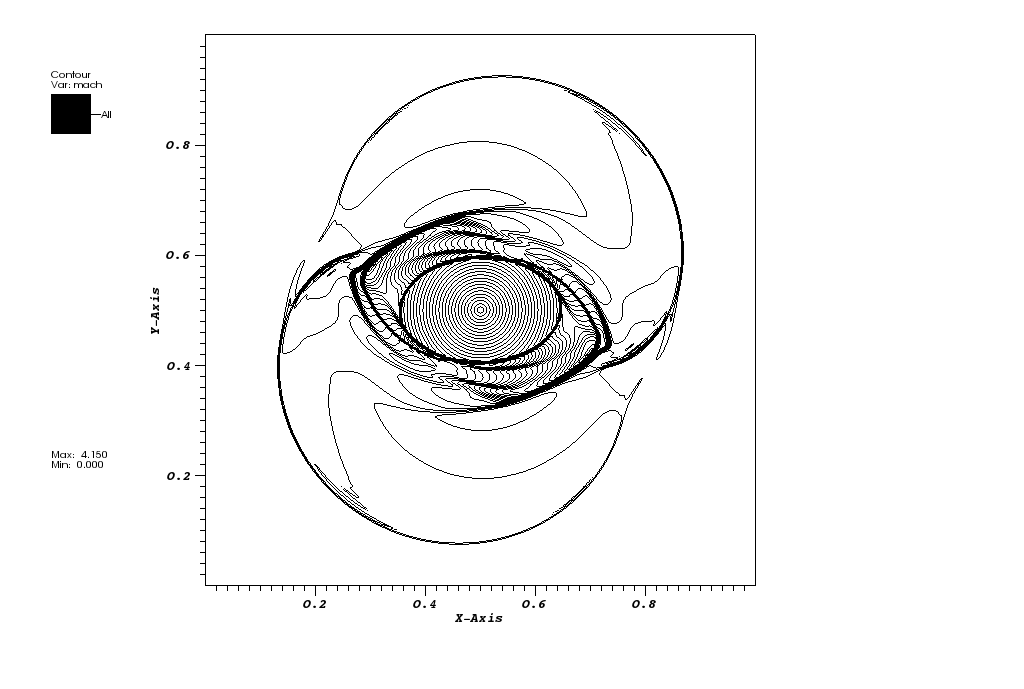}}\\
\vspace{-0.05in}
\subfigure[][]{
\includegraphics[width=3.0in, trim= 0.7in 1.1in 2.in 0.4in,clip=true]{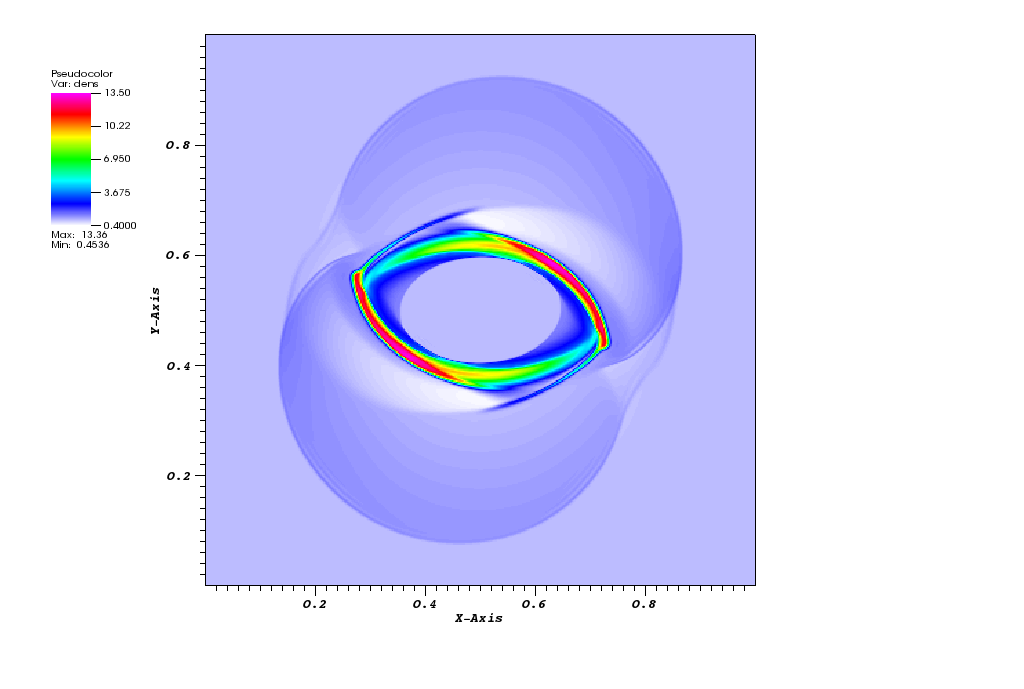}}
\hspace{-0.5in}
\subfigure[][]{
\includegraphics[width=2.85in, trim= 2.3in 1.1in 1.in 0.4in,clip=true]{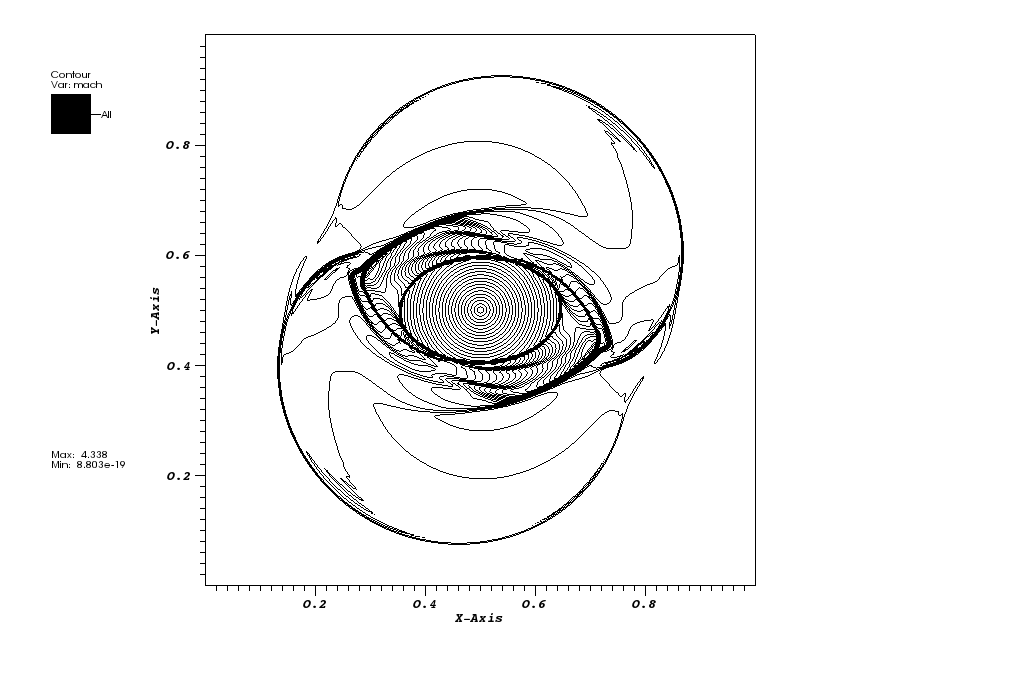}}
\end{tabular}
\caption{2D MHD Rotor problem. The panels on the left column show the density at $t=0.15$, and 
the panels on the right show the 30 contour lines of the Mach number at the same time.
(a) $\rho$ using PPM, (b) Mach number using PPM,
(c) $\rho$ using WENO-JS + CharTr, (d) Mach number using WENO-JS + CharTr,
(e) $\rho$ using PCM, and (f) Mach number using PCM.}
\label{Fig:MHD_rotor}
\end{figure}

Next, we consider the MHD rotor problem
\cite{Balsara1999,Toth2000}. 
As the problem has been discussed by various people we rather focus on discussing the solution of PCM here.
We use the same setup conditions as described in \cite{lee2009unsplit}.
Exhibited in Fig. \ref{Fig:MHD_rotor}(e) and Fig. \ref{Fig:MHD_rotor}(f) are respectively the density and the 30 contour lines
of the Mach number on $400\times 400$ cells, both at $t=0.15$. With minor discrepancies, we see that the PCM solution successfully
demonstrates its ability to solve MHD flows in multiple space dimensions. 
To test PCM for multidimensional MHD flows, we integrated the PCM algorithm in
the MHD scheme \cite{lee2009unsplit, lee2013solution} 
of the FLASH code \cite{fryxell2000flash,dubey2009extensible,dlee_flash}.
Of noteworthy point is that the contour lines of the Mach number in Fig. \ref{Fig:MHD_rotor}(f)
remain concentric in the central region without any distortion from the near-perfect symmetry. 
In all runs the HLLD Riemann solver \cite{miyoshi2005multi} was used with a fixed value $C_{\mbox{cfl}}=0.8$.
The PPM run used the MC slope limiter for monotonicity.

\subsection{3D Tests}

Lastly, for 3D cases, we have selected three test problems in MHD in such a way that we can
fully quantify the performance of PCM both for assessing its convergence rate in 3D
and for verifying its code capability in discontinuous flows.
\subsubsection{3D Convergence Test}

\paragraph{\underline{Alfv\'{e}n Wave Convergence Test -- UG}}
We solve the circularly polarized Alfv\'{e}n Wave propagation problem \cite{lee2013solution,gardiner2008unsplit}
as  our first 3D test problem to quantify the PCM's order of accuracy in full 3D. 
The computational domain is resolved on $2N_x \times N_y \times N_z$ grid cells, where 
we choose $N_x = N_y = N_z = 8, 16, 32$ and $64$ for the grid convergence study.
As in \cite{lee2013solution} we ran the same two configurations of the wave mode that are
the standing wave mode and the traveling wave mode until $t=1$. 
In both we choose the Roe Riemann solver with $C_{\mbox{cfl}}=0.95$. 

\begin{figure}[pbht!]
\centering
\begin{tabular}{ccc}
\subfigure[][]{
\includegraphics[width=2.5in, trim= 0.5in 0.0in 0.in 0.0in,clip=true]{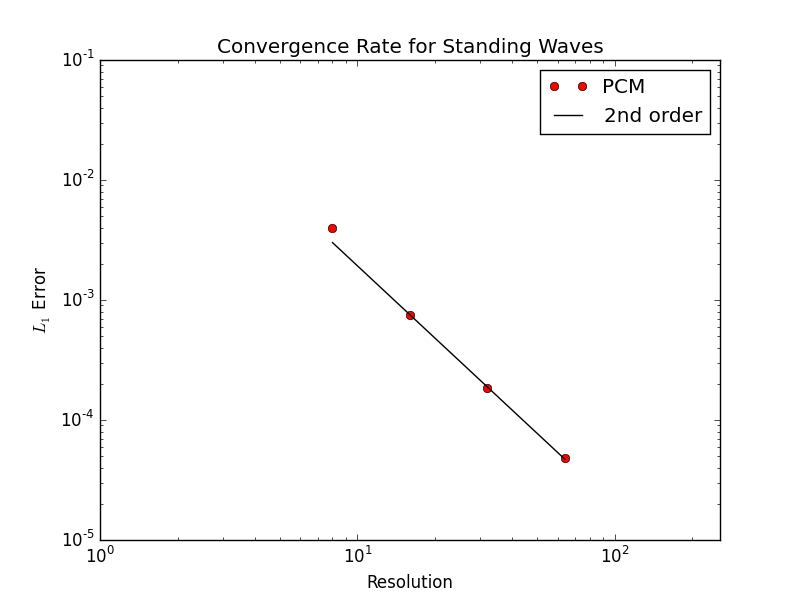}}
\subfigure[][]{
\includegraphics[width=2.5in, trim= 0.5in 0.0in 0.in 0.0in,clip=true]{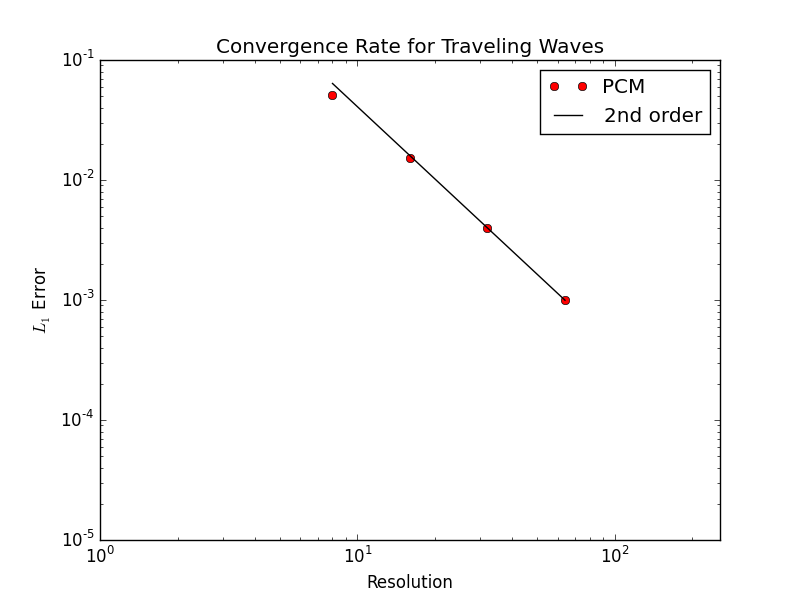}}
\end{tabular}
\caption{The circularly polarized Alfv\'{e}n wave convergence rate for both
the standing and traveling wave problems using PCM combined with the Roe Riemann solver.}
\label{Fig:3D_conv}
\end{figure}

Respectively, Fig. \ref{Fig:3D_conv}(a) and Fig. \ref{Fig:3D_conv}(b) are the $L_1$ numerical errors on a logarithmic scale
for the standing wave case and the traveling wave case. We observe that the rate of PCM convergence 
in  3D is second-order as expected, which agrees with the results reported in \cite{lee2013solution}.
One difference is noted in the standing wave case in Fig. \ref{Fig:3D_conv}(a) that 
PCM's $L_1$ error in each grid resolution
is much lower than those obtained 
with PPM + HLLD + F-CTU with $C_{\mbox{cfl}}=0.95$
in \cite{lee2013solution}. However, the magnitudes of the PCM error in the traveling wave case 
in Fig. \ref{Fig:3D_conv}(b) look pretty much similar to the equivalent run in \cite{lee2013solution}.

\subsubsection{3D Discontinuous Tests}

\paragraph{\underline{(a) 3D MHD Blast -- UG}}
\begin{figure}[pbht!]
\centering
\begin{tabular}{ccc}
\subfigure[][]{
\includegraphics[width=2.5in, trim= 0.4in 0.0in 0.in 0.5in,clip=true]{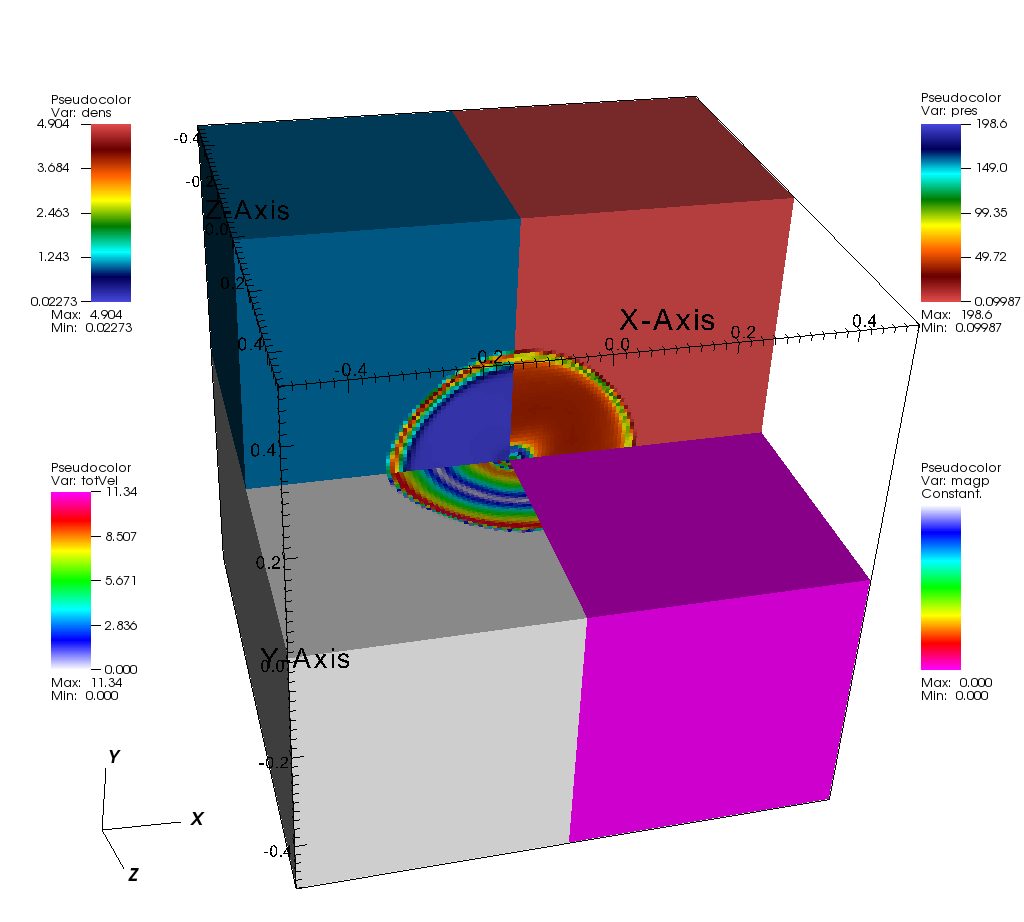}}
\subfigure[][]{
\includegraphics[width=2.5in, trim= 0.4in 0.0in 0.in 0.5in,clip=true]{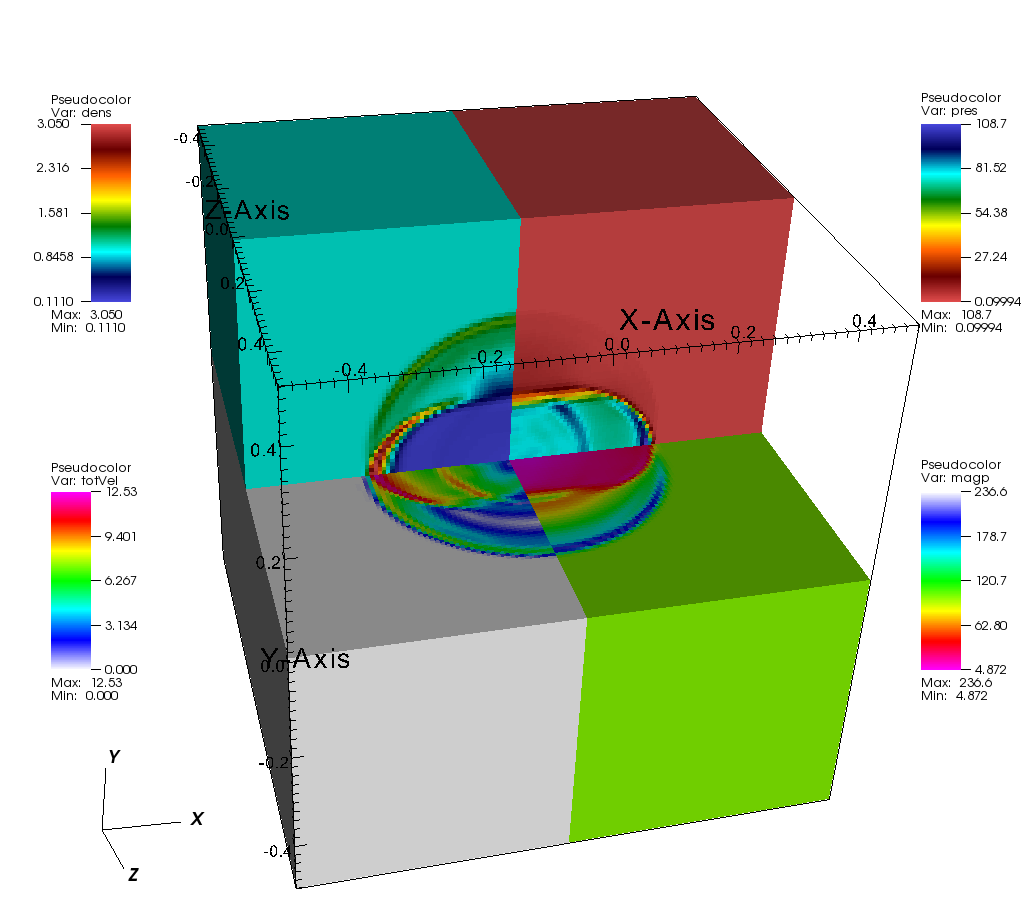}}\\
\subfigure[][]{
\includegraphics[width=2.5in, trim= 0.4in 0.0in 0.in 0.5in,clip=true]{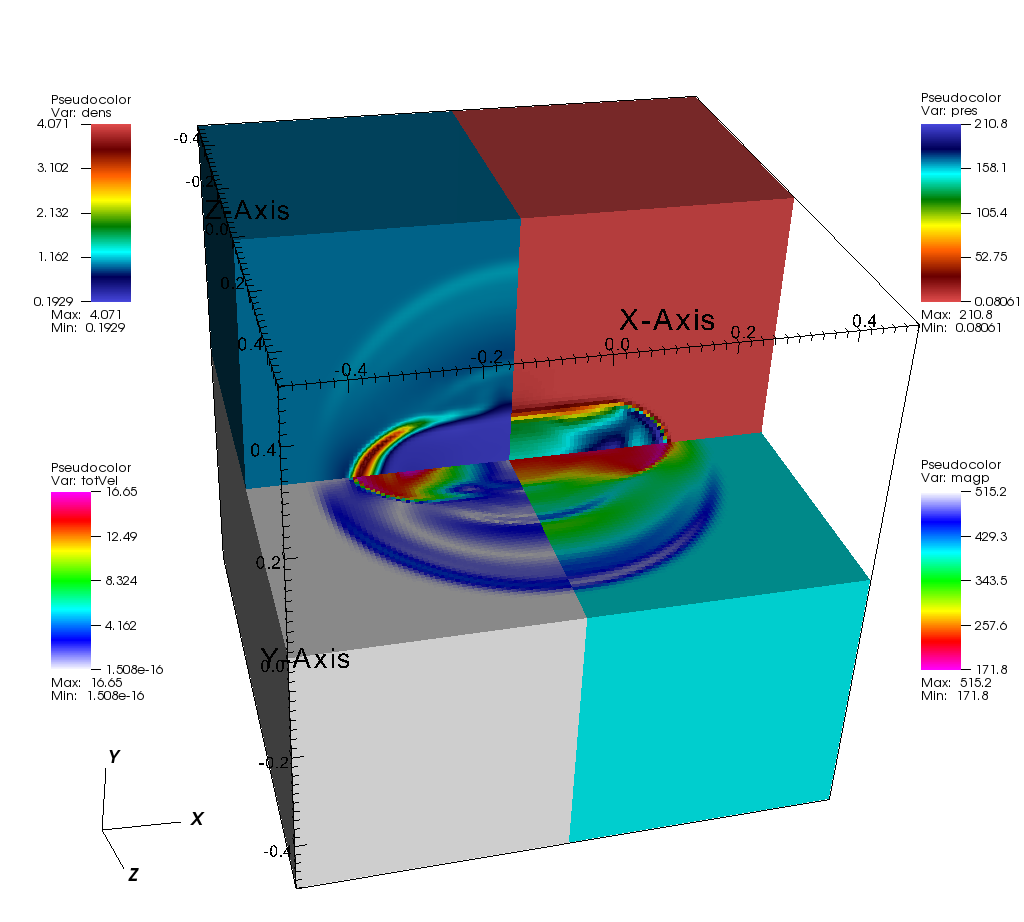}}
\end{tabular}
\caption{The 3D MHD blast problem. All results used the PCM scheme with the hybrid Riemann solver 
using $C_{\mbox{cfl}}=0.8$. 
(a) $B_x=0$, (b) $B_x=\frac{50}{\sqrt{4\pi}}$, and (c) $B_x=\frac{100}{\sqrt{4\pi}}$.
In each panel, we show four different plasma quantities, gas pressure in the top right quadrant, density in top left, 
total velocity in bottom left, and magnetic pressure in bottom right. Each corresponding color bar is shown immediately next to the
corresponding quantity.}
\label{Fig:3D_blastBS}
\end{figure}

We consider the 3D variant of the MHD blast problem by adopting the setup conditions in
\cite{lee2013solution} to demonstrate the three-dimensional
propagation of strong MHD shocks using the PCM algorithm.
The original 2D version of the spherical blast wave problem was studied in \cite{koessl1990numerical},
and later various people adopted the similar setup conditions 
\cite{stone2008athena,lee2009unsplit,lee2013solution,Balsara1999,zachary1994higher,balsara2015divergence,
ziegler2011semi,Mignone2010a,li2008high,kawai2013divergence,londrillo2000high} 
for their code verifications in strongly magnetized shock flows.

We display four different fluid variables in each panel in Fig. \ref{Fig:3D_blastBS}. 
From the top right quadrant to the bottom right quadrant in counter clockwise direction, 
we show the gas pressure $p$, the density $\rho$, the total velocity $U=\sqrt{u^2+v^2+w^2}$, and
the magnetic pressure $B_p$, all plotted at $t=0.01$. The grid resolution
$128\times 128\times 128$ as well as all other parameters are 
chosen as same as in \cite{lee2013solution}
in order to provide a direct comparison.

We tested PCM in three different plasma conditions defined by the three different 
strengths of $B_x=0, \frac{50}{\sqrt{4\pi}}$, and $\frac{100}{\sqrt{4\pi}}$, as displayed in
Fig. \ref{Fig:3D_blastBS}(a)  $\sim$ Fig. \ref{Fig:3D_blastBS} (c). Of particular interest to note is 
with the initial low plasma $\beta$ conditions in the last two cases, $\beta=1\times 10^{-3}$
and $2.513\times 10^{-4}$, respectively. 
On the other hand, the first setup in Fig. \ref{Fig:3D_blastBS}(a) produces the 
non-magnetized plasma flow, hence it allows us to test the PCM algorithm
in the pure hydrodynamical limit in 3D.
As clearly seen, all results have produced confidently accurate solutions.
We also note that PCM has produced larger values of extrema in each variable than those
reported in \cite{lee2013solution}, without exhibiting any unphysical oscillations.
This test demonstrate that the PCM algorithm is well-suited for simulating low-$\beta$ flows
in full 3D.

\paragraph{\underline{(b) Magnetic Field Loop Advection -- UG}}
\begin{figure}[pbht!]
\centering
\begin{tabular}{ccc}
\subfigure[][]{
\includegraphics[width=2.5in, trim= 0.5in 0.0in 0.in 0.5in,clip=true]{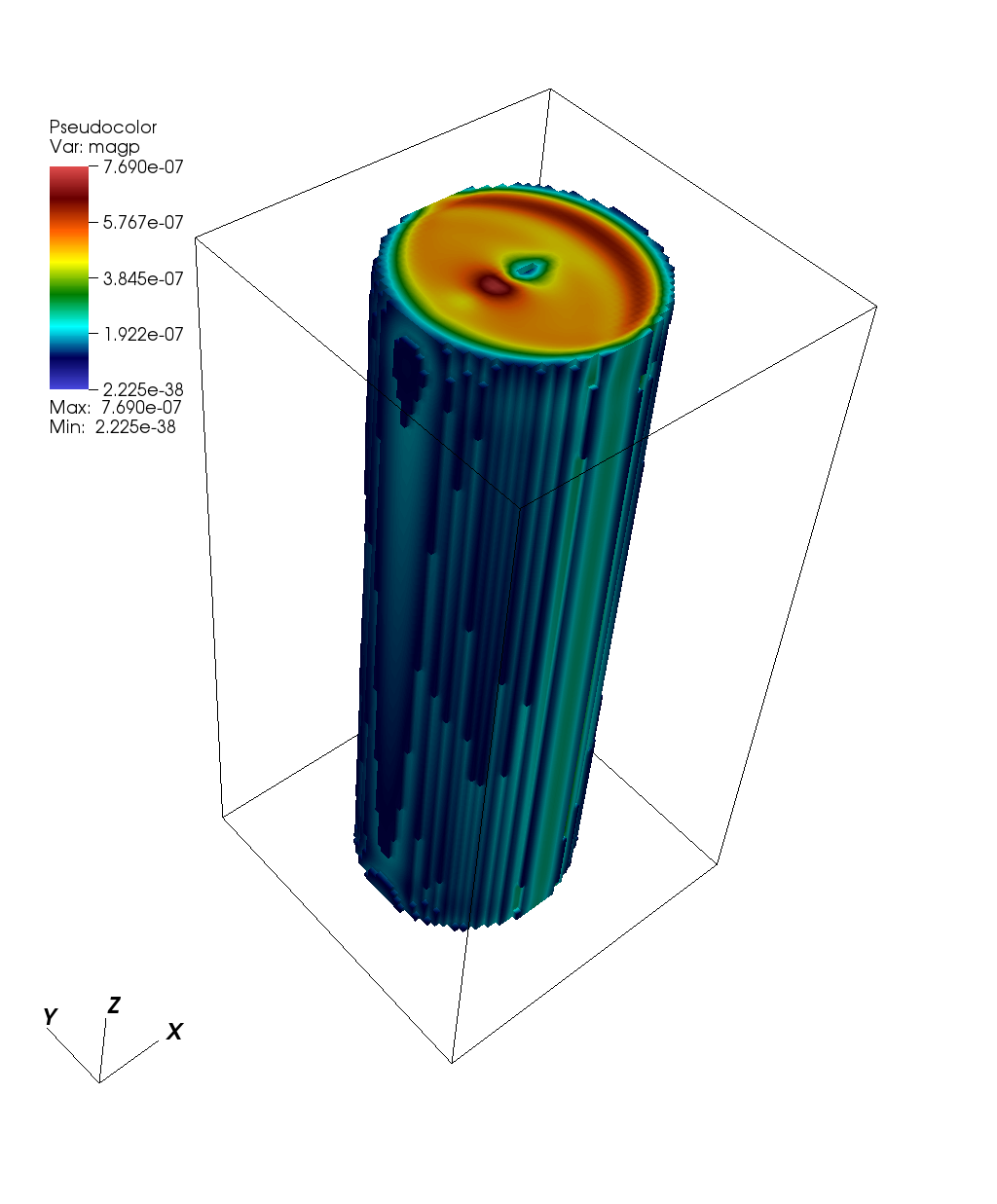}}
\subfigure[][]{
\includegraphics[width=2.5in, trim= 0.5in 0.0in 0.in 0.5in,clip=true]{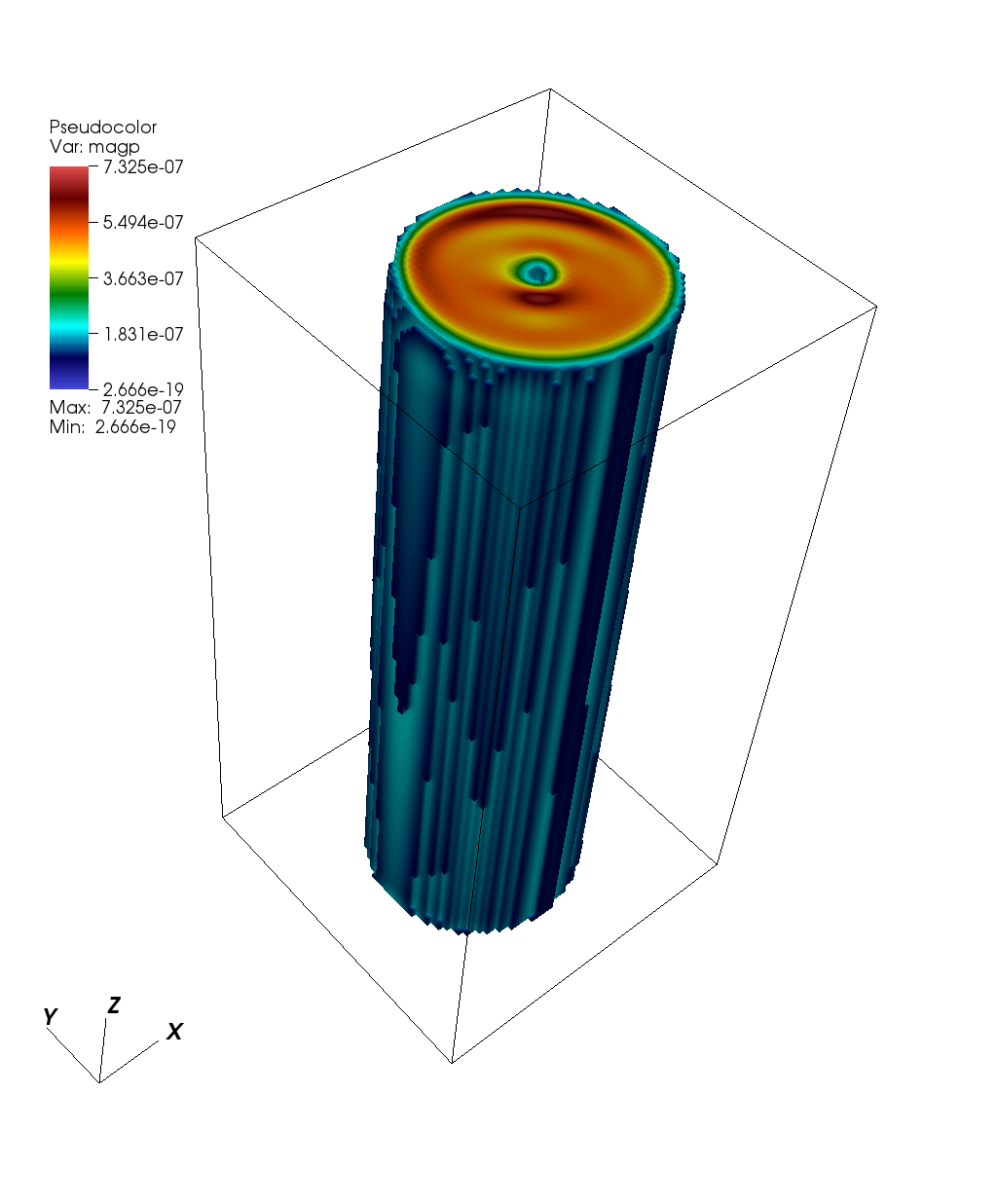}}\\
\subfigure[][]{
\includegraphics[width=2.5in, trim= 0.5in 0.0in 0.in 0.5in,clip=true]{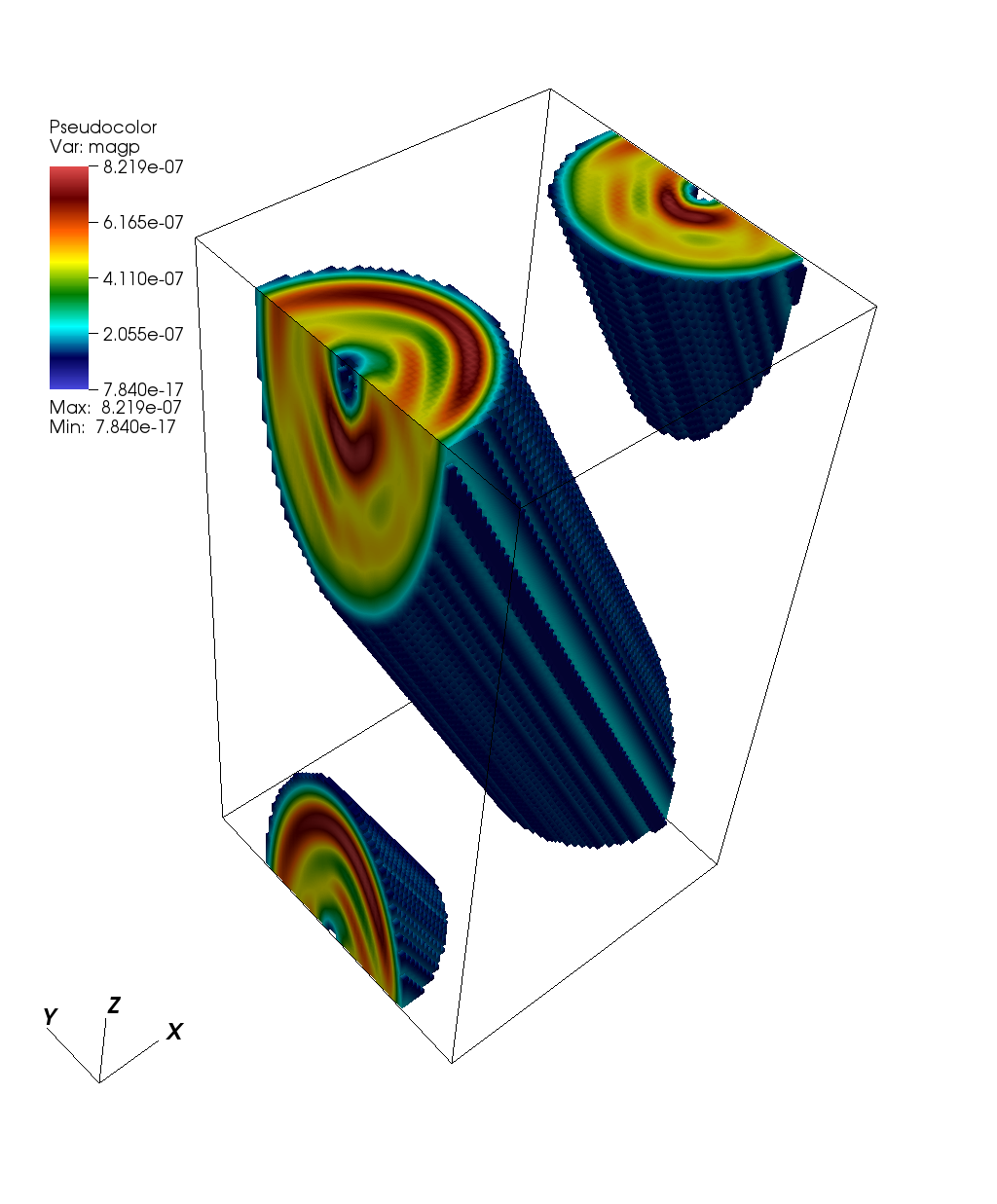}}
\end{tabular}
\caption{The 3D field loop advection using PCM.
(a) the small angle advection with $\theta \approx 0.537^\circ$ at $t=2$,
(b) the large angle advection at $t=2$ using $\bU=(1,1,2)^T$, and
(c) the standard Gardiner-Stone advection at $t=1$. All runs were calculated on
$64 \times 64 \times 128$ cells using the upwind-MEC.}
\label{Fig:3D_fieldloop}
\end{figure}

Since this problem was originally studied and reported in \cite{gardiner2005unsplit}, the problem has become
a popular benchmark case among various code developers 
to demonstrate their MHD algorithms' capabilities in advecting the initial
field loop which is weakly magnetized with a very high plasma $\beta=2\times 10^6$. 
The problem is known to be challenging \cite{gardiner2008unsplit,gardiner2005unsplit}, however, 
many have demonstrated that their codes can successfully produced  comparable results
\cite{lee2009unsplit,lee2013solution,balsara2010multidimensional,li2008high,balsara2013efficient,
kappeli2011fish,stone2009simple}.
In addition to the two original setups \cite{gardiner2008unsplit,gardiner2005unsplit}
where the initial field loops advect with the angle diagonal to the domain,
Lee \cite{lee2013solution} recently reported that a small-angle advection
is much more challenging. As an example, Lee adopted the advection angle 
$\theta \approx 0.573^\circ$ relative to the $x$-axis
for the small-angle advection case in both 2D and 3D. 
The study found that a proper amount of multidimensional numerical dissipation plays a key role in
maintaining the clean small-angle advection, and designed the algorithm called
{\it{upwind-MEC}}.
Here we repeat all three configurations (two large-angle advection cases
and one small-angle advection case) by following the same setups in \cite{lee2013solution}.
All the results in Fig. \ref{Fig:3D_fieldloop} were obtained using PCM and the Roe Riemann solver
with $C_{\mbox{cfl}}=0.8$ on $64\times 64 \times 128$ cells.

First, Fig. \ref{Fig:3D_fieldloop}(a) shows the small-angle advection 
with $\theta \approx 0.573^\circ$ relative to the $x$-axis with the velocity fields
given by $\bU=(\cos\theta, \sin\theta, 2)^T$. 
Compared to this, in Fig. \ref{Fig:3D_fieldloop}(b), we use 
$\bU=(1,1,2)^T$ which yields the large-angle advection. In both cases
the tilt angle $\omega$ (see \cite{lee2013solution} for details) is set to be same as $\theta$.
As manifested, both runs cleanly preserve the initial geometry of the field loop, 
convincing us that the PCM algorithm is robust and accurate in this challenging problem.
As a final test we also perform the standard field loop advection setup of
Gardiner and Stone \cite{gardiner2008unsplit}. The result is shown in Fig. \ref{Fig:3D_fieldloop}(c).
We see clearly that the PCM algorithm has produced well-behaving, accurate and confident
solutions in this test. The results in Fig. \ref{Fig:3D_fieldloop} can be directly compared to the results
reported in \cite{lee2013solution}.

\section{Conclusions}
\label{Sec:conclusions}
We summarize key features of the PCM algorithm studied in this paper.
\begin{itemize}
\item We have presented a new high-order finite volume scheme for the solutions of the compressible
gas dynamics and ideal MHD equations in 1D. This baseline 1D algorithm uses piecewise cubic polynomials
for spatial reconstruction by adopting the non-oscillatory approximations of the fifth order WENO schemes
to determine the unique piecewise cubic polynomial on each cell. To provide the nominal fifth-order accuracy
in space, we have developed a new non-oscillatory WENO-type reconstruction for $q'_{C,i}$ approximation.
The new approach makes use of the two parabolic polynomials, termed as {\it{PPM-Build}}, to achieve
fourth-order accuracy in establishing $q'_{C,i}$ approximation.
\item We have formulated a new fourth-order temporal updating scheme, all integrated in PCM by design, 
based on the simple predictor-corrector type characteristic tracing approach.
The overall solution accuracy of the baseline 1D PCM scheme, combining both spatially and temporally, 
seemingly converges with fifth-order. We show the PCM scheme compares greatly with the spatially fifth-order WENO
integrated with RK4, demonstrating even smaller $L_1$ errors in PCM. 
\item A comparison of the computational expenses of PCM, PPM and WENO-JS + RK4 in 1D
reveals that PCM has a superior advantage over the fifth-order counterpart WENO-JS + RK4
by a factor of 1.71. 
\item We have integrated the baseline 1D PCM algorithm for multidimensional cases by adopting
the simple dimension-by-dimension approach. As anticipated, this approach yields at most
second-order accurate solutions in multidimensional simulations of smooth flows. 
In the presence of  flow discontinuities and shocks, however, 
the results obtained with the present simple multidimensional PCM extension 
shows a great level of confidence in predicting numerical solutions of hydrodynamics and MHD.
An approach to extend the fifth-order property of the baseline 1D PCM to multiple spatial dimensions
will be further investigated in our future work.
\end{itemize}

\section{Acknowledgements}
The software used in this work was in part developed by the DOE NNSA-ASC OASCR Flash Center at the University of Chicago.
D. Lee also gratefully acknowledges the FLASH group for supporting the current work.
\newpage
\section*{References}

\bibliography{mybibfile}


\end{document}